\documentclass[11pt,a4paper]{article}
\usepackage{authblk}
\usepackage{graphicx}
\usepackage{caption}
\usepackage{float}
\usepackage{algorithm}
\usepackage[noend]{algorithmic}
\graphicspath{figures/}
\usepackage{times}
\usepackage{epsfig}
\usepackage{subfigure}
\usepackage{multirow}
\usepackage{amsmath}
\usepackage{mathtools}
\usepackage{amsfonts}
\usepackage{amssymb}
\usepackage{amscd}
\usepackage{relsize}
\usepackage{balance}
\usepackage{soul,xcolor}
\setstcolor{red}
\newfloat{algpseudocode}{t}{lop}
\newtheorem{theorem}{Theorem}[section]

\newtheorem{definition}{Definition}[section]
\newtheorem{proposition}{Proposition}[section]

\makeatletter
\DeclareRobustCommand*\cal{\@fontswitch\relax\mathcal}
\makeatother

\newcommand{\comments}[1]{}

\newtheorem{mydef}{Definition}

\newtheorem{mylemma}{Lemma}

\newcommand{\leftsemijoin}{\mbox{$\mathrel{\raise1pt\hbox{\vrule height5pt
depth0pt width0.6pt\hskip-1.5pt$>$\hskip -2.5pt$<$}}$}}

\begin{document}
\date{}

\title{Evaluating Hybrid Graph Pattern Queries Using Runtime Index Graphs}

%\title{Efficient Evaluation of Hybrid Graph Pattern Queries Using Lightweight Index Structures}

%\title{Exploiting Summaries to Efficiently Evaluate Hybrid Patterns on Large Data Graphs\thanks{The research of the first author was supported by the National Natural Science Foundation of China under Grant No. 61872276.}}
%\title{Leveraging Graph Simulations for Evaluating Tree Pattern Queries on Large Graphs}

\author[1]{Xiaoying Wu\thanks{xiaoying.wu@whu.edu.cn}}
\author[2]{Dimitri Theodoratos\thanks{dth@njit.edu}}
\author[3]{Nikos Mamoulis\thanks{nikos@cs.uoi.gr}}
\author[2]{Michael Lan\thanks{mll22@njit.edu}}
\affil[1]{School of Computer, Wuhan University, China}
\affil[2]{New Jersey Institute of Technology, USA}
\affil[3]{University of Ioannina, Greece}
%\affil[2]{New Jersey Institute of Technology, USA}

\maketitle

%\textbf{Abstract.}

\begin{abstract}

Graph pattern matching is a fundamental operation for the analysis and exploration of data graphs. In this paper, we present a novel approach for efficiently finding homomorphic matches for hybrid graph patterns, where each pattern edge may be mapped either to an edge or to a path in the input data, thus allowing for higher expressiveness and flexibility in query formulation.  A key component of our approach is a lightweight index structure that leverages graph simulation to compactly encode the query answer search space.  The index can be built on-the-fly during query execution and does not have to persist on disk. Using the index, we design a multi-way join algorithm to enumerate query solutions without generating any potentially exploding intermediate results. We demonstrate through extensive experiments that our approach can efficiently evaluate a wide range / broad spectrum of graph pattern queries and greatly outperforms existing approaches and recent graph query engines/systems.

\end{abstract}

%\category{H.2.4}{Database Management}{Systems}[Query processing,Textual databases]
%\terms{Algorithms}
%\keywords{mixed pattern, pattern matching, graph simulation, graph homomorphism}

%\vspace{1mm}
%\noindent
%{\bf Categories and Subject Descriptors:} H.2.4 {[Database Management]}: {Systems{\em $-$query processing, textual  databases}}

%\vspace{1mm}
%\noindent
%{\bf General Terms:} Algorithms.

%\vspace{1mm}
%\noindent
%{\bf Keywords:} XPath query evaluation, XML

\section{Introduction}
\label{sec:intro}

Graphs model complex relationships between entities in a multitude of modern applications. A fundamental operation for querying, exploring and analyzing graphs is
finding the matches of a query graph pattern in the data graph.  Graph  matching  is  a building block of search and analysis tasks  in  many  application  domains  of data science,  such  as  social network analysis \cite{ChingEKLM15}, protein interaction analysis \cite{PrzuljCJ06}, cheminformatics \cite{SmalterHJL10}, knowledge bases \cite{dbpedia,SuchanekKW07} and road network management \cite{netrepository}.

Existing approaches are characterized by: (a) the type of edges the patterns have, and (b) the type of morphism used to map the pattern to the data graph. An edge in a  query  pattern  can  be  either  a {\em direct edge}, which represents a direct relationship in the data graph (edge-to-edge  mapping) \cite{Ullmann76,CordellaFSV04,AbergerTOR16,BhattaraiLH19,MhedhbiS19,Sun020,SunSC0H20,MhedhbiKS21,KimCP0HH21},  or  a {\em reachability edge}, which represents a node reachability relationship in the data graph (edge-to-path mapping) \cite{FanLMWW10,ChengYY11,ZengH12,LiangZJZH14,FletcherPP16}. The morphism determines how a pattern is mapped to the data graph and, in this context, it can be an isomorphism (injective function) \cite{Ullmann76,BhattaraiLH19,MhedhbiS19,Sun020} or a homomorphism (unrestricted function) \cite{FanLMWW10,ChengYY11,LiangZJZH14,AbergerTOR16,MhedhbiKS21}. Graph simulation \cite{HenzingerHK95} and its variants \cite{FanLMTWW10,MaCFHW14} are other ways to match patterns to data graphs.

Earlier contributions considered isomorphisms and edge-to-edge mappings, %\cite{Ullmann76,CordellaFSV04}
while more recent ones focus on homomorphic mappings. %\cite{FanLMWW10,ChengYY11,LiangZJZH14}.
By allowing edge-to-path mapping on graphs, patterns with reachability edges are able to extract matches “hidden” deeply within large graphs which might be missed by patterns with only direct edges. On the other hand, the patterns with direct edges can discover important direct connections in the data graph which can be missed by patterns with only reachability edges. We propose, in this paper, a general framework that considers patterns which allow both direct and reachability edges, which are called {\em hybrid} graph patterns. This framework incorporates the benefits from both types of edges.

\begin{figure}[!t]
    \centering%
     \scalebox{.7}{ \epsfig{file=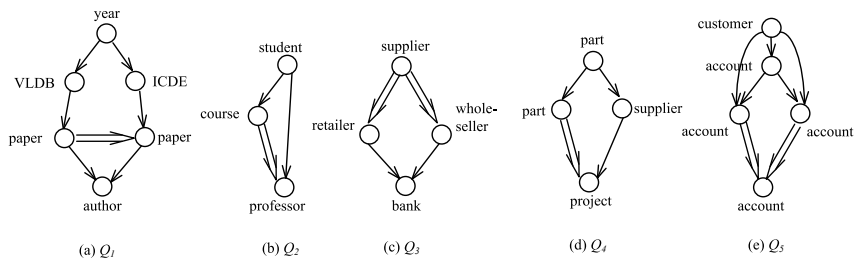}}%
       \vspace*{-1ex}
     \caption{Hybrid graph pattern examples.}
      \label{fig:introeg}
\vspace*{-3ex}
\end{figure}

Fig. \ref{fig:introeg} shows five example hybrid graph patterns from different application domains.  Double line edges denote reachability edges, while single line edges denote direct edges. The pattern graph in Fig. \ref{fig:introeg}(a) is a pattern query over a citation network which categorizes paper by years and then by venues/journals. It finds those authors who at the given year have a VLDB paper that directly or indirectly cites a ICDE paper at the same year by that same author. The pattern graph in Fig. \ref{fig:introeg}(b) is a pattern query over a university data graph. It finds students who is a teaching assistant of a course, and that course has a prerequisite course taught by a professor who is the advisor of that student. The pattern graph in Fig. \ref{fig:introeg}(c) is a query over a service provider data graph searching for a {\em supplier}, a {\em retailer}, a {\em whole-seller}, and a {\em bank} such that the {\em supplier}  directly or indirectly supplies products to the {\em retailer} and the {\em whole-seller}, and both of them receive directly services from the same bank. The pattern graph in Fig. \ref{fig:introeg}(d) finds parts which have subparts used by a project, and those parts are provided by a supplier who serves the same project. The pattern graph in Fig. \ref{fig:introeg}(e) is a query over a service provider data graph looking for individuals who performed a pattern of direct and indirect (sequences of) money transfers between legal or illegal accounts that can suggest a money laundering activity.

Graph pattern matching is an NP-hard problem, even for isomorphic matching of patterns with only direct edges \cite{GareyJ79}. Finding the homomorphic matches of query patterns which involve reachability edges on a data graph is more challenging (technically, a homomorphism is defined for edge-to-edge mapping but we generalize the term later so that it refers also to edge-to-path mapping). Reachability edges in a query pattern increase the number of results since they are offered more chances to be matched to the data graph compared to direct edges. Furthermore, finding matches of reachability edges to the data graph is an expensive operation and requires the use of a node readability index \cite{CohenHKZ03,JinXRF09,SuZWY17}. Despite the use of reachability indexes, evaluating reachability  edges remains a costly operation. Existing approaches for evaluating  pattern queries with reachability relationships produce a huge number of intermediate results (that is, results for subgraphs of the query graph which do not appear in any result for the query). As a consequence, existing approaches  do not scale satisfactorily when the size of the data graph increases.

Existing graph pattern matching algorithms can be broadly classified into the following two approaches: the {\em join-based} approach ({\em JM}) \cite{ChengYY11,AbergerTOR16,MhedhbiKS21} and the {\em tree-based} approach ({\em TM}) \cite{ZengH12,BiCLQZ16,HanKGPH19,BhattaraiLH19,SunSC0H20}. Given a graph pattern query $Q$, {\em JM} first decomposes $Q$ into a set of subgraphs. The query is then evaluated by matching each subgraph against the data graph and joining together these individual matches. Unlike {\em JM}, {\em TM} first decomposes or transforms $Q$ into one or more tree patterns using various methods, and then uses them as the basic processing unit. Both {\em JM} and {\em TM} suffer from a potentially exploding number of intermediate results %(that is, results for subgraphs of the query graph which do not appear in any result for the query)
which can be substantially larger than the output size of the query, thus spending a prohibitive amount of time on examining false positives. As a consequence, existing approaches  do not scale satisfactorily when the size of the data graph increases. %Our experimental results also reveal that
Also, query engines of contemporary graph DBMS are unable to handle efficiently graph pattern queries containing reachability \mbox{relationships  \cite{WuTSL22}.}

In this paper, we address the problem of evaluating hybrid graph patterns using homomorphisms over a data graph. This is a general setting for graph pattern matching.  We develop a new graph pattern matching framework, which consists of two phases: (a) the \emph{summarization phase}, where a query-dependent summary graph is built on-the-fly to serve as a compact search space for the given query,  and (b) the \emph{enumeration phase}, where query solutions are produced using the summary graph.

\vspace*{1ex}\noindent{\bf Contribution}. The main contributions of the paper are as follows:
\begin{list}{}{\setlength{\leftmargin}{\parindent}\setlength{\parsep}{0cm}%
\setlength{\partopsep}{0cm}\setlength{\itemsep}{0cm}\setlength{\parskip}{0cm}%
\setlength{\labelwidth}{\parindent}%
\setlength{\topsep}{0cm}}

\item[$\bullet$] We propose the concept of {\em runtime index graph} (RIG) to encode all possible homomorphisms from a query pattern to the data graph. By losslessly summarizing the occurrences of a given pattern,  a RIG represents results more succinctly. A RIG graph can serve as a search space for the query answer. It can be efficiently built {\em on-the-fly} and does not have to persist on disk.

\item[$\bullet$] We develop a novel simulation-based technique called double simulation for identifying and excluding nodes of the data graph which do not participate in the query answer. We design an efficient algorithm to compute double simulations.  Using this filtering method, we build a refined RIG graph to further reduce the query answer search space.  We also present tuning strategies to improve the performance of double simulation computation and RIG construction.

\item[$\bullet$]  We design an effective join-based search ordering strategy for searching query occurrences. The search ordering strategy takes into account both the query graph structure and data graph statistics.

\item[$\bullet$]  We develop a novel algorithm for enumerating occurrences of graph pattern queries.  In order to compute the results, our algorithm performs multiway joins by intersecting node lists  and node adjacency lists in the runtime index graph. Unlike both the {\em JM} and {\em TM} methods, it avoids generating a potentially exploding number of intermediate results and has a small memory footprint.

\item[$\bullet$]  We integrate the above techniques to design a graph pattern matching algorithm, called \mbox{{\em GM}} and we run extensive experiments to evaluate its performance and scalability  on real datasets.  We compare {\em GM} with the {\em JM} and {\em TM} approaches. The results show that {\em GM} can efficiently evaluate graph pattern queries with varied structural characteristics and with tens of nodes on data graphs, and that it outperforms by a wide margin both {\em JM} and {\em TM}.

\end{list}

\begin{figure}[!t]
    \centering%
     \scalebox{.7}{ \epsfig{file=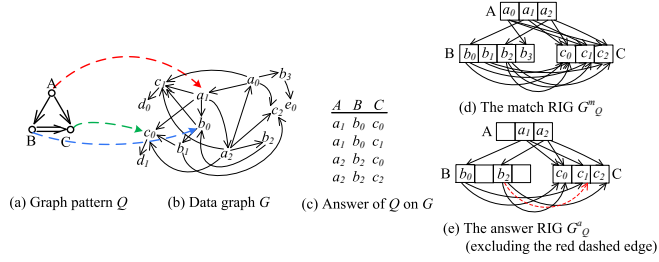}}%
       %\vspace*{-2ex}
     \caption{A hybrid graph pattern query $Q$, a data graph $G$, a homomorphism from $Q$ to $G$, the answer of $Q$ on $G$, and two runtime index graphs of $Q$ on $G.$}
     \label{fig:eg}
%\vspace*{-2ex}
\end{figure}
   	% \label{sec:intro
\section{Problem Definition}
\label{sec:problem}

In this section, we present the data model, graph pattern queries, edge-to-edge and edge-to-path mappings and homomorphisms. We also present related concepts that are needed for the results presented later.

\vspace*{1ex}\noindent{\bf Data Graph.} We assume that the data is presented in the form of a data graph defined below. We focus on directed, connected, and node-labeled graphs. All techniques in this paper can be readily extended to handle more general cases, such as disconnected graphs and multiple labels on nodes/edges.

\begin{definition}[Data Graph]
	A {\em data graph} is a directed node-labeled graph $G=(V,E)$ where $V$ denotes the set of nodes and $E$ denotes the set of edges (ordered pairs of nodes). Let ${\mathcal L}$ be a finite set of node labels. Each node $v$ in $V$ has a label $label(v) \in {\mathcal L}$ associated with it.
\end{definition}

%Graph $G$ is {\em dense} when the ratio $|E|/|V|$ is high, {\em sparse} otherwise.
Given a label $a$ in ${\mathcal L}$, the inverted list $I_a$ is the list of nodes in $G$ whose label is $a$. Fig. \ref{fig:eg}(b) shows a data graph $G$ with labels $a, b$, and $c$. Label subscripts are used to distinguish nodes with the same label. The inverted list of label $a$ in $G$ is  $I_a =\{a_0, a_1, a_2\}.$

\begin{definition}[Node reachability]
\label{def:reachability}
	A node $u$ is said to {\em reach} node $v$ in $G$, denoted by $u\prec v$, if there exists a path from $u$ to $v$ in $G$. Clearly, if ($u$, $v$) $\in E$,  then $u\prec v$. Abusing tree notation, we refer to $v$ as a {\em child} of $u$ (or $u$ as a {\em parent} of $v$) if ($u$, $v$) $\in E$, and $v$ as a {\em descendant} of $u$ (or $u$ is an {\em ancestor} of $v$) if $u\prec v$. \hfill$\Box$
\end{definition}

Given two nodes $u$ and $v$ in $G$, in order to efficiently check whether $u\prec v$, graph pattern matching algorithms use some kind of reachability indexing scheme. %In most reachability indexing schemes the data graph node labels are the entries in the index for the data graph \cite{SuZWY17}.
Our approach can flexibly use any indexing scheme to check node reachability. In order to check if $v$ is a child of $u$, some basic access method can be used; for example, adjacency lists of $G$.

\begin{definition}[Graph Pattern Query] A query is a connected directed graph $Q$. Every node $q$ in $Q$ has a label $label(q)$ from ${\mathcal L}$. There can be two types of edges in $Q$. %A {\em direct} (resp. {\em reachability}) edge denotes a direct (resp. reachability) structural relationship between the respective two nodes.
A graph pattern that contains both direct and reachability edges is a {\em hybrid} graph pattern. \hfill$\Box$ %We assume $Q$ is connected.
\end{definition}

\vspace*{1ex}\noindent{\bf Queries.} We consider  graph pattern queries that involve direct and/or reachability edges.

\begin{definition}[Graph Pattern Query] A query is a connected directed graph $Q$. Every node $q$ in $Q$ has a label $label(q)$ from ${\mathcal L}$. There can be two types of edges in $Q$. A {\em direct} (resp. {\em reachability}) edge denotes a direct (resp. reachability) structural relationship between the respective two nodes. A graph pattern that contains both direct and reachability edges is a {\em hybrid} graph pattern. \hfill$\Box$ %We assume $Q$ is connected.
\end{definition}

Intuitively, a direct edge in the query represents an edge in the data graph $G$. A reachability edge in the query represents the existence of a path in $G$. Fig. \ref{fig:eg}(a) shows a hybrid graph pattern query $Q$. Single line edges denote direct edges while double line edges denote reachability edges.

\vspace*{1ex}\noindent{\bf Homomorphisms.} Queries are matched to the data graph using homomorphisms.

\begin{definition}[Graph Pattern Homomorphism to a Data Graph] Given a graph pattern $Q$ and a data graph $G$, a {\em homomorphism} from $Q$ to $G$ is a function $h$ mapping the nodes of $Q$ to nodes of $G$, such that: (1) for any node $q\in Q$, $label(q)$ = $label(h(q))$; and (2) for any edge $e$=($p, q$) $ \in Q$, if $e$ is a direct edge, ($h(p),h(q)$) is an edge of $G$, while if $e$ is a reachability edge, $h(p) \prec h(q)$ in $G$.  \hfill$\Box$
\end{definition}

Fig. \ref{fig:eg}(a,b) shows a homomorphism $h$ of query $Q$ to the data graph $G$. Query edges $(A, B)$ and $(A, C)$ which are direct edges, are mapped by $h$ to an edge in $G$. Edge $(B, C)$ is reachability edge, which is  mapped by $h$ to a path of edges in $G$ (possibly consisting of a single edge).

\vspace*{1ex}\noindent{\bf Query Answer.} We call an {\em occurrence} of a pattern query $Q$ on a data graph $G$ a tuple indexed by the nodes of $Q$ whose values are the images of the nodes in $Q$ under a homomorphism from $Q$ to $G$.

\begin{definition}[Query Answer]
The {\em answer} of $Q$ on $G$, denoted as $Q(G)$, is a relation whose schema is the set of nodes of $Q$, and whose instance is the set of the occurrences of $Q$ under all possible homomorphisms from $Q$ to $G$.  \hfill$\Box$	
\end{definition}

Fig.\ref{fig:eg}(c) shows the answer of $Q$ on $G$.

\vspace*{1ex}\noindent\textbf{Problem statement.} Given a large directed graph $G$ and a pattern query $Q$, our goal is to efficiently find the answer of $Q$ on $G.$

   	% \label{sec:problem}

%\vspace*{-2ex}
\section{Transitive Reduction of Graph Pattern Queries}
\label{sec:reduction}

Edges in a hybrid graph pattern query $Q$ denote reachability relationships between pattern nodes. A reachability edge expresses a structural constraint more relaxed than a direct edge.  The reachability relationship implied by a reachability edge $e$ may possibly be derived from a path of edges  in $Q$ which does not include $e$. In this context, edge $e$ is redundant; it is not needed for determining the answer of $Q$.

\begin{figure}[!t]
    \centering%
     \scalebox{.6}{ \epsfig{file=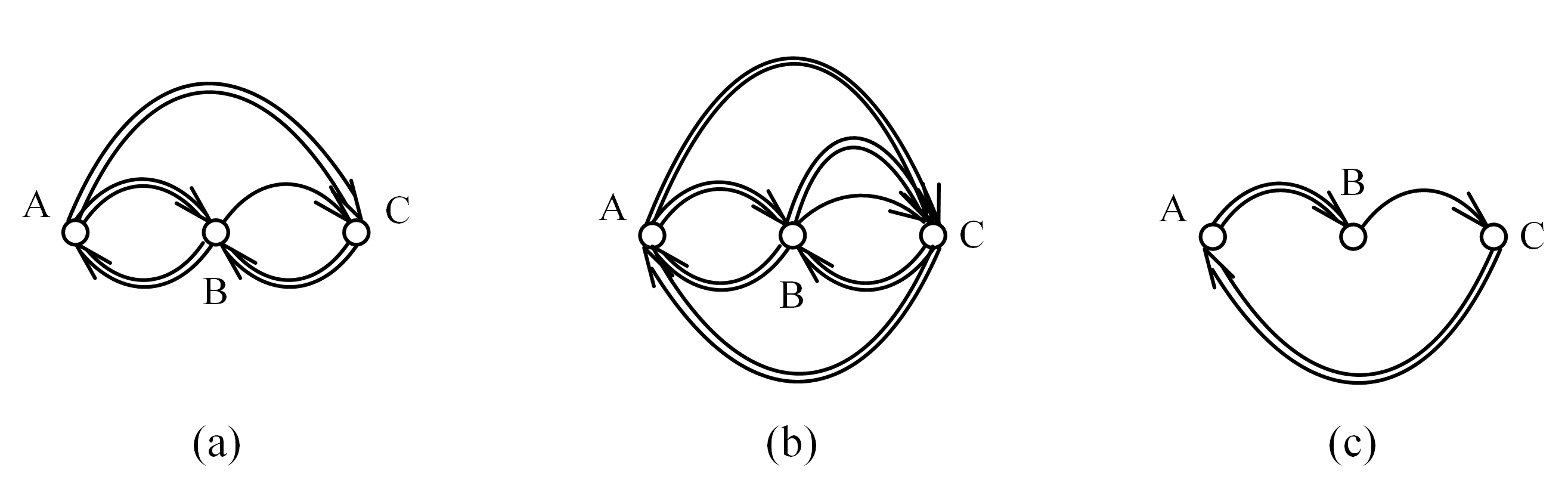}}%
       %\vspace*{-3ex}
     \caption{(a) A graph pattern query $Q$,  (b) the transitive closure of $Q$, (c) a transitive reduction of $Q$.}
      \label{fig:treg}
%\vspace*{-2ex}
\end{figure}

We say that two queries are {\em equivalent} if they have the same answer on any data graph. An edge in a query $Q$ is {\em redundant} if its removal from $Q$ results in a query which is equivalent to $Q$. A reachability edge $e = (x, y)$ in a query $Q$ is called {\em transitive} if there is a simple directed path from $x$ to $y$ (other than edge $e$) in $Q$. This path can contain direct and reachability edges. Clearly, transitive edges are redundant. Consider the graph pattern query $Q$ in Fig.~\ref{fig:treg}(a). Since there is a path $A \rightarrow B\rightarrow C$ from node $A$ to node $C$, reachability edge $(A, C)$ is transitive and, therefore, redundant.

Finding the occurrences of reachability edges requires  edge-to-path mappings in the data graph. This operation is more expensive than the edge-to-edge mappings required for finding the occurrences of direct edges.
% subgraph isomorphisms \cite{FanLMWW10}.
Given a graph pattern query, detecting redundant reachability edges and producing an equivalent query of minimal size without redundant edges can thus speedup the query evaluation process. For query evaluation purposes, we compute the transitive reduction of pattern queries, defined next. %In this section, we address the problem of pattern query reduction.

\begin{definition}
\label{def:transRed}

A query pattern graph $Q'$ is a {\em transitive
reduction} of a query pattern graph $Q$ whenever the following two conditions hold: (a) $Q'$ is equivalent to $Q$ and there is a reachability edge from node $x$ to node $y$ in $Q'$ if and only the reachability edge $(x, y)$ is the only path from $x$ to $y$ in $Q$, and (b) there is no equivalent query pattern graph with fewer reachability edges than $Q'$ satisfying condition (a).
\end{definition}

If $Q$ is acyclic, there is a unique transitive reduction of $Q$ which can be obtained by removing the transitive edges. This might not be the case if $Q$ has cycles.

To compute the transitive reduction of $Q$, we need the {\em transitive closure} of $Q$ which is a graph pattern query equivalent to $Q$ having a reachability edge (x, y) from node $x$ to node $y$ whenever $x \prec y$ in $Q$. The transitive closure of $Q$ can be constructed from $Q$ by exhaustively applying to $Q$ the following two inference rules:

\begin{list}{}{\setlength{\leftmargin}{1.2cm }\setlength{\parsep}{0cm}%
\setlength{\partopsep}{0cm}\setlength{\itemsep}{0cm}\setlength{\parskip}{0cm}%
\setlength{\labelwidth}{1.2cm }%
\setlength{\topsep}{0cm}}

  \item[(IR1) ]  If the child edge $(x, y)$ belongs to $Q$  then add the reachability edge $(x, y)$ to $Q$.
%  \item[(IR1) ]  $x/y\ \vdash\ x//y$
%  \item[(IR2) ]  $x//y,\ y//z\ \vdash\ x//z$
\item[(IR2) ]  If the reachability edges $(x, y)$ and $(x, z)$ belong to $Q$ then add the reachability edge $(x, z)$ to $Q$.
\end{list}

%We say that a query is in {\em full form} if it is equal to its closure.

%\vspace*{-1ex}

%\noindent

Computing a transitive reduction of a query can be done efficiently from its transitive closure by removing  transitive edges  \cite{AhoGU72}.

Fig. \ref{fig:treg}(b) shows the transitive closure of query $Q$ shown in Fig.  \ref{fig:treg}(a). Fig. \ref{fig:treg}(c) shows a transitive reduction of $Q.$

In the following discussion, we assume that graph pattern queries are equal to one of their transitive reductions.

  % \label{sec:reduction}
\section{A Lightweight Index as Compact Search Space}
\label{sec:rigGraph}

\subsection{Runtime Index Graph}
\label{subsec:rig}
% checking child relationship
%\vspace{-1ex}

Given a pattern query $Q$ and a data graph $G$,  we propose the concept of {\em runtime index graph} to serve as a search space of the answer of $Q$ on $G.$ We start be providing some definitions.
%compactly encode all possible homomorphisms of $Q$ in $G$.

The {\em match set} $ms(q)$ of a node $q$ in $Q$ is the inverted list $I_{label(q)}$ of the label of node $q$. A {\em match} of an edge $e = (p, q)$ in $Q$ is a pair $(u, v)$ of nodes in $G$ such that $label(p) = label (u)$, $label(q) = label (v)$ and: (a)~$u \prec v$ if $e$ is a reachability edge, while (b)~$(u, v)$ is an edge in $G$ if $e$ is a direct edge. The {\em match set} $ms(e)$ of $e$ is the set of all the matches of $e$ in $G$.

If $q$ is a node in $Q$ labeled by label $a$, an {\em occurrence} of $q$ in $G$ is the image $h(q)$ of $q$ in $G$ under a homomorphism $h$ from $Q$ to $G$. The {\em occurrence set of $q$ on $G$}, denoted as $os(q)$, is the set of all the occurrences of $q$ on $G$. This is a subset  of the match set $ms(q)$  %(the inverted list $I_a$)
containing only those nodes that occur in the answer of $Q$ on $G$ for $q$ (that is, nodes that occur in the column $q$ of the answer). For instance, the occurrence set of node $A$ of query $Q$ in Fig. \ref{fig:eg} is $\{a_1, a_2$\}. If $e = (p, q)$ is an edge in $Q$, an {\em occurrence} of $e$ in $G$ is a pair $(u, v)$ of nodes from $G$ such that $u = h(p)$ and $v = h(q)$, where $h$ is a homomorphism from $Q$ to $G$. The {\em occurrence set of $e$ on $G$}, denoted as $os(e)$, is the set of all the occurrences of $e$ on $G$. This is %the set of pairs $(u, v)$ of nodes in $G$ such that there is an occurrence $t$ of $Q$ on $G$ with $t.p = u$ and $t.q = v$ (that is, $os(e)$ is in
the projection of the answer of $Q$ on $G$ on the columns $p$ and $q$. Clearly, $os(e) \subseteq ms(e)$. In the example of Fig. \ref{fig:eg}, the occurrence set of the edge $(A, B)$ of query $Q$ is $\{(a_1, b_0), (a_2, b_2)\}$.

\begin{definition}[Runtime Index Graph]
A {\em runtime index graph (RIG)} of pattern query $Q$ over data graph $G$ is a $k$-partite graph $G_Q$ where $k$ is the number of nodes in $Q$.  For every node $q \in Q$, graph $G_Q$ has an independent node set, denoted $cos(q)$, such that $os(q) \subseteq cos(q) \subseteq ms(q)$. %Every node in $cos(q)$ is incident to an edge in $G_Q$ if $q$ is incident to an edge in $Q$.
Set $cos(q)$ is called the {\em candidate occurrence set} of $q$ in $G_Q$. For every edge $e_Q = (p, q)$ in $Q$, graph $G_Q$ has a set $cos(e_Q)$ of edges from nodes in $cos(p)$ to nodes in $cos(q)$ such that $os(e_Q) \subseteq cos(e_Q) \subseteq ms(e_Q)$. Set $cos(e_Q)$ is called the {\em candidate occurrence set} of $e_Q$ in $G_Q$.

\end{definition}

By definition, we can have many RIGs of a given query $Q$ on data graph $G$. Among them, the largest one is called the {\em match} RIG of $Q$ on $G$, denoted as $G_Q^m$, and  the smallest one is called the {\em answer} RIG of $Q$ on $G$, denoted as $G_Q^a$. For any edge $e$ in $Q$, the candidate occurrence set for $e$ in $G_Q^a$ is the occurrence set $os(e)$, while the candidate occurrence set for $e$ in $G_Q^m$ is the match set $ms(e)$. Figs.~\ref{fig:eg}(d) and (e), respectively, show the match RIG and the answer RIG for query $Q$ on the data graph $G$ in Fig. \ref{fig:eg}(b).

A RIG $G_Q$ losslessly summarizes all the occurrences of $Q$ on $G$ as shown by the proposition below.

\begin{proposition}
\label{prop:rig}
Let $G_Q$ be a RIG of a pattern query $Q$ over a data graph $G$. Assume that there is a homomorphism from $Q$ to $G$ which maps nodes $p$ and $q$ of $Q$ to nodes $v_p$ and $v_q$, respectively, of $G$. Then, if $(p, q)$ is an edge in $Q$, $(v_p, v_q)$ is an edge of $G_Q.$
\end{proposition}

By Proposition \ref{prop:rig}, $G_Q$ encodes all the homomorphisms from $Q$ to $G$, thus it represents a search space of the answer of $Q$ on $G.$ Similarly to factorized representations of query results studied in the context of classical databases and probabilistic databases \cite{OlteanuS16}, $G_Q$ exploits computation sharing to reduce redundancy in the representation and computation of query results. Besides recording candidate occurrences sets for the edges of query $Q$, a RIG also records how the edges in the candidate occurrence sets can be joined to form occurrences for query $Q$. We later present an algorithm for enumerating the results of $Q$ on $G$ from a RIG $G_Q$.

\vspace*{.5ex}\noindent\textbf{RIG vs. other query related auxiliary data structures.} A number of recent graph pattern matching algorithms also use query related auxiliary data structures to represent the query answer search space \cite{FanLMTWW10,HanLL13,BiCLQZ16,HanKGPH19,BhattaraiLH19}. These auxiliary data structures are designed to support searching for (an extension of) graph simulation \cite{FanLMTWW10} or
subgraph isomorphisms \cite{HanLL13,BiCLQZ16,HanKGPH19,BhattaraiLH19}. Unlike RIG, they are subgraphs of the data graph, hence they do not contain reachability information between data nodes, and consequently, they are not capable of compactly encoding edge-to-path homomorphic matches.

\subsection{Refining RIG using Double Simulation}
\label{subsec:fbsim}

\vspace*{0.5ex}\noindent\textbf{Motivation.} While a RIG $G_Q$ is typically much smaller than $G$,  it can still be quite large for efficiently computing the query answer. The reason is that it may contain redundant nodes (and their incident edges). Redundant nodes are nodes which are not part of any occurrence of query $Q$.  To further reduce the query answer search space, we would like to refine a RIG $G_Q$ as much as possible by pruning redundant nodes and edges. Ideally, we would like to build the answer RIG $G^a_Q$ before computing the query answer, however when $Q$ is a general graph, finding $G^a_Q$ is a NP-hard problem even for isomorphisms and edge-to-edge mapping \cite{BhattaraiLH19}. Therefore, we focus on constructing a refined $G_Q$ whose independent node sets contain much less redundant nodes while maintaining its ability of serving as a search space for the answer of $Q$ on $G$. We do so by providing next an efficient data graph node filtering algorithm.

Most existing data node filtering methods are either simply based on query node labels \cite{AbergerTOR16,MhedhbiKS21}, or apply an approximate subgraph isomorphism algorithm \cite{HeS08} on query edge matches, or use a BFS tree of the query to filter out data nodes violating children or parent structural constraints of the tree \cite{HanKGPH19,BhattaraiLH19}. They are unable to prune nodes violating ancestor/descendant structural constraints of the input query. While the recent node pre-filtering method \cite{WuTSL22} can prune nodes violating ancestor/descendant structural constraints, it is unable to prune data nodes violating children or parent structural constraints. Moreover, that pruning technique does not capture the specific structure among those ancestors and descendants.

Inspired by the graph simulation technique used in \cite{MiloS99,kaushiketc02} which constructs a covering index for queries over tree data, we propose to extend the traditional graph simulation to construct a refined runtime index graph. The refined runtime index graph serves as a compact search space for queries over graphs.

\vspace*{0.5ex}\noindent\textbf{Double simulation.} As opposed to a homomorphism, which is a function, a graph simulation is a binary relation on the node sets of two directed graphs. Simulation provides one possible notion of structural equivalence between the nodes of the two graphs.

Since the structure of a node is determined by its incoming and outgoing paths, we define a type of simulation called {\em double simulation}, which handles both the incoming and the outgoing {\em paths} of the graph nodes. Double simulation extends dual simulation \cite{MaCFHW14},  since the later permits only edge-to-edge mappings between nodes of the input query pattern and the data graph (thus it handles only the incoming and outgoing edges of the graph nodes).

\begin{mydef}[Double Simulation]
\label{def:dsim}
The {\em double simulation} ${\cal FB}$ of a query $Q=(V_Q, E_Q)$ by a directed data graph $G=(V_G, E_G)$ is the largest binary relation $S$ $\subseteq V_Q \times V_G$ such that, whenever $(q,v)\in S$, the following conditions hold:
\begin{enumerate}
   \item $label(q)$ = $label(v)$.
   \item For each edge $e_Q=(q, q') \in E_Q$, there exists $v' \in V_G$ such that $(q',v') \in S$ and $(v, v')\in ms(e_Q).$
  \item  For each edge $e_Q=(q', q) \in E_Q$, there exists $v' \in V_G$ such that $(q',v') \in S$ and $(v', v)\in ms(e_Q).$
\end{enumerate}
\end{mydef}

For $q\in V_Q$, let ${\cal FB}$($q$) denote the set of all nodes of $V_G$ that double simulate $q$. Clearly, we have ${\cal FB}$($q$) $\subseteq ms(q)$. We have also $os(q)\subseteq {\cal FB}$($q$), as the structural constraints imposed by a homomorphism from $Q$ to $G$ imply those imposed by the ${\cal FB}$ simulation of $Q$ by $G$.   Using ${\cal FB},$ we will show how to construct a refined RIG of $Q$ on $G$ in Section \ref{subsec:rigBuild}.

The double simulation of $Q$ by $G$ is unique, since there is exactly one largest binary relation $S$ satisfying the above three conditions.  This can be proved by the fact that, whenever we have two binary relations $S_1$ and $S_2$ satisfying the three conditions between $Q$ and $G$, their union $S_1 \cup S_2$ also satisfies those conditions.

We call the largest binary relation that satisfies the  conditions 1 and 2 above {\em forward simulation} of $Q$ by $G$, while the largest binary relation that satisfies conditions 1 and 3 above is called {\em backward simulation}.  While the double simulation preserves both incoming and outgoing edge types (direct or reachability) between $Q$ and $G$, the forward and the backward simulation preserve only outgoing and incoming edge types, respectively.

Table~\ref{tab:dsim} shows the simulations ${\cal F}$, ${\cal B}$, and ${\cal FB}$ of the query $Q$ on the graph $G$ of Fig.~\ref{fig:eg}. In particular, for the reachability query edge $(B, C)$ of $Q$, the matches considered for double simulation are $(b_0, c_0)$, $(b_0, c_1)$, $(b_1, c_0)$, $(b_1, c_2)$, $(b_2, c_0)$, $(b_2, c_1)$, $(b_2, c_2)$.

\begin{table}[t!]
\caption{Forward (${\cal F}$), backward (${\cal B}$), and double (${\cal FB}$) simulation of the query $Q$ on the graph $G$ of Fig.~\ref{fig:eg}.}
\label{tab:dsim}
 \begin{center}
\begin{tabular}{|c|c|c|c|}
\hline \bfseries $q$ &
\bfseries ${\cal F}(q)$ & \bfseries ${\cal B}(q)$ & \bfseries ${\cal FB}(q)$\\
\hline\hline
$A$ & $\{a_1,a_2\}$ & $\{a_0,a_1,a_2\}$ & $\{a_1,a_2\}$\\
$B$ & $\{b_0,b_1,b_2\}$ & $\{b_0,b_2,b_3\}$ & $\{b_0,b_2\}$\\
$C$ & $\{c_0,c_1,c_2\}$ & $\{c_0,c_1,c_2\}$ & $\{c_0,c_1,c_2\}$\\
\hline
\end{tabular}
\end{center}
\vspace*{-2ex}
\end{table}

\subsection{A Basic Algorithm for Computing Double Simulation}
\label{subsec:fbsimbas}

To compute ${\cal FB},$ we present first a basic algorithm called {\em FB\-SimBas} (Algorithm \ref{alg_fbsimBas}). Algorithm {\em FB\-SimBas} is based on an extension of a naive evaluation strategy originally designed for comparing graphs of unknown sizes \cite{HenzingerHK95,MaCFHW14}. While the original method works for edge-to-edge mappings between the given two graphs, {\em FB\-SimBas} allows edge-to-path mappings from a reachability edge in the pattern graph to a path in the data graph.

Given a query $Q$ and a data graph $G,$  {\em FB\-SimBas} implements the following strategy: starting with the largest possible relation between the node sets of $Q$ and $G,$ it incrementally disqualifies pairs of nodes violating the conditions of Definition \ref{def:dsim}. The process terminates when no more node pairs can be disqualified.

\begin{algorithm}[!t]
\caption{Algorithm {\em FB\-SimBas} for computing double simulation.}
\label{alg_fbsimBas}
%  \rule{\linewidth}{.5pt}

\begin{flushleft}
%\emph{Input:} Data graph $G$, reachability index \emph{bfl}, pattern query $Q$
\emph{Input:} Data graph $G$, pattern query $Q$

\emph{Output:} Double simulation  ${\cal FB}$ of $Q$ by $G$
%\vspace*{-2ex}
\begin{small}
 \algsetup{linenodelimiter=.}
  \begin{algorithmic}[1]
%\STATE Let $root$ denote the root node of $Q$;
%\STATE Let $G_Q$ denote the runtime graph of $Q$ on $G$;
\STATE Let $FB$ be an array structure indexed by nodes of $Q$;
\STATE Initialize $FB(q)$ to be $ms(q)$ for every node $q$ in $V_Q$;

\WHILE{($FB$ has changes)}
\STATE forwardCheck();
\STATE backwardCheck();
\ENDWHILE
\RETURN $FB$;
\end{algorithmic}

\vspace*{1ex}
\textbf{Procedure} forwardPrune()
%\vspace*{-3ex}
\begin{algorithmic}[1]
\FOR{(each edge $e_Q=(q_i, q_j)\in E_Q$ and each node $v_{q_i}\in FB(q_i)$)}
\IF{(there is no $v_{q_j}\in FB(q_j)$ such that $(v_{q_i}, v_{q_j})\in ms(e_Q)$)}
\STATE delete $v_{q_i}$ from $FB(q_i)$;
\ENDIF
\ENDFOR
\end{algorithmic}

\vspace*{1ex}
\textbf{Procedure} backwardPrune()
%\vspace*{-3ex}
\begin{algorithmic}[1]

\FOR{(each edge $e_Q=(q_i, q_j)\in E_Q$ and each node $v_{q_j}\in FB(q_j)$)}
\IF{(there is no $v_{q_i}\in FB(q_i)$ such that $(v_{q_i}, v_{q_j})\in ms(e_Q)$)}
\STATE delete $v_{q_j}$ from $FB(q_j)$;
\ENDIF
\ENDFOR

\end{algorithmic}

\end{small}
\end{flushleft}
% \rule{\linewidth}{.5pt}
%\vspace*{-4ex}
% \end{figure}
\end{algorithm}

More concretely, {\em FB\-SimBas} works as follows. Let $FB$ be an array structure indexed by the nodes of $Q.$ The algorithm initializes $FB$ %to be the largest possible relation between the nodes sets $V_Q$ and $V_G.$   This is realized
by setting  $FB(q)$ to be equal to the match set $ms(q)$ of $q$,  for each $q\in V_Q$. The main process consists of two procedures  which iterate on the edges of $Q$ and check the conditions of Definition \ref{def:dsim} in different directions. The first procedure, called {\em forwardPrune}, checks the satisfaction of the forward condition in Definition \ref{def:dsim} by visiting each edge $e_Q=(q_i, q_j)\in E_Q$ from the tail node $q_i$ to the head node $q_j.$  Specifically, {\em forwardPrune} removes each $v_{q_i}$ from $FB(q_i)$ if there exists no $v_j \in FB(q_j)$ such that $(v_i, v_j)$ is in $ms(e_Q).$  The second procedure, called {\em backwardPrune}, checks the satisfaction of the backward condition in Definition \ref{def:dsim} by visiting each edge in the opposite direction. The above process is repeated until $FB$ becomes stable, i.e., no more changes can be made to $FB.$

\begin{figure}[t]
\centering
  \begin{minipage}{0.4\textwidth}
  \centering
    \scalebox{.8}{ \epsfig{file=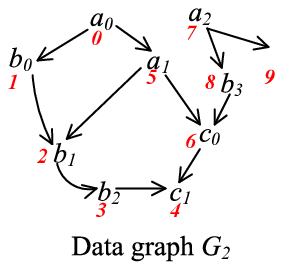}}%
   \end{minipage}
  \begin{minipage}{0.56\textwidth}

   \scalebox{.7}{\begin{tabular}{|c|c c c|c c c c|c c c|}
\hline
Step &  \multicolumn{3}{|c|}{$FB(A)$} &	\multicolumn{4}{|c|}{$FB(B)$} &\multicolumn{3}{|c|}{$FB(C)$}\\

   &   $a_0$&$a_1$ & $a_2$ & $b_0$ & $b_1$ & $b_2$ & $b_3$ & $c_0$ & $c_1$ & $c_2$\\	
\hline\hline
1	&$\times$ & & & & & & & & &  \\
2	& & & &$\times$ & & & & &$\times$ &$\times$  \\
3	& & &$\times$ & & $\times$&$\times$ & & & &  \\
4	& & & & & & &$\times$ & & &  \\
5	& &$\times$ & & & & & & & &  \\
6	& & & & & & & &$\times$ & &  \\
\hline
\end{tabular}}
 \end{minipage}

 \caption{Nodes pruning of {\em FB\-SimBas} for query $Q$ of Fig. \ref{fig:eg}(a) on graph $G_2$.}
 \label{fig:fbsimbas}
\end{figure}

\vspace*{.5ex}\noindent\textbf{An example.} Fig.~\ref{fig:fbsimbas} shows the node pruning steps performed by Algorithm {\em FB\-SimBas} for the query $Q$ of Fig. \ref{fig:eg}(a) on the graph $G_2$. We assume that the edges of $Q$ are considered in the order: $(A, B), (A, C)$, and $(B, C)$. The first column shows the step number. Odd numbers correspond to Procedure  {\em forwardPrune} while even numbers correspond to Procedure {\em backwardPrune}. The other three columns show the nodes pruned at each step from the candidate $FB$ sets for the query nodes $A$, $B$, and $C$. An `$\times$' symbol indicates that the corresponding node is pruned. Notice that $Q$ has an empty answer on $G$. Algorithm {\em FB\-SimBas} detects and prunes all the redundant nodes and hence $Q$ has an empty RIG. This demonstrates one advantage of our RIG-based query evaluation approach: an empty RIG allows an early termination of the query evaluation process.

\vspace*{.5ex}\noindent\textbf{Complexity.} Let $|V_Q|$ denote the cardinality of $V_Q$ and $|I_{max}|$ denote the size of the largest inverted list of $G.$  As there are $V_Q$ pattern nodes for each of which at most $|I_{max}|$ graph nodes can be removed, {\em FB\-SimBas} executes at most $|V_Q|\times |I_{max}|$ passes (that is, an execution of {\em forwardPrune} followed by an execution of {\em backwardPrune}). In each pass, it takes $O(|V_Q|\times |I_{max}|^2\times R)$ to check the conditions of forward simulation and backward simulation, where $R$ denotes the time for checking if a pair of nodes in $G$ is a query edge match. Therefore, {\em FB\-SimBas} has a combined complexity of $O(|V_Q|^2\times |I_{max}|^3\times R).$ Since pattern $Q$ is typically much smaller than data graph $G$, {\em FB\-SimBas} has a worst-case runtime of $O(|I_{max}|^3\times R)$.% in terms of data complexity.

\subsection{Exploiting Pattern Structures for Double Simulation Computation}
\label{subsec:fbsimdag}

Recall that {\em FB\-SimBas} picks an arbitrary order to process/evaluate the query edges. It has been shown in \cite{MennickeKNKB19}, and also verified by our experimental study, that the order in which the edges are evaluated has an impact on the overall runtime similar to the impact of join order on join query evaluation.     We would like to explore the pattern structure to design a more efficient algorithm for computing relation ${\cal FB}$ for $Q$ by $G$, which not only converges faster because of a reduced number of iteration passes of {\em FB\-SimBas} but also reduces the computation cost. In order to do so, we first describe an algorithm for computing ${\cal FB}$ for directed acyclic graph (dag) pattern queries.

\vspace*{.5ex}\noindent\textbf{A ${\cal FB}$  algorithm for dag patterns.}   We leverage the acyclic nature of the query pattern, and develop a multi-pass algorithm called {\em FB\-SimDag} based on dynamic programming.   As in {\em FB\-SimBas},  ${\cal FB}$ is initially set to be the largest possible relation between the nodes sets $V_Q$ and $V_G.$ Unlike {\em FB\-SimBas} which visits edges of $Q$ in two directions, in each pass, {\em FB\-SimDag} traverses nodes of $Q$ by their topological order two times, first bottom-up (reverse topological order) and then top-down (forward topological order).   During each traversal, pairs of nodes (and their affiliated edges) of $G$ violating the conditions of Definition \ref{def:dsim} are disqualified from nodes ${\cal FB}.$  As we will show later, the bottom-up traversal computes a forward simulation of $Q$ by $G$, while the top-down traversal computes a backward simulation of $Q$ by $G$.  Such a property is however not satisfied by {\em FB\-SimBas} which traverses pattern edges in an arbitrary order. The algorithm terminates when no more node pairs can be disqualified from $FB.$

\vspace*{.5ex}\noindent\textbf{An  algorithm for computing  ${\cal FB}$ for dag patterns.}   Leveraging the acyclic nature of the dag query pattern, we develop a multi-pass algorithm called {\em FB\-SimDag} which is based on dynamic programming.   As with {\em FB\-SimBas},  $FB$ is initially set to be the largest possible relation between the nodes sets $V_Q$ and $V.$ Unlike {\em FB\-SimBas} which visits each edge of $Q$ in two directions in each pass, {\em FB\-SimDag} traverses nodes of $Q$ by their topological order two times, first bottom-up (reverse topological order) and then top-down (forward topological order). During each traversal, nodes of $G$ violating the conditions of Definition \ref{def:dsim} are removed from $FB$.    %During each traversal, pairs of nodes (and their affiliated edges) of $G$ violating the conditions of Definition \ref{def:dsim} are removed from ${\cal FB}.$******
As we will show later, the bottom-up traversal computes a forward simulation of $Q$ by $G$, while the top-down traversal computes a backward simulation of $Q$ by $G$.  In contrast, {\em FB\-SimBas}  traverses pattern edges in an arbitrary order. The algorithm {\em FB\-SimDag} terminates when no more nodes can be removed from $FB.$

% \begin{figure}[!t]
\begin{algorithm}[!t]
\caption{Algorithm {\em FB\-SimDag} for computing double simulation.}
\label{alg_fbsimDag}
%  \rule{\linewidth}{.5pt}

\begin{flushleft}
%\emph{Input:} Data graph $G$, reachability index \emph{bfl}, pattern query $Q$
\emph{Input:} Data graph $G$, dag pattern query $Q$

\emph{Output:} Double simulation  ${\cal FB}$ of $Q$ by $G$
%\vspace*{-2ex}
\begin{small}
 \algsetup{linenodelimiter=.}
  \begin{algorithmic}[1]
%\STATE let $root$ denote the root node of $Q$;
%\STATE let $G_Q$ denote the runtime graph of $Q$ on $G$;
%\STATE Let $FB$ be an array structure indexed by nodes of $Q$;
%\STATE Initialize $FB(q)$ to be $I_q$ for every node $q$ in $V_Q$;
\STATE Lines 1-2 in Algorithm {\em FB\-SimBas};
\WHILE{($FB$ has changes)}
\STATE forwardSim();
\STATE backwardSim();
\ENDWHILE
\RETURN $FB$;
\end{algorithmic}

\vspace*{1ex}
\textbf{Procedure} forwardSim()
%\vspace*{-3ex}
\begin{algorithmic}[1]
\FOR{(each $q\in V_Q$ in a reverse topological order and each $v_q\in FB(q)$)}
%\IF{($\exists q_i \in \emph{ch}(q)$ where no $v_{q_i}\in {\cal F}(q_i)$ s.t. $(v_q, v_{q_i})$  is an occurrence of $(q, q_i)\in E_Q$)}
\IF{($\exists q_i \in children(q)$ where no $v_{q_i}\in FB(q_i)$ s.t. $(v_q, v_{q_i})\in ms(e_Q)$, where $e_Q=(q, q_i)\in E_Q$)}
\STATE delete $v_q$ from $FB(q)$;
\ENDIF
\ENDFOR
\end{algorithmic}

\vspace*{1ex}
\textbf{Procedure} backwardSim()
%\vspace*{-3ex}
\begin{algorithmic}[1]

\FOR{(each $q \in V_Q$ in a topological order and each $v_q\in FB(q)$)}
\IF{($\exists q_i \in parents(q)$ where no $v_{q_i}\in FB(q_i)$ s.t. $(v_{q_i},v_q)\in ms(e_Q)$, where $e_Q=(q_i,q)\in E_Q$)}
\STATE delete $v_q$ from $FB(q)$;
\ENDIF
\ENDFOR

\end{algorithmic}

\end{small}
\end{flushleft}
% \rule{\linewidth}{.5pt}
%\vspace*{-4ex}
% \end{figure}
\end{algorithm}

Algorithm \ref{alg_fbsimDag} shows the pseudocode of {\em FB\-SimDag}. The algorithm first invokes procedure {\em forwardSim} to check for nodes $v_q\in FB(q)$ which satisfy the forward simulation condition. Procedure {\em forwardSim} considers outgoing query edges of $q$ by traversing nodes of $Q$ in a bottom-up way.  When $q\in V_Q$ is a sink node in $Q$, $v_q$ trivially satisfies the forward simulation condition. Otherwise, if edge $e_Q=(q, q_i)\in E_Q$ but there is no $v_i \in FB(q_i)$ such that $(v, v_i)$ is in $ms(e_Q),$ $v_q$ is removed from $FB(q).$

When the bottom-up traversal terminates,  {\em FB\-SimDag} proceeds to do a top-down traversal of $Q$  using procedure {\em backwardSim}.  This procedure checks whether nodes $v_q\in FB(q)$  satisfy the backward simulation condition by considering incoming query edges of $q$.  When $q\in V_Q$ is a source node in $Q$, $v_q$ trivially satisfies the backward simulation condition. Otherwise, if edge $e_Q=(q_i, q)\in E_Q$ but there is no $v_i \in {\cal FB}(q_i)$ such that $(v_i, v)$ is in $ms(e_Q),$ $v_q$ is removed from $FB(q).$

The above process is repeated until $FB$ stabilizes, i.e., no $FB(q)$, $q\in V_Q$, can be further reduced. When $Q$ is a tree pattern, a single pass (i.e., one bottom-up traversal and one top-down traversal on $Q$) is sufficient for $FB$ to stabilize \cite{WuTSL20}.

\begin{theorem}%
%\begin{proposition}
%\begin{mylemma}
\label{prop:dsim}
Algorithm {\em FB\-SimDag} correctly computes double simulation ${\cal FB}$ of dag pattern $Q$ by data graph $G$.
%\end{mylemma}
% \end{proposition}
\end{theorem}

% \begin{figure}[!t]
\begin{algorithm}[!t]
\caption{Algorithm {\em FB\-Sim} for computing double simulation.}
\label{alg_fbsim}
%  \rule{\linewidth}{.5pt}

\begin{flushleft}
%\emph{Input:} Data graph $G$, reachability index \emph{bfl}, pattern query $Q$
\emph{Input:} Data graph $G$, pattern query $Q$

\emph{Output:} Double simulation  ${\cal FB}$ of $Q$ by $G$
%\vspace*{-2ex}
\begin{small}
 \algsetup{linenodelimiter=.}
  \begin{algorithmic}[1]

\IF{$Q$ is a dag}
\RETURN {\em FB\-SimDag}($Q$, $G$)

\ENDIF
%\STATE Initialize ${\cal FB}(q)$ to be $I_q$ for every node $q$ in $V_Q$;
\STATE Lines 1-2 in Algorithm {\em FB\-SimBas};
\STATE Decompose query graph $Q$ to a dag $Q_{dag}$ and a back edge set $E_{bac}.$
\WHILE{(${\cal FB}$ has changes)}
\STATE lines 2-4 in {\em FB\-SimDag}($Q_{dag}$, $G$);
\STATE lines 2-4 in {\em FB\-SimBas}($E_{bac}$, $G$);
\ENDWHILE
\RETURN $FB$;
\end{algorithmic}

\end{small}
\end{flushleft}
% \rule{\linewidth}{.5pt}
%\vspace*{-1ex}
% \end{figure}
\end{algorithm}

\vspace*{.5ex}\noindent\textbf{ {\em FB\-SimDag} vs. {\em FB\-SimBas}.} While both {\em FB\-SimDag} and {\em FB\-SimBas} have the same theoretical worst case time complexity, our empirical study shows that {\em FB\-SimDag} converges faster than {\em FB\-SimBas} in practice.   The main difference of the two algorithms is that  {\em FB\-SimDag} considers query nodes in a (forward and backward) topological order, whereas {\em FB\-SimBas} considers query nodes in an arbitrary order. For every query node $q$ in $Q$,  in order to check the satisfaction of forward (resp. backward) simulation condition of nodes in $FB(q)$,   {\em FB\-SimDag} scans $FB(q)$ only once in each pass, whereas {\em FB\-SimDag} has to scan $FB(q)$ multiple times, since it checks $FB$ associated with each child (resp. parent) node of $q$ separately. Also, in each pass, once {\em FB\-SimDag} finishes checking the forward simulation condition for nodes of $FB(q)$ with a bottom-up traversal of $Q$,  nodes retained in $FB(q)$ remain to forward simulate $q$ in the current pass, since the subsequent processing for other query nodes in the same pass will not touch $FB(q).$ This property holds also for the backward simulation checking process of {\em FB\-SimDag}. In contrast, {\em FB\-SimBas} does not enjoy this property since it computes forward (resp. backward) simulations for query nodes of $Q$ in an arbitrary order.

\begin{figure}[t]
\centering
  \begin{minipage}{0.4\textwidth}
  \centering
    \scalebox{.8}{ \epsfig{file=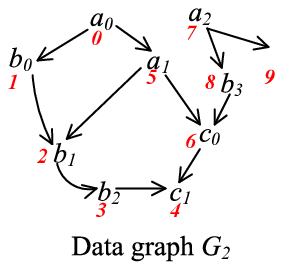}}%
   \end{minipage}
  \begin{minipage}{0.56\textwidth}

   \scalebox{.7}{\begin{tabular}{|c|c c c|c c c c|c c c|}
\hline
Step &  \multicolumn{3}{|c|}{$cs(A)$} &	\multicolumn{4}{|c|}{$cs(B)$} &\multicolumn{3}{|c|}{$cs(C)$}\\

   &   $a_0$&$a_1$ & $a_2$ & $b_0$ & $b_1$ & $b_2$ & $b_3$ & $c_0$ & $c_1$ & $c_2$\\	
\hline\hline
1	&$\times$ & & & & & & & & &  \\
2	& & & &$\times$ & & & & &$\times$ &$\times$  \\
3	& &$\times$ &$\times$ & & $\times$&$\times$ & & & &  \\
4	& & & & & & &$\times$ &$\times$ & &  \\
\hline
\end{tabular}}
 \end{minipage}
\caption{Nodes pruning of {\em FB\-SimDag} for query $Q$ of Fig. \ref{fig:eg}(a) on graph $G_2$.}
 \label{fig:fbsimdag}
 \vspace*{-3ex}
\end{figure}

\vspace*{.5ex}\noindent\textbf{An example.} Fig.~\ref{fig:fbsimdag} shows the node pruning steps performed by Algorithm {\em FB\-SimDag} for the query $Q$ of Fig.~\ref{fig:eg}(a) on the graph $G_2$. Comparing the result shown in Fig.~\ref{fig:fbsimbas}, one can see that it takes {\em FB\-SimDag} fewer steps than {\em FB\-SimBas} to converge.

\vspace*{0.5ex}\noindent\textbf{Dag+$\Delta$: an efficient ${\cal FB}$ algorithm.} Based on {\em FB\-SimDag}, we design a new ${\cal FB}$ algorithm called {\em FB\-Sim} (Algorithm \ref{alg_fbsim}). % \cite{abs-2112-08638}. %(Algorithm \ref{alg_fbsim}) that works for general graph patterns.
The algorithm first decomposes the input graph pattern $Q$ into a dag $Q_{dag}$ and a set $E_{bac}$ of back edges ($\Delta$). The main body of the algorithm has two phases: it first calls {\em FB\-SimDag} to compute {\em FB} on $Q_{dag}.$ After that, it calls {\em FB\-SimBas} on $E_{bac}$ to update {\em FB}. The above process is repeated until {\em FB} becomes stable.

While {\em FB\-Sim} has the same worst case complexity as {\em FB\-SimBas}, our experimental study in Section \ref{sec:experiments} demonstrates that our Dag+$\Delta$ approach for computing double simulations runs faster than {\em FB\-SimBas} in many cases.

In the next section, we will describe several optimization techniques to further boost the ${\cal FB}$ computation.

\subsection{Building A Refined RIG}
\label{subsec:rigBuild}

%\setlength{\textfloatsep}{2pt}% Remove \textfloatsep
% \begin{figure}[!t]
\begin{algorithm}[!t]
\caption{Algorithm {\em BuildRIG} for building a refined RIG}
\label{alg_buildRIG}
%  \rule{\linewidth}{.5pt}

\begin{flushleft}
\emph{Input:} Data graph $G$, pattern query $Q$

\emph{Output:} RIG $G_Q$ of $Q$ on $G$
\end{flushleft}

%\vspace*{1.5ex}
\begin{small}
 \algsetup{linenodelimiter=.}
  \begin{algorithmic}[1]

\STATE select();
\FOR{(each edge $(q_i, q_j)\in E_Q$)}
\STATE expand($q_i$, $q_j$);
\ENDFOR
\RETURN $G_Q$;
\end{algorithmic}

\begin{flushleft}
 %\vspace*{1ex}
\textbf{Procedure} select()
 %\end{flushleft}
 \begin{algorithmic}[1]

\STATE Use Algorithm {\em FB\-SimBas} or {\em FB\-Sim} to compute ${\cal FB}$ of $Q$ by $G$;
\STATE Initialize $G_Q$ as a $k$-partite graph without edges having one independent set $cos(q)$ for every node $q \in V_Q$, where $cos(q)$ = ${\cal FB}(q)$ ;

\end{algorithmic}

\begin{flushleft}
 %\vspace*{1ex}
\textbf{Procedure} expand($p$, $q$)
 \end{flushleft}
%\vspace*{-3ex}
\begin{algorithmic}[1]
\FOR{(each $v_p\in cos(p)$)}
\FOR{(each $v_q\in cos(q)$)}
\IF{($(v_p, v_q) \in ms(e_Q)$, where $e_Q=(p, q)\in E_Q$)}
\STATE Connect $v_p$ to $v_q$ with a directed edge;
\ENDIF
\ENDFOR
\ENDFOR
\end{algorithmic}
\end{flushleft}
\end{small}
% \rule{\linewidth}{.5pt}
%\vspace*{-3ex}
\end{algorithm}{}
% \end{figure}

We now present Algorithm {\em BuildRIG} (Algorithm \ref{alg_buildRIG}) for building a refined RIG in two phases: in the \emph{node selection} phase (line 1), all the RIG nodes are obtained by pruning redundant data nodes. This is achieved by computing the double simulation relation. In the \emph{node expansion} phase (lines 2-3),  the RIG nodes are expanded with incident edges to construct the final RIG graph. During the RIG construction,  once node $v_q \in cos(q)$ has been expanded, the outgoing and incoming edges of $v_q$ are indexed by the parents and children of query node $q.$ This allows efficient intersection operations of adjacency lists of selected nodes in the RIG graph. These efficient intersection operations are useful in the phase of query occurrence enumeration as we will show in Section \ref{sec:alg}.

As an example, consider building a refined RIG for query $Q$ on graph $G$ in Fig. \ref{fig:eg} using Algorithm \ref{alg_buildRIG}. After the node selection phase, we obtain the following ${\cal FB}$ relation: $FB(A)=\{a_1, a_2\}$, $FB(B)=\{b_0, b_2\}$ and $FB(C)=\{c_0, c_1, c_2\}$. The RIG generated from the node expansion phase is shown in Fig.~\ref{fig:eg}(e). The RIG has one more edge than the answer RIG (shown by a red dashed line), but it has fewer nodes and edges than the match RIG (Fig.~\ref{fig:eg}(d)).

\vspace*{.5ex}\noindent\textbf{Speedup convergence for simulation computation.} As described in Section \ref{subsec:fbsim}, the computation of ${\cal FB}$ terminates only when no more nodes can be pruned from the candidate occurrences of the query nodes during the multi-pass process. This process can be costly since we need to repeatedly check the candidate occurrence sets of the query nodes. We  describe below optimizations to speedup the convergence of the simulation process.

First,  if no change is made to candidate occurrences corresponding to a subquery of $Q$ in the previous pass,  then the computation on that subquery for the current pass can be skipped. To achieve this,  we associate with each query node $q$ a flag indicating whether nodes were pruned from its candidate occurrence set $FB(q)$ during the last pass. The flags are consulted in the current pass to decide whether the computation can be skipped.

Second, the node selection enforces the {\em existence} semantics.  A data node $v$ is retained in its candidate occurrence set as long as there exist nodes in the candidate occurrence sets of the corresponding parent and child query nodes that make $v$ satisfy the ${\cal FB}$ conditions of Definition \ref{def:dsim}.  Checking node $v$ in the current pass can be skipped  if the nodes guaranteeing its existence are not pruned in the previous pass. We design an index on the nodes in the candidate occurrence sets of the query nodes. Specifically, the index records for each data node $v\in FB(q)$ of query node $q$ those nodes in the candidate sets of $q$'s parents and children in $Q$ that guarantee $v$'s existence in $FB(q)$. The index structure is maintained throughout the multi-pass process.

Finally, our empirical study show that most of redundant nodes are detected and pruned during the first two or three passes of the iterative process. To reduce the cost of RIG construction, we can approximate double simulation relations by stopping the ${\cal FB}$ computation after N passes of iterations, or when the total pruned nodes number in the current pass is below a specified threshold. In our experimental evaluation (Section \ref{sec:experiments}), we apply the first strategy and fix N to be three.

\vspace*{.5ex}\noindent\textbf{Early expansion termination for dags.} When expanding a node $v_p\in cos(p)$ by edges (lines 2-4 in Procedure expand), it is not always necessary to scan the entire set $cos(q)$ for every child node $q$ of $p$. When the input data graph $G$ is a dag, we can associate each node $u$ in $G$ with an interval label, which is an integer pair (\emph{begin}, \emph{end}) denoting the first discovery time of $u$ and its final departure time in a depth-first traversal of $G.$ Nodes in $cos(q)$ are accessed in ascending order of the $begin$ value of their interval labels. Interval labelling guarantees that node $v_p$ does not reach node $v_q$ if $v_p.end < v_q.begin.$ Therefore, once such a node $v_q$ is encountered, the scanning over $cos(q)$ can be safely terminated since all the subsequent nodes of $v_q$ in $cos(q)$ have a $begin$ value which is larger than $v_q.begin$ (and $v_p.end$). Our empirical study showed that the early expansion termination technique can improve the performance by up to 30\%.

\vspace*{.5ex}\noindent\textbf{Batch checking direct connectivity constraints.} Checking whether $(v_{q_i}, v_{q_j})$ is in $cos(e_Q)$ of the query edge $e_Q=(q_i, q_j)$ is a core operation in the two phases of Algorithm {\em BuildRIG}. Two cases can be distinguished: (a) $e_Q$ is a reachability edge and (b) $e_Q$ is a direct edge.

For case (a), we can use a graph reachability index to check whether $v_{q_i}\prec v_{q_j}$. For case (b), we can use the adjacency lists of $G$ to check the direct connectivity relationship from $v_{q_i}$ to $v_{q_j}.$ A straightforward method for (b) is to sort nodes of adjacency lists (by node id for example), and use a binary search to check if $v_{q_j}$ is in the forward adjacency list of $v_{q_i}$ (checking the forward simulation condition for $v_{q_i}$), or $v_{q_i}$ is in the backward adjacency list of $v_{q_j}$ (checking the backward simulation condition for $v_{q_j}$).  Processing a large number of such $v_{q_i}'s$ or $v_{q_j}'s$ requires repeated launching of a binary search against adjacency lists.  This can be costly since a binary search incurs random memory access.

We develop methods to quickly check the direct connectivity constraint between nodes. Let $adj_f$ and $adj_b$ denote the functions mapping a graph node to its forward and backward adjacency lists, respectively.

The node selection phase of Algorithm {\em BuildRIG} targets on quickly identifying nodes satisfying the double simulation conditions.  Our techniques apply to general graph patterns, for the simplicity of description, we assume $Q$ to be a dag. Consider $e_Q=(q_i, q_j)$ be a direct edge of $Q.$ Let $FB_n(q_i)$ denote the simulation of query node $q_i$ obtained during the $n$th pass of ${\cal FB}$ computation.  Suppose $k$ to be the current pass number and we are now in the process of computing backward simulation for node $q_j.$ We need to find all $v_j\in FB_{k-1}(q_j)$ such that there exists $v_i\in FB_k(q_i)$ and that $(v_i, v_j)$ is in $cos(e_Q).$ Instead of processing each $v_j$ individually,  we do the following batch operation: $FB_{k-1}(q_j)\cap \bigcup_{v_i\in FB_k(q_i)}adj_f(v_i).$  The batch operation first computes the union of the forward adjacency lists of nodes in ${\cal FB}_k(q_i)$, then intersects the result with $FB_{k-1}(q_j).$ This way, we get rid of all the nodes from $FB_{k-1}(q_j)$ violating the direct connectivity constraint w.r.t $e_Q$ in one step.  Checking direct connectivity constraints for the forward simulation computation is similar.

For each node $v_{q_i} \in cos(q_i),$ the node expansion phase of Algorithm {\em BuildRIG} targets on quickly identifying all the nodes in $cos(q_j)$ that expand $v_{q_i},$ i.e., nodes having a direct relationship with $v_{q_i}$ in $G.$  We again convert the direct connectivity constraint checking into a set intersection operation $adj_f(v_{q_i}) \cap cos(q_j)$. Clearly, every element $v$ in the intersection is connected to $v_{q_i}$ by an outgoing edge from $v_{q_i}$ in $G$. This way, we obtain all the nodes in $cos(q_j)$ having a direct connectivity relationship with $v_{q_i}$ in one step.

\begin{figure}[!t]
    \centering%
     \scalebox{.8}{ \epsfig{file=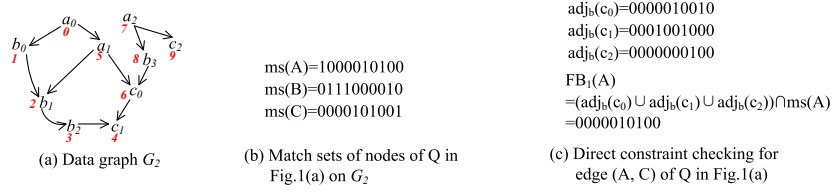}}%
     \caption{Checking direct connectivity constraint for edge $(A, C)$ of query $Q$ in Fig.~\ref{fig:eg}(a) against graph $G_2$ in the node selection phase of Algorithm {\em BuildRIG}.}
      \label{fig:eg2}
\end{figure}

As an example, consider checking the direct connectivity constraint for edge $(A, C)$ of $Q$ in Fig.~\ref{fig:eg}(a) against graph $G_2$ in the node selection phase of Algorithm {\em BuildRIG}. The red digits associated nodes of $G_2$ are node IDs.  Fig.~\ref{fig:eg2}(b) shows match sets (i.e., the $FB_0$) of query nodes represented as bit vectors. Each bit vector is indexed by node IDs, and a '1' bit denotes the existence of the corresponding node in the associated node set. Likewise, the adjacency lists of $G_2$ are represented also in bit vectors. Fig.~\ref{fig:eg2}(c) shows how the direct connectivity constraint checking is performed (using the bitwise operations) for edge $(A, C)$ during the first pass of forward simulation computation. The result bit vector for $FB_1(A)$ has '1' at the $5^{th}$ and $7^{th}$ bits, corresponding to nodes $a_1$ and $a_2$, respectively.         % \label{sec:rigGraph}
\section{A Multi-way Intersection-based Enumeration Algorithm}
\label{sec:alg}

\subsection{The Algorithm}
\label{subsec:alg}

We now present our graph pattern answer enumeration algorithm, called {\em MJoin}, which is shown in Algorithm~\ref{alg_mijoin}. Before giving the details of the algorithm, we provide its high level idea below.

\vspace*{.5ex}\noindent\textbf{High level idea.} Given a query $Q$ and data graph $G$, relation $cos(e)$ contains the candidate occurrences of query edge $e$ on $G.$
%Given a query $Q$ and a data graph $G$, let $R_e$ denote the relation containing the occurrences of each query edge $e$ on $G.$
Conceptually,  {\em MJoin} produces occurrences of $Q$ by joining multiple such relations at the same time. Instead of using standard query plans that join one relation (i.e., query edge) at a time, {\em MJoin} considers a new style of multi-way join plans which join one join key (i.e., query node in graph terms) at a time.  A query-node-at-a-time style join plan considers only the distinct join key values if a specific join key value occurs in multiple tuples. Hence, it can avoid enumerating large intermediate results that typically occur with Selinger-style binary-joins (query-edge-at-a-time joins in graph terms) \cite{NguyenABKNRR14}. %Selinger-style binary-joins \cite{NguyenABKNRR14}.

The new style multi-way joins have been exploited in recent graph matching algorithms \cite{Veldhuizen14,AbergerLTNOR17,FreitagBSKN20}. The main difference between {\em MJoin} and those algorithms lies in the implementation of the new style multi-way joins, discussed in detail in Section \ref{sec:implementations}. Briefly, {\em MJoin} exploits the runtime index graph $G_Q$, the factorized representation of query answer of $Q$ on $G$, to perform multi-way joins. We show below how this can be done by multi-way intersecting node adjacency lists of $G_Q$.

%\begin{figure}[!t]
\begin{algorithm}[!t]
 \rule{\linewidth}{.5pt}

 \begin{flushleft}
\emph{Input:} Data graph $G$, pattern query $Q,$ and runtime index graph $G_Q$ of $Q$ on $G.$

\emph{Output:} The answer of $Q$ on $G.$
\end{flushleft}

%\vspace*{-2ex}
\begin{small}
 \algsetup{linenodelimiter=.}
  \begin{algorithmic}[1]

%\STATE Build the compactruntime index graph $G_Q$ of $Q$ on $G$ using Algorithm \ref{alg_fbsim};
\STATE Pick an order $q_1, \ldots, q_n$ of nodes of $Q$, where $n=|V(Q)|$;
\STATE Let $t$ be a tuple where $t[i]$ is initialized to be $null$ for $i\in [1,n]$;
%\STATE Let $S_i$ denote the node set of $q_i$ in $G_Q$;
%\FOR{(every $v\in S_1$)}
%\STATE $t[1]:=v$;
%\STATE  MIJoin(2, $t$);
%\ENDFOR
\STATE  enumeration(1, $t$);
\end{algorithmic}

 \begin{flushleft}
 \vspace*{1ex}
\textbf{Procedure} enumeration(index $i$, tuple $t$)\\
 \end{flushleft}
 \begin{algorithmic}[1]

\IF{($i=|V(Q)|+1$)}
\RETURN $t$;
\ENDIF
\STATE $N_i:=\{q_j\mid(q_i,q_j)\in E(Q)~or~(q_j,q_i)\in E(Q),j\in[1,i-1]\}$
\STATE $cos_i$ := $cos(q_i)$;
\FOR{(every $q_j\in N_i$)}
\STATE $cos_{ij}:=\{v_i\in cos_i\mid (v_i, t[j])$ or $(t[j],v_i)$ is an edge of $G_Q\}$;
\STATE $cos_i$ := $cos_i \cap cos_{ij}$;
\ENDFOR
\FOR{(every node $v_i \in cos_i$)}
\STATE $t[i]$ := $v_i$;
\STATE enumeration($i+1$, $t$);
\ENDFOR

\end{algorithmic}

\end{small}
%\rule{\linewidth}{.5pt}
%\vspace*{-3ex}
\caption{Algorithm {\em MJoin}.}
\label{alg_mijoin}
%\vspace*{-4ex}
%\end{figure}
\end{algorithm}

\vspace*{.5ex}\noindent\textbf{The algorithm.} The core of Algorithm {\em MJoin} is a backtracking search procedure which extends partial query occurrences  by matching one query node at a time using multi-way intersections.   More concretely, {\em MJoin} first picks a {\em search order} to search solutions. This is a linear order of the query nodes. A search order heavily influences the query evaluation performance.  We will discuss how to choose a good search order later. Then, the algorithm performs a recursive backtracking search to find occurrences of the query nodes iteratively, one at a time by the given order, before returning any query occurrences.

Let's assume that the chosen search order is $q_1, \ldots, q_n.$ Let $Q_i$ denote the subquery of $Q$ induced by the nodes $q_1, \ldots, q_i$, where $i\in[1,n]$.  Algorithm {\em MJoin} calls a recursive function \emph{enumerate} which searches for potential occurrences of a single query node $q_i$ in each recursive step. The index $i$ of the current query node is passed as a parameter to \emph{enumerate}. When $i>0$, the backtracking nature of \emph{enumerate} entails that a specific occurrence for the subquery $Q_{i-1}$ has already been considered in the previous recursive steps.  The second parameter $t$ of \emph{enumerate} is a tuple of length $n,$ where $t[1:i]$ is an occurrence of $Q_i$. Initially, $i$ is set to 0 and all the values of $t$ are set to $null$.

At each search step $i$, function \emph{enumerate} generates for node $q_i$ a local candidate occurrence node set $cos_i$, which is a subset of $cos(q_i)$ in $G_Q$, based on the current search state. This allows \emph{enumerate} to focus on those nodes that can contribute to a solution. Specifically, \emph{enumerate} first determines query nodes that have been considered in a previous recursive step and are adjacent to the current node $q_i.$ It then initializes $cos_i$ to be equal to $cos(q_i).$ Next, \emph{enumerate} conducts the multi-way intersection: for each $q_j\in N_i$, it intersects $cos_i$ with the forward adjacency list of $t[j]$ in $G_Q$ when $(q_i, q_j)$ is an edge of $Q$, or with the backward adjacency list of $t[j]$ when $(q_j, q_i)$ is an edge of $Q$ (lines 5-7). If the resulted $cos_i$ is not empty, then \emph{enumerate} iterates over the nodes in $cos_i$ (line 8). In every iteration, a node of $cos_i$ is assigned to $t[i]$ (line 9) and \emph{enumerate} proceeds to the next recursive step (line 10). If $cos_i$ is empty or all the nodes in $cos_i$ have been considered,  \emph{enumerate} backtracks to the last matched query node $q_{i-1}$, reassigns an unmatched node (if any) from $cos_{i-1}$ to $t[i-1]$, and recursively calls \emph{enumerate}. In the final recursive step, when $i=n+1$, tuple $t$ contains one specific occurrence for all the query nodes and is returned as an occurrence of $Q$ (line 2).

\vspace*{.5ex}\noindent\textbf{Example.} In our running example, let $G_Q$ be the refined RIG, i.e., the graph of Fig.~\ref{fig:eg}(e) including the red dashed edge. Assume the matching order of $Q$ is $A, B, C$. When $i=1$, Algorithm {\em MJoin} first assigns $a_1$ from $cos_1$ (=$cos(A)$) to tuple $t[1]$, then recursively calls \emph{Enumerate}(2, $t$).  The intersection of $a_1$'s adjacency list with $cos_2$ (=$cos(B)$) is \{$b_0$\}. Node $b_0$ is then assigned to $t[2].$ At $i=3$,  {\em MJoin} computes $cos_3$, which is the intersection of forward adjacency lists of $a_1$ and $b_0$ with $cos(D)$ is \{$c_0, c_1$\},  assigns $c_0$ and $c_1$ to $t[4]$, and returns two tuples \{$a_1, b_0, c_0$\} and \{$a_1, b_0, c_2$\} in that order. Then, {\em MJoin} backtracks till $i=1$, assigns the next node $a_2$ from $cos(A)$ to $t[1]$ and proceeds in the same way. It finally returns another two tuples \{$a_2, b_2, c_0$\} and \{$a_1, b_0, c_2$\}. Note that edge $(b_2, c_1)$ (the red dashed edge in Fig.~\ref{fig:eg}(e)) is not filtered out by the double simulation pruning process and its redundancy is detected only after {\em MJoin} is run.

\subsection{Search Order}
\label{subsec:searchOrder}

%\vspace*{.5ex}\noindent\textbf{Search order.}
A search order $\sigma$ is a permutation of query nodes that is chosen for searching query solutions.   %The search procedure is equivalent to multiple joins.
The performance of a query evaluation algorithm is heavily influenced by join orders \cite{NeumannR18}. As the number of all possible search orders is exponential in the number of query nodes, it is expensive to enumerate all of them.

The search order $\sigma$ for query $Q$ is essentially a left-deep query plan \cite{HeS08}. The traditional dynamic programming technique would take $O(2^{|V_Q|})$ time to generate an optimized join order. This is not scalable to large graph patterns, as verified by our experimental evaluation in Section \ref{sec:experiments}.

We therefore use a greedy method to find a search order for $Q$ leveraging statistics of $G_Q.$ Our greedy method is based on the join ordering strategy proposed in \cite{HeS08}. We refer to this method as $JO$.
% for the graph pattern matching problem.
$JO$ selects as a start node of $\sigma$ a node $q$ in $V_Q$ with the smallest candidate occurrence set $cos(q)$ in $G_Q$ among the nodes in $V_Q.$ Subsequently, $JO$ iteratively selects as the next node in $\sigma$ a node $q'$ of $Q$  which  satisfies the following two conditions: (a) $q'$ is a new node adjacent in $Q$ to some node in $\sigma$, and (b) $cos(q')$ is the smallest among all the nodes $q'$ satisfying condition (a). The rationale here is to enforce connectivity to reduce unpromising intermediate results caused by redundant Cartesian products \cite{BiCLQZ16} as well as to minimize (estimated) join costs. Different from the original method which uses the cardinality of the inverted lists of the data graph $G$  \cite{HeS08}, $JO$ uses the cardinality of the candidate occurrence sets of a refined RIG $G_Q$,  which provide a better cost estimation for generating an effective search order.

In our experiments, we also implemented a well known  ordering method  called  $RI$ \cite{BonniciGPSF13}. Unlike $JO$, $RI$ generates $\sigma$  based purely on the topological structure of the given query, independently of any target data graph. The rationale of $RI$ is to introduce as many edge constraints as possible and as early as possible in the ordering. Roughly speaking, vertices that are highly connected with vertices previously present in the ordering tend to come early in the final ordering. In our \emph{enumerate} procedure,  edge constraints will translate into intersection operations to produce candidate occurrence sets for the query nodes under consideration. Intuitively, the search order chosen by $RI$ is likely to reduce the computation cost, since it tends to ensure the search space of \emph{enumerate} would be reduced significantly after each iteration. We examine this intuition and compare the effectiveness of $RI$ with $JO$ in the experiments.

\subsection{The Complexity}
\label{subsec:algcomplexity}

We adapt the complexity analysis results recently developed for multi-way join processing on relational data \cite{NgoRR13,FreitagBSKN20} to graph pattern query processing on graph data.

\vspace*{1ex}\noindent\textbf{Complexity.}  Given graph pattern query $Q$, let $n$ and $m$ denote the number of nodes and edges of $Q$ respectively and $G_Q$ denote a runtime index graph of $Q$ on data graph $G.$
A {\em fractional edge cover} of $Q$ is a vector \textbf{x} = ($x_1,\ldots,x_m$), $\textbf{x} \in \mathbf{R}^{m}$, in correspondence to the edges $(e_1, \ldots, e_m)$ of Q,  such that $x_j>$ 0 for all $j\in [1, m]$ and $\sum_{e_j\in E(Q):v_i\in e_j}x_j\geq 1,$ for all $v_i\in V(Q)$ \cite{NgoRR13}. Let $(x_1,\ldots,x_m)$ be a fractional cover of $Q$ that minimizes the product $\prod_{e_j\in E(Q)}|R_j|^{x_j}$, where $R_j$ denotes the candidate occurrence set of the edge $e_j$ of $Q$ on $G_Q.$

\begin{theorem}
\label{them:complexity}
\em The time complexity of Algorithm {\em MJoin} is in $O(nm\prod_{e_j\in E(Q)}|R_j|^{x_j})$ and its space complexity is in $O(n\times MaxCos)$, where $MaxCos$= $max\{|cos(q)|\}$, for all query nodes $q\in V(Q)$, and $cos(q)$ is the candidate occurrence set of node $q$ in $G_Q.$
\end{theorem}

The proof of Theorem \ref{them:complexity} can be found in Appendix. %\ref{sec:appendix}.

\vspace*{1ex}\noindent\textbf{Worst-case optimal join.}  %Let $Q$ be a graph pattern query, where $m = |E(Q)|$ and $n = |V(Q)|.$
Let $R^G_j$ be the match set of the edge $e_j$ on data graph $G$. Let \textbf{x} = \{$x_1,\ldots,x_m$\} be an arbitrary fractional edge cover of $Q.$ %Consider the natural join $R^G_1\Join \ldots \Join R^G_m$ described by $Q.$
Atserias, Grohe, and Marx \cite{AtseriasGM13} (AGM) derived a tight bound on the number of output tuples $|Q|$ of $Q$ in terms of the sizes of the input relations and the fractional edge cover. This bounds states that

\begin{equation}
\label{eq:7}
    |Q|~\leq~\displaystyle\prod_{e_j\in E(Q)}|R^G_j|^{x_j}
\end{equation}

The bound above concerns with data complexity as the size of $Q$ is generally much smaller than the size of $G.$ The worst-case output size of $Q$ can be determined by minimizing the right-hand size of Inequality (\ref{eq:7}) \cite{NgoRR13}.  A join algorithm for computing $Q$ is defined to be {\em worst-case optimal} if its runtime is proportional to this worst-case output size \cite{NgoRR13}.  We have the following theorem.

\begin{theorem}
\label{them:wco}
Algorithm {\em MJoin} is a worst-case optimal join algorithm.
\end{theorem}

\noindent{\bf Proof.} Let $G_Q$ denote a runtime index graph of $Q$ on data graph $G.$  Let $R_j$ denote the candidate occurrence set of the edge $e_j$ of $Q$ on $G_Q.$ Clearly, $|R_j|~\leq~|R^G_j|$. By Theorem \ref{them:complexity}, the time complexity of Algorithm {\em MJoin} is bounded by $\displaystyle\prod_{e_j\in E(Q)}|R_j|^{x_j}$, hence it is also bounded by $\displaystyle\prod_{e_j\in E(Q)}|R^G_j|^{x_j}$. This completes the proof. \hfill$\Box$

   	% \label{sec:alg}
\section{Set Intersection Implementation and Graph Indexes}
\label{sec:implementations}

Algorithms {\em BuildRIG} and {\em MJoin} employ set operations for checking direct connectivity constraints and for performing multi-way joins, respectively.  Set intersection in particular is the core operation in join algorithms of recent graph database systems \cite{AbergerLTNOR17,FreitagBSKN20,MhedhbiKS21}. Efficient set intersection (and set operations in general) are already supported in standard libraries (e.g., in C++ through STL). In this section, we first present data structure design considerations and implementation details of multi-way intersections for our graph pattern matching approach. Then, we give a brief discussion on the index structures used by recent index-based graph matching algorithms.

\vspace*{.5ex}\noindent\textbf{Multi-way set intersection.}  Typical data structures used by set intersection algorithms include sorted arrays and hash sets. Compared to a sorted array, a hash set representation is more efficient. One can compute the intersection between two sorted arrays in time $O(n_1 + n_2)$, and between two hash sets in expected $O(min(n_1, n_2))$ time, where $n_1$ and $n_2$ are the cardinalities of the two inputs. However, a hash set has the following two problems: (1) it uses more memory than a sorted array, and (2) it incurs repeated random accesses to memory for data accessing, in contrast, a sorted array allows the more efficient sequential access.

A bitmap enjoys the benefits of both: the constant-time random access of a hash set, and  the good locality of a sorted array. However, a bitmap suffers from impractical memory usage when the universe size is too large compared to the cardinality of the sets. We therefore decide to represent sets using compressed bitmaps. Specifically, we used RoaringBitmap \cite{ChambiLKG16}, which have been shown to outperform conventional compressed bitmaps such as WAH, EWAH or Concise \cite{WangLPS17}.

We store adjacency lists of graph nodes as well as candidate occurrence sets as instances of RoaringBitmap. To get data out of a bitmap, we use batch iterators given by the RoaringBitmap API. Our empirical studies show that batch iterators run much faster (2-10x) than standard iterators. We implement multi-way intersections using the FastAggregation utility methods from the RoaringBitmap API.

Our {\em MJoin} is the first join algorithm for graph pattern matching that exploits a compressed bitmap data structure to efficiently implement multi-way intersections. Nevertheless, it can be enhanced in the following two ways, which are left as future work.

First, the bitmap data structure can be naturally split into chunks which facilitates parallelising aggregations like bitwise OR and XOR. We can exploit this property to design a parallel graph pattern evaluation algorithm that works with multiple threads. Second, recent set intersection algorithms, such as \cite{DingK11,LemireKKDOSK18,Han0Y18,ZhangLSF20}, focus on designing new data structures for set intersection and/or exploiting single-instruction-multiple-data (SIMD) instructions to enhance the performance of a data structure. {\em EmptyHeaded} \cite{AbergerLTNOR17}, a relational engine for graph processing, exploited these recent work and implemented SIMD set intersections for the trie data structure. We would like to incorporate SIMD-based set intersection algorithms for compressed bitmaps on modern CPUs in our graph pattern matching approach.

%Besides facilitating the set operations, another benefit of using the bitmap data structure is that it can be naturally split into chunks which facilitates parallelise aggregations like bitwise OR and XOR. Nevertheless we leave the investigation of parallel evaluation of graph pattern queries as future work.

%Recent set intersection algorithms, such as \cite{DingK11,LemireKKDOSK18,ZhangLSF20}, focus on designing new data structures for set intersection and/or exploiting single-instruction-multiple-data (SIMD) instructions to enhance the performance of a data structure. {\em EmptyHeaded} \cite{AbergerLTNOR17}, a relational engine for graph processing, implemented SIMD set intersections for the trie data structure. It is interesting to incorporate SIMD-based set intersection algorithms for compressed bitmaps on modern CPUs in our graph pattern matching approach, which we leave as future work.

% We implement a hybrid set intersection method: if the cardinalities of two sets are similar, we use the merge-based method; otherwise, we adopt the Galloping algorithm [1]. The cost of the set intersection is proportional to the cardinality of the smallest set. merge-join = bitwise Anding

%comparison with enumeration without RIG
%{\em GM-N} does not construct a RIG but uses the simulation relation ${\cal FB}$ to directly compute the query occurrences.
% do the intersection with merge-join
% described in GF and Rapid

\vspace*{.5ex}\noindent\textbf{Graph join indexes.} %We note that set intersections on graphs are essentially joins of nodes with their adjacent nodes.
To perform fast matching, many recent graph matching algorithms and contemporary database management systems \cite{ZhangLY09,ZhaoH10,ChengYY11,Veldhuizen14,AbergerTOR16,RiveroJ17,FreitagBSKN20,MhedhbiKS21} build indexes on the input data graph or relations. %Proposed graph index structures include structural pattern-based indexes \cite{ZhangLY09,ZhaoH10,ChengYY11,RiveroJ17}, (hash) trie indexes \cite{AbergerLTNOR17,FreitagBSKN20}, and compressed sparse row (CSR)-based indexes \cite{MhedhbiGKS21}.
A majority of  these index-based graph matching algorithms construct indexes on the input data and evaluate all queries with the assistance of the indexes.   Previous performance studies \cite{KatsarouNT15,Sun019,FreitagBSKN20} show that these index-based methods have severe scalability issues due to the index construction.

For the structural pattern-based algorithms \cite{ZhangLY09,ZhaoH10,ChengYY11,RiveroJ17}, since the number of structural patterns (such as paths, trees, and general subgraphs) is superlinear to the size of a graph, the increase of the number of graph nodes and edges leads to a detrimental increase in the indexing time \cite{KatsarouNT15}.

Graph matching algorithms that adopt worst-case optimal (WCO) style joins usually build a tree-like index structure such as a trie \cite{Veldhuizen14,AbergerLTNOR17,FreitagBSKN20}. These methods either fix a global attribute order \cite{Veldhuizen14,FreitagBSKN20} or consider all permutations of attributes that can partake in a join \cite{AbergerLTNOR17}. Then, they build an index structure for each input relation consistent with a chosen attribute order.  Earlier methods \cite{Veldhuizen14,AbergerLTNOR17} require to build index structures and persist them into disk prior to query processing. The precomputation is expensive and the persistence of index structures entails an enormous storage and maintenance overhead \cite{FreitagBSKN20}.  A recent method \cite{FreitagBSKN20} sorts the input data on-the-fly during query processing.

Unlike the index structures used by the aforementioned algorithms, our light\-weight RIG index structure is query-based, and can be built efficiently by exploiting compressed bitmaps.  Instead of indexing each input relation separately by a given attribute order \cite{Veldhuizen14,AbergerLTNOR17,FreitagBSKN20},  our RIG structure is holistic by taking into account the whole query structure and it is not built according to any specific node order. A RIG compactly encodes all the homomorphisms of the query to the data graph, whose size (total number of nodes and edges) is typically much smaller than the query answer (total number of occurrences). Our experimental evaluation in Section \ref{sec:experiments} shows that the size of RIG over  the size of data graph is under 1\%. Also,  since a RIG is built on-the-fly during query processing and does not have to persisted to disk, it can handle changing data transparently.

% \label{sec:implementations}
%comparison method
% -binary join, left-deep order
% -tree-join algorithm

\section{Experimental Evaluation}
\label{sec:experiments}

We conduct extensive performance studies to evaluate the effectiveness and efficiency of our proposed RIG-based graph pattern matching approach. % and the multi-way intersection-based join algorithm.

\subsection{Setup}
\label{subsec:setup}

We implemented our approach, abbreviated as {\em GM}, the join-based approach ({\em JM}), and the tree-based approach ({\em TM}), for finding homomorphisms of graph pattern queries on data graphs.  Our implementation was coded in Java.

For {\em JM}, we first compute the occurrences for each edge of the input query on the data graph, then find an optimized left-deep join plan through dynamic programming, and finally use the plan to evaluate the query as a sequence of binary joins. For {\em TM}, we first transform the graph pattern query into a tree query,  evaluate the tree query using a tree pattern evaluation algorithm, and filter out occurrences of the tree query that violate the reachability relationships specified by the missing edges of the original query. For the {\em TM} approach, we implemented the recent algorithm for evaluating tree patterns on graphs  described in \cite{WuTSL20}, which has been shown to outperform other existing algorithms. We applied the node pre-filtering technique described in \cite{chen05dags,ZengH12} to both approaches, {\em JM} and {\em TM}, in our implementation.

The above three graph matching algorithms were implemented using a recent efficient scheme, called {\em Bloom Filter Labeling} (BFL) \cite{SuZWY17}, for reachability checking which was shown to greatly outperform most existing schemes  \cite{SuZWY17}.

In addition to pure algorithms, we also experiment with query engines and systems. Specifically, we compared with: (1) a relational engine for graph processing called EmptyHeaded \cite{AbergerTOR16}, referred to as {\em EH}; (2) a recent graph query engine\footnote{https://github.com/queryproc/optimizing-subgraph-queries-combining-binary-and-worst-case-optimal-joins.} {\em GraphflowDB} \cite{MhedhbiKS21}, referred to here as {\em GF}; (3) a recent subgraph query processing engine {\em RapidMatch} \footnote{https://github.com/RapidsAtHKUST/RapidMatch.}, referred to as {\em RM};  and (4) a full graph system Neo4j.  The above four engines/systems were all designed to process child edge-only graph pattern queries (so they do not need reachability indexes).

Our implementation was coded in Java. All the experiments reported here were performed on a workstation running Ubuntu 16.04 with 32GB memory and 8 cores of Intel(R) Xeon(R) processor  (3.5GHz). The Java virtual machine memory size was set to 16GB.

\begin{table}[!t]
\caption{Key statistics of the graph datasets used.}
\label{tab:data_stat}
\begin{center}
\scalebox{.7}{\begin{tabular}{|l|c|c|c|c|c|c|}
\hline
Domain &  Dataset  & $|V|$ &	$|E|$	&  $|L|$	& 	$d_{avg}$\\
\hline\hline
Biology	& Yeast (\em yt) &3.1K	&12K&	71&	8.05\\
        & Human (\em hu) &4.6K	&86K&	44&	36.9\\
        & HPRD (\em hp) &9.4K	&35K&	307&	7.4\\
\hline
Social	& Epinions (\em ep) &76K	&509K&	20&	6.87\\
	    & DBLP (\em db) &317K	&1049K&	20&	6.62\\
\hline
Communi.	& Email (\em em) &265K	&420K&	20&	2.6\\
\hline
Product	& Amazon (\em am) &403K	&3.5M&	3&	6.29\\
\hline
Web	& BerkStan (\em bs) &685K	&7.6M&	5&	11.76\\
	& Google (\em go) &876K	&5.1M&	5&	6.47\\
\hline

\end{tabular}}
\end{center}
\vspace*{-2ex}
\end{table}

\vspace*{.5ex}\noindent\textbf{Datasets.}  We ran experiments on nine real-world graph datasets from the Stanford Large Network Dataset Collection which have been used extensively in previous work \cite{MhedhbiS19, HanKGPH19, Sun020, FreitagBSKN20}. The datasets have different structural properties and come from a variety of application domain: biology, social networks, communication networks, the Web, and product co-purchasing. Table~\ref{tab:data_stat} lists the properties of the datasets.

% We transform undirected graphs into directed graphs by replacing each edge connecting two vertices with two edges.

\begin{figure}[!t]
    \centering%
     \scalebox{.7}{ \epsfig{file=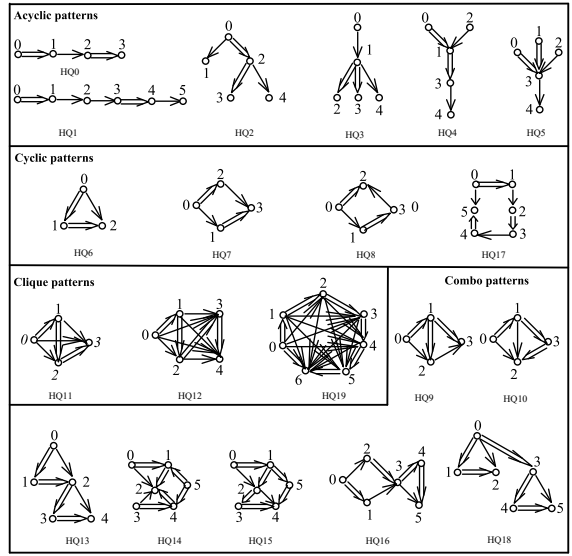}}%
       %\vspace*{-1ex}
     \caption{Hybrid graph pattern queries used for evaluation.}
      \label{fig:graphQs}
\vspace*{-3ex}\end{figure}

%\vspace*{0.5ex}
\noindent\textbf{Queries.} For each dataset, we used three query sets distinguished by the type of queries contained: child edge-only, hybrid and descendant edge-only, abbreviated by C, H, and D, respectively.

Each query set for the biology datasets contains 10 queries. The 10 C-queries were selected from randomly generated queries provided by \cite{Sun020}, consisting of 6 dense queries and 4 sparse queries.  The number of nodes of each query ranges ranging from 4 to 20 for {\em hu}, and from 4 to 32 for {\em hp} and {\em yt}. We constructed D-queries (resp. H-queries) by turning all (resp. with 50\% probability) the edges of C-queries into descendant edges .

Each of the remaining six datasets in Table \ref{tab:data_stat} used three designed query sets (of type C, H and D in respective).  Each query set has 20 queries which are grouped into four classes: acyclic, cyclic, clique, and combo patterns. We call a graph pattern query {\em acyclic} if its corresponding undirected graph is acyclic, and {\em cyclic} otherwise. A pattern is called {\em combo} if its undirected graph contains more than two cycles. A pattern is called {\em clique} if its undirected graph is complete.

The templates of H-queries are shown in Figure \ref{fig:graphQs}, where double line edges denote descendant edges, while single line edges denote child edges. The number associated with each node of a query template denotes the node id. Query instances are generated by assigning labels to nodes. C-queries and D-queries have the same structure as H-queries but replacing edges by child and descendant edges, respectively. Many of the C-queries and D-queries were used in previous work \cite{ChengYY11,MhedhbiKS21}.

\vspace*{1ex}\noindent\textbf{Metrics.} We measured the runtime of individual queries in a query set. For query listing, this includes two parts: (1) the matching time, which consists of the time spent on filtering vertices, building auxiliary data structures such as runtime index graphs (RIGs), and generating query plans (or search order), and (2) the result enumeration time, which is the time spent on enumerating occurrences. The number of occurrences for a given query on a data graph can be quite large. Following usual practice \cite{HanKGPH19,SunSC0H20,Sun020}, we terminated the evaluation of a query after finding a limited number of ) matches (we set it to be $10^7$ in the experiments) covering as much of the search space as time allowed. We stopped the execution of a query if it did not complete within 10 minutes, so that the experiments could be completed in a reasonable amount of time. %We refer to these queries as unsolved.
We recorded the elapsed time of these stopped queries as 10 minutes.

\begin{figure*}[!t]
	\center
    \subfigure[{\em em}]{ \scalebox{0.46}{ \label{fig:emhtime} \epsfig{file=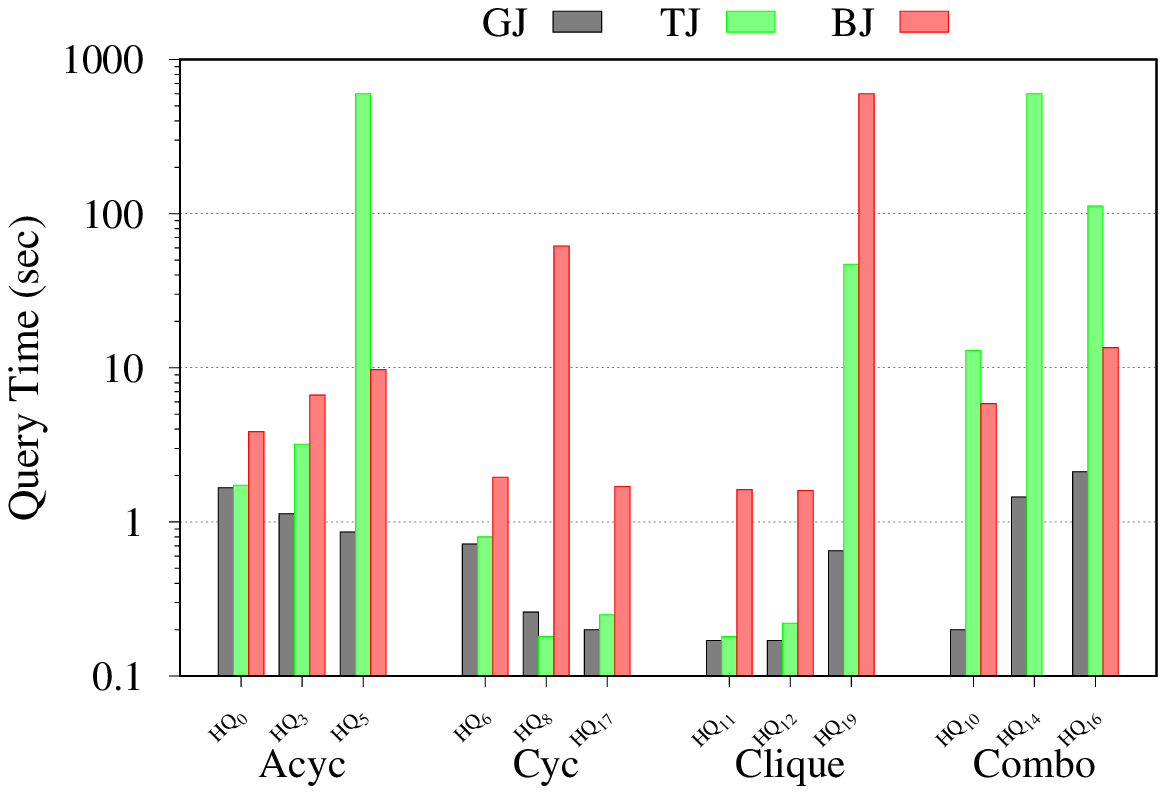} } }
    \subfigure[{\em ep}]{ \scalebox{0.46}{ \label{fig:ephtime} \epsfig{file=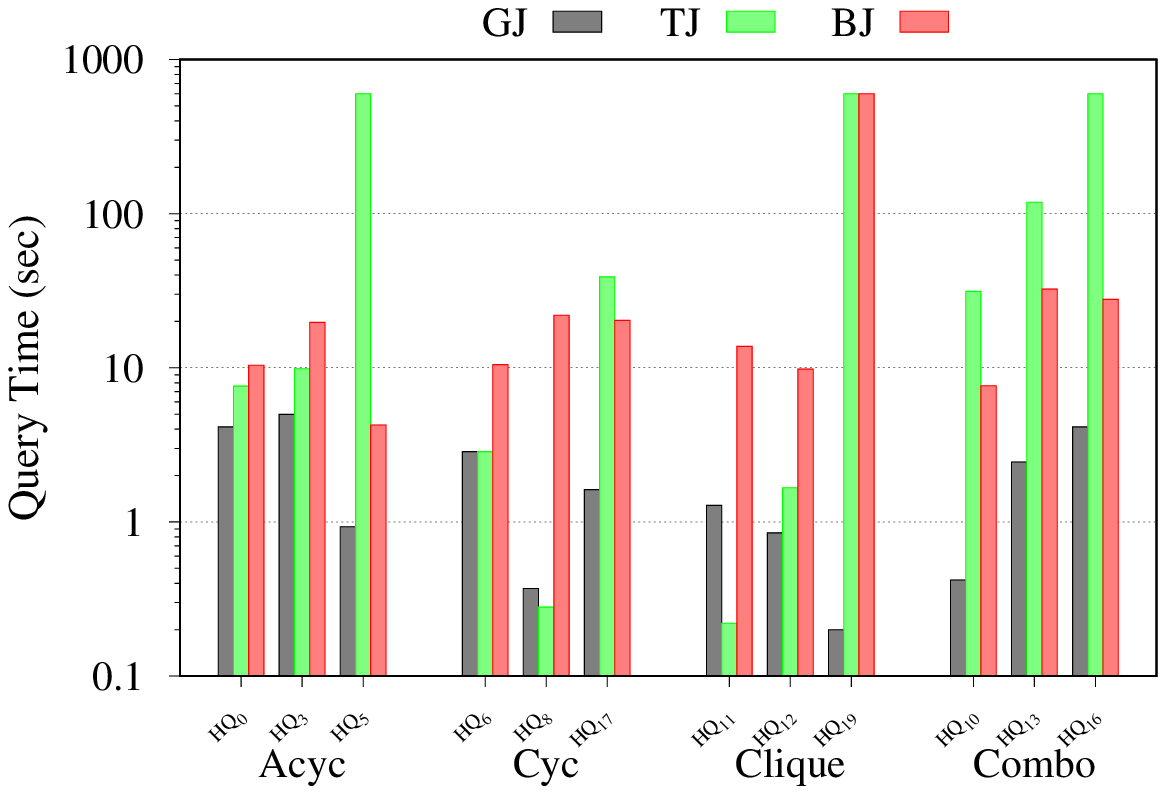} } }
    \subfigure[{\em hp}]{ \scalebox{0.46}{ \label{fig:hphtime} \epsfig{file=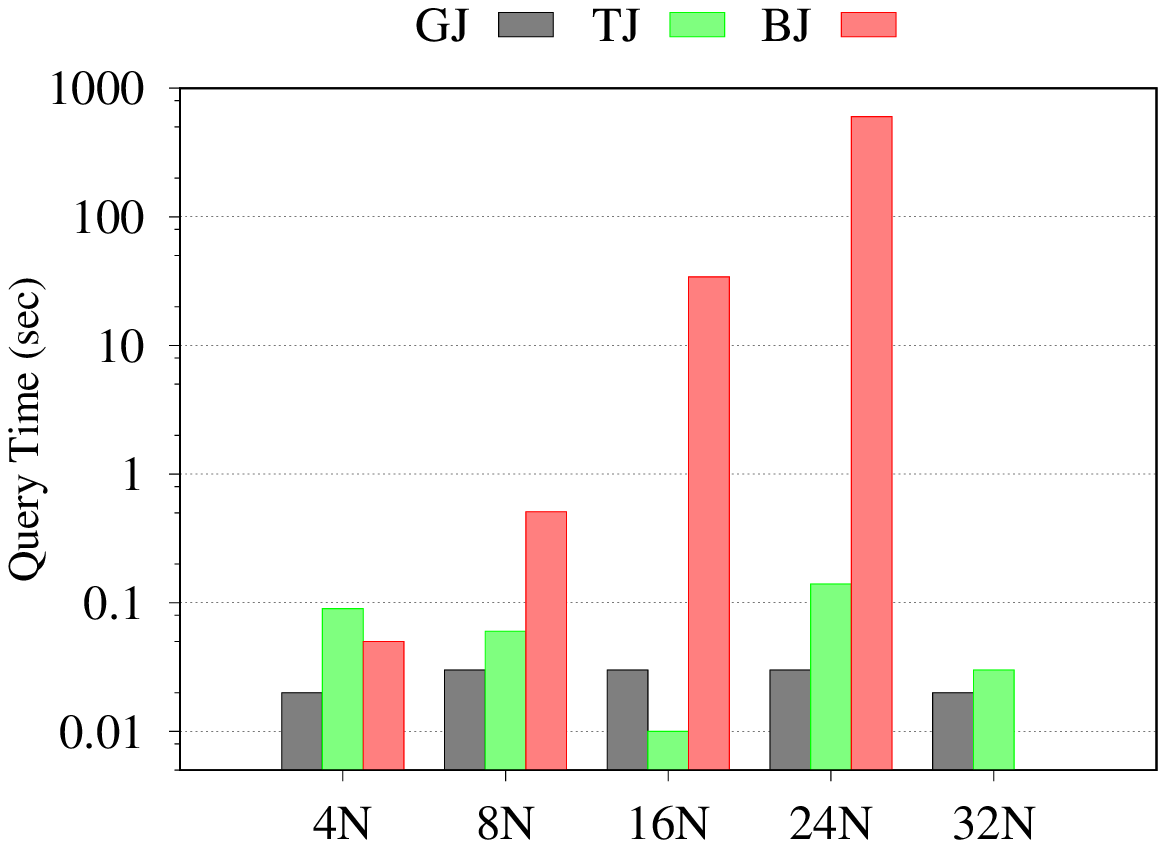} } }
      \subfigure[{\em yt}]{ \scalebox{0.46}{ \label{fig:ythtime} \epsfig{file=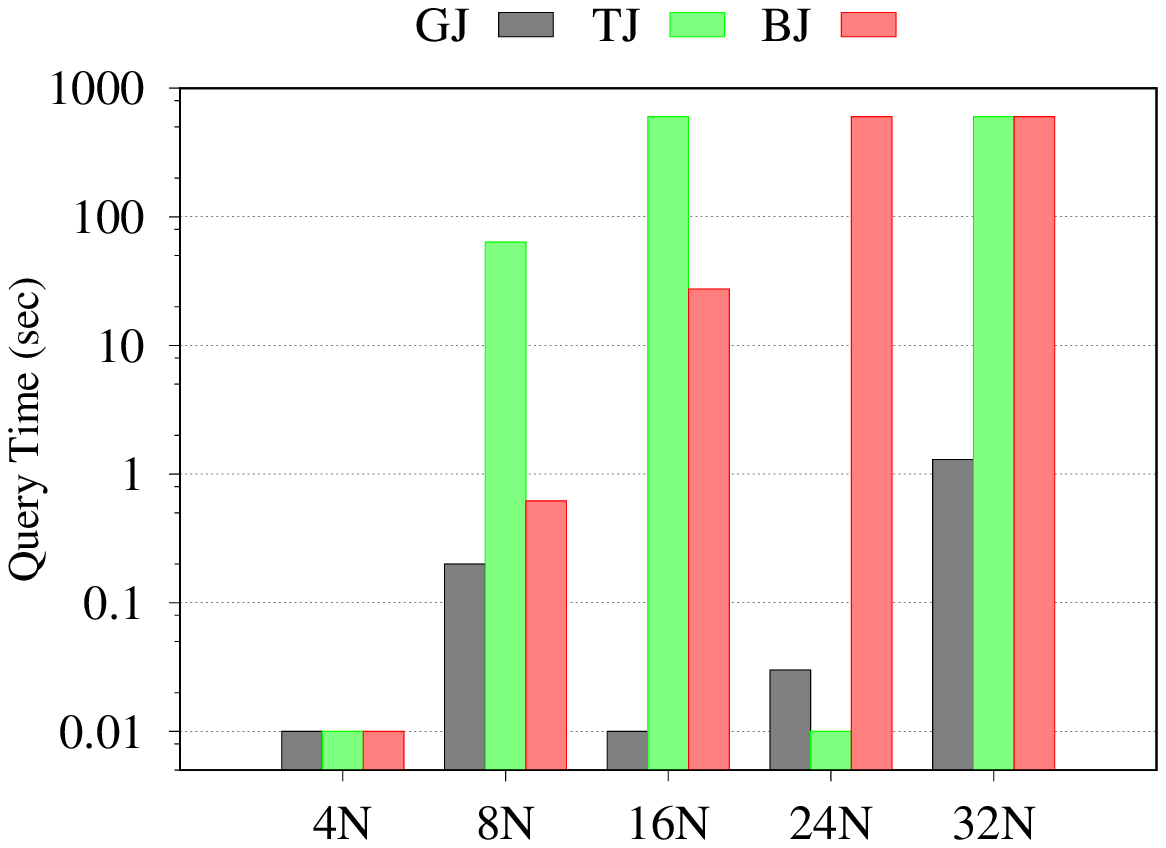} } }
      \subfigure[{\em hu}]{ \scalebox{0.46}{ \label{fig:huhtime} \epsfig{file=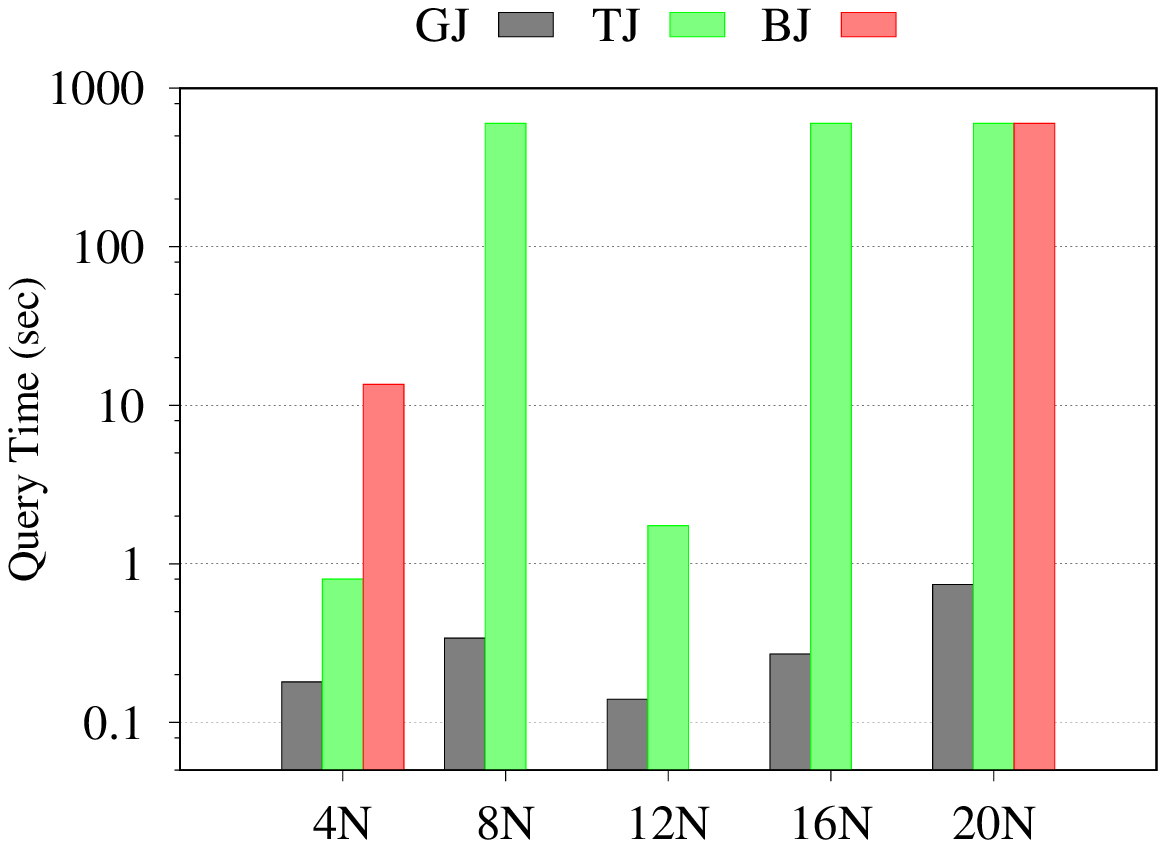} } }
    \caption{H-query evaluation time of {\em GM, TM} and {\em JM} on {\em em, ep, yt, hu} and {\em hp}.}
	\label{fig:htime}
\end{figure*}

\subsection{Time Performance}

We now compare time performance of the three algorithms {\em JM, TM, } and {\em GM} on evaluating different types of graph pattern queries. Unless otherwise stated,  the pattern transitive reduction (see Section \ref{sec:reduction}) was applied to each query before it was evaluated by the algorithms. Node pre-filtering \cite{chen05dags,ZengH12} was applied in all cases except for {\em GM} on C-queries where it is not beneficial.

\vspace*{1ex}\noindent\textbf{Evaluating H-queries.} Fig. \ref{fig:htime} shows the execution time of the three algorithms on evaluating hybrid queries on different data graphs.  Specifically,  Fig. \ref{fig:emhtime} and \ref{fig:ephtime} show the results of evaluating hybrid query instances of the query templates of Fig. \ref{fig:graphQs} on graphs {\em em} and {\em ep}, respectively. Due to space limit, the figures only show the results of three queries from each of the acyclic, cyclic, clique, and combo pattern classes. Fig.\ref{fig:hphtime}, \ref{fig:ythtime},  and \ref{fig:huhtime} show the results of evaluating random hybrid queries on graphs {\em hp}, {\em yt} and {\em hu}, respectively. The x-axis represents number of nodes of each query. Again, we chose 5 queries for each data graph due to space limit.

The overall best performer is {\em GM}. It is able to solve all the queries in all cases (including different pattern classes, dense queries, and large queries. It outperforms {\em TM} and {\em JM} by up to two and three orders of magnitude, respectively.  Both {\em TM} and {\em JM} were unable to solve all the queries. For example, both failed to solve the combo pattern query $HQ_{14}$ on {\em em} and the 7-clique query $HQ_{19}$ on {\em ep}. {\em JM} in particular had a high percentage of unresolved cases on large queries (which have more than 10 nodes) on {\em hp}, {\em yt} and {\em hu}. {\em TM} has relatively good performance on {\em hp}, because it has small candidate tuples to compute; but it failed for more than half of the times on the dense dataset {\em hu}.

A large percentage of the failures of {\em JM} is due to the out-of-memory error, since it generates a large number of intermediate results during the query evaluation. Another cause of the inefficiency of {\em JM} is due to the join plan selection. As described in \cite{ChengYY11}, in order to select an optimized join plan, {\em JM} uses dynamic programming to exhaustively enumerate left-deep tree query plans. For queries with more than 10 nodes, the number of enumerated query plans can be huge. For example, for a query with 24 nodes on {\em hp}, {\em JM} enumerates 2,384,971 query plans in total.

Most of the failures of {\em TM} is due to a time out error. Recall that {\em TM} works by evaluating a tree query of the original graph query. For each tuple of the tree query, it checks the non-tree edges for satisfaction. Hence, its performance is tremendously affected when the number of tree solutions is very large.

The experiment confirms the advantages of our proposed RIG-based graph pattern matching techniques over both {\em TM} and {\em JM}.

\begin{figure*}[!t]
	\center
    %\subfigure[{\em em}]{ \scalebox{0.5}{ \label{fig:emctime} \epsfig{file=plots/emctime.eps} } }
    \subfigure[{\em ep}]{ \scalebox{0.46}{ \label{fig:epctime} \epsfig{file=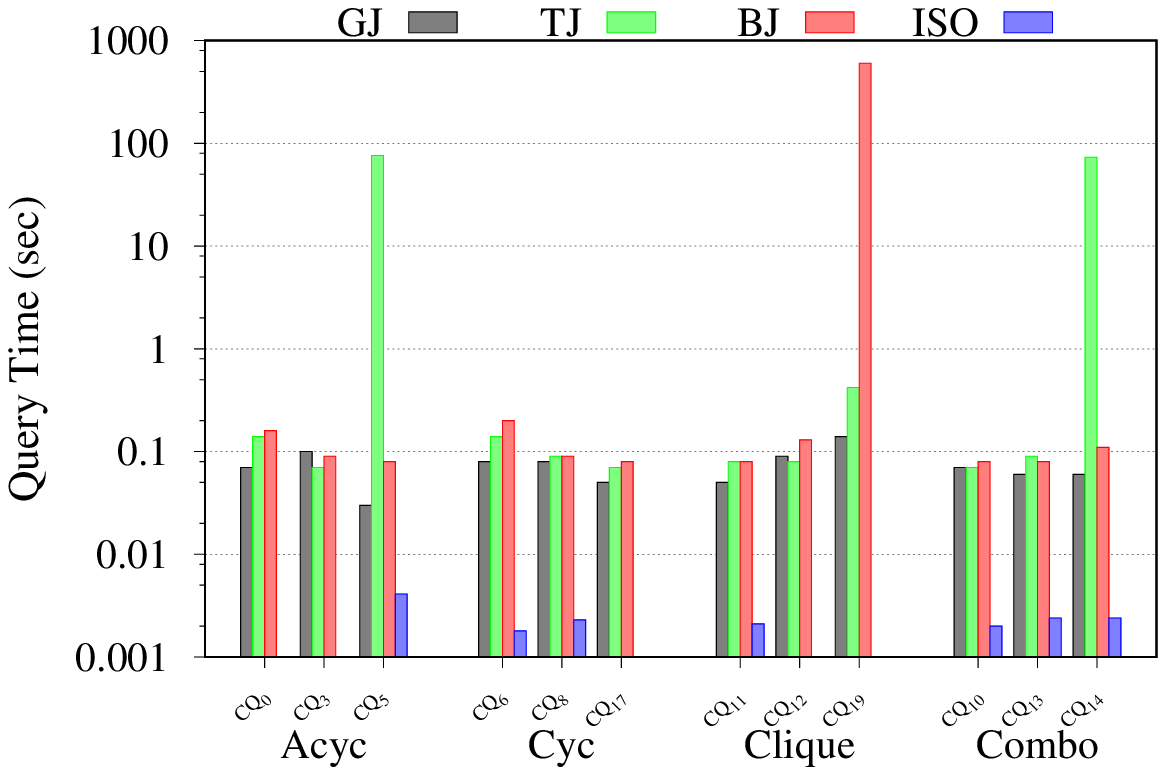} } }
     \subfigure[{\em bs}]{ \scalebox{0.46}{ \label{fig:bsctime} \epsfig{file=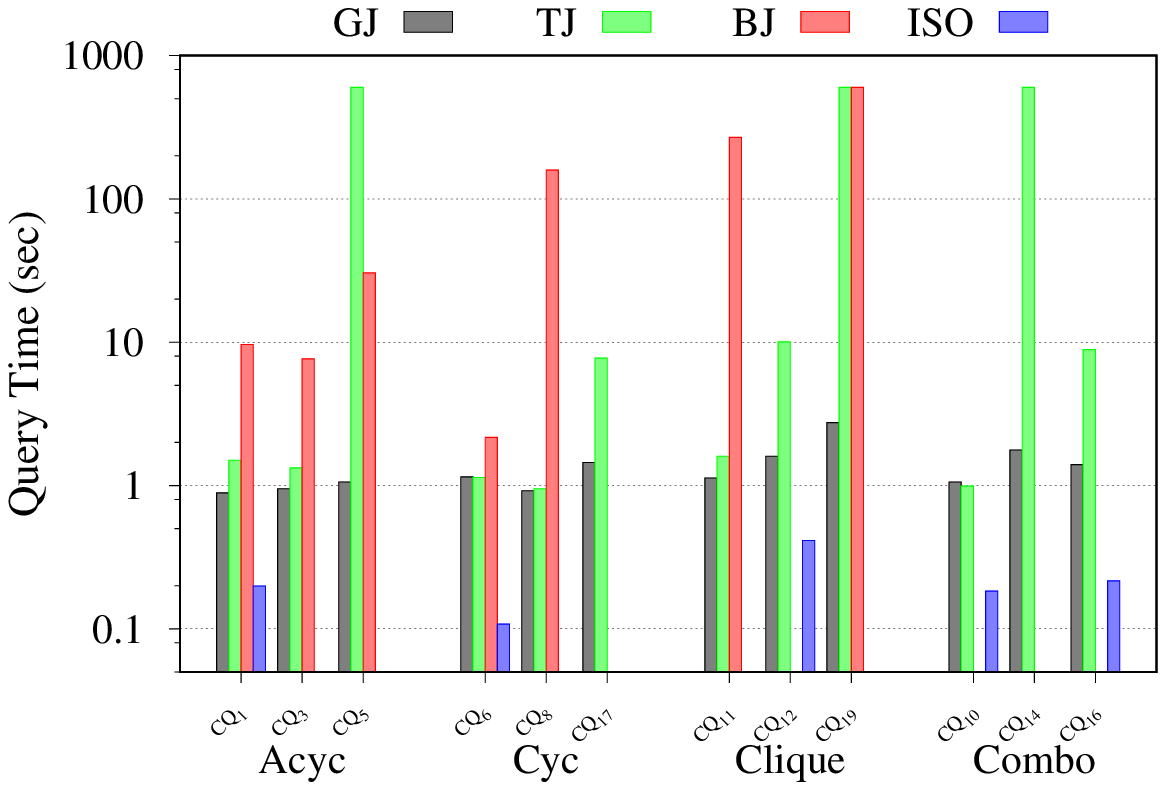} } }
    %\subfigure[{\em yt}]{ \scalebox{0.44}{ \label{fig:ytctime} \epsfig{file=plots/ytctime.eps} } }
    \subfigure[{\em hu}]{ \scalebox{0.46}{ \label{fig:huctime} \epsfig{file=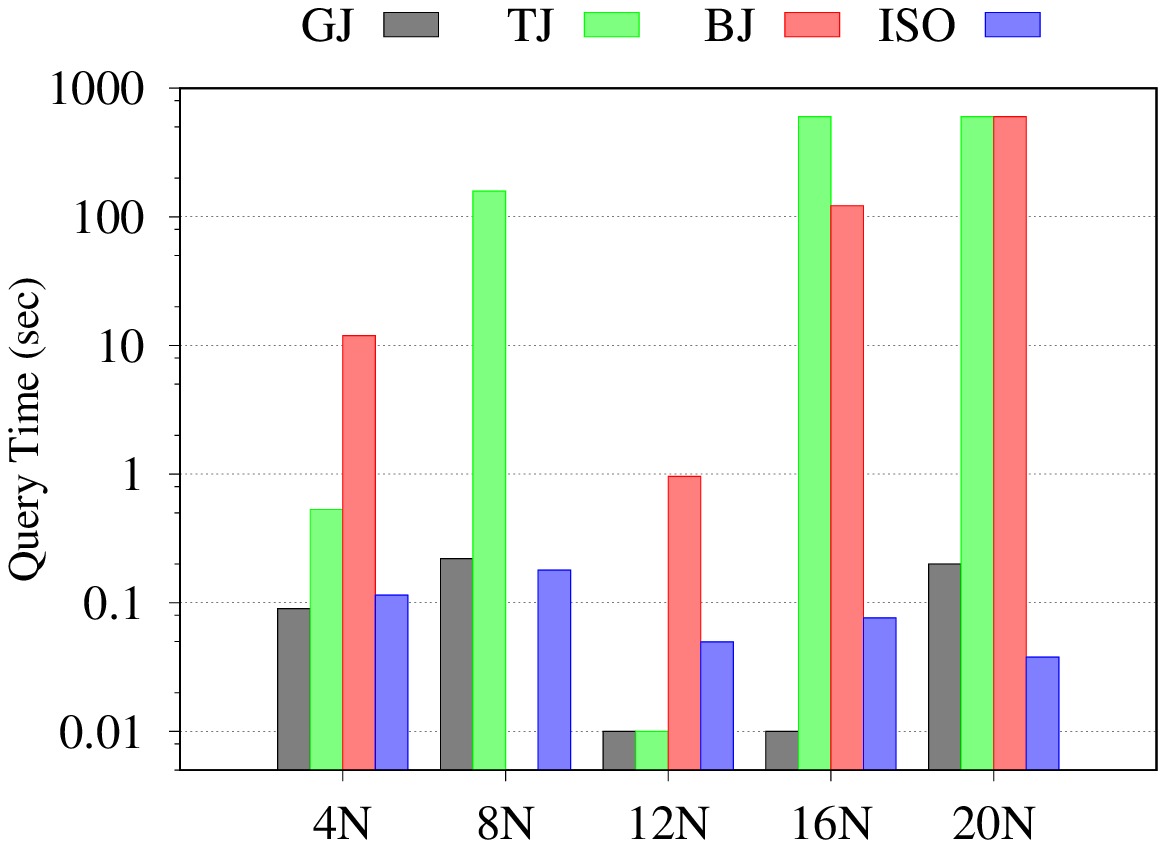} } }
    %\subfigure[{\em hp}]{ \scalebox{0.44}{ \label{fig:hpctime} \epsfig{file=plots/hpctime.eps} } }

    %\subfigure[{\em go}]{ \scalebox{0.44}{ \label{fig:goctime} \epsfig{file=plots/goctime.eps} } }
    %\subfigure[{\em am}]{ \scalebox{0.44}{ \label{fig:amctime} \epsfig{file=plots/amctime.eps} } }
    %\caption{C-query evaluation time of {\em GJ, TJ, BJ} and {\em ISO}  on {\em em, ep, yt, hu, hp, am, bs,} and {\em go}.}
    \caption{C-query evaluation time of {\em GM, TM, JM} and {\em ISO} on {\em ep, bs} and {\em hu}.}
	\label{fig:ctime}
\end{figure*}

\vspace*{1ex}\noindent\textbf{Evaluating C-queries.}  Fig. \ref{fig:epctime}, \ref{fig:bsctime} and \ref{fig:huctime} show the results of evaluating C-queries on graphs {\em ep}, {\em bs}, and {\em hu}, respectively. Queries for the first two are instances of query templates described above, while queries for the third one are random queries originally used in \cite{Sun020}.

Again, {\em GM} shows the best average performance and it can solve all the queries in all the cases. While {\em JM} has slightly better performance than {\em GM} on {\em ep}, on {\em hu} (resp. {\em bs}), it is unable to solve 40\% (resp. 50\%) of the queries and is up to 4 orders magnitude (resp. 84$\times$) slower than {\em GM} on its solved queries. On both {\em bs} and {\em hu}, {\em TM} solves more queries  than {\em JM}, and it outperforms {\em JM} by up to 166$\times$ on the solved queries, but on {\em ep}, its average performance is 119$\times$ slower than {\em JM}.

\vspace*{1ex}\noindent\textbf{Isomorphism vs homomorphism.} Subgraph isomorphism is more restrict than subgraph homomorphism, since the former is defined by injective mapping from pattern nodes to data nodes while the latter is defined by a function. Hence, in addition to respect the structural constraint specified by the query, a subgraph isomorphism algorithm needs also to deal with the one-to-one mapping constraint.

Fig. \ref{fig:ctime} shows the results of the subgraph isomorphism algorithm {\em ISO} \cite{Sun020} evaluating the same queries. {\em ISO} integrates techniques from state-of-the-art subgraph isomorphism algorithms \cite{Sun020}.  Its implementation has been highly optimized to aggressively prune nodes (that violate the two constraints) from the search space.   From Fig. \ref{fig:ctime},  we observe that, on {\em ep} and {\em bs}, {\em ISO} runs faster than the three homomorphism algorithms on a number of queries thanks to optimization techniques applied. However, it reports the segmentation fault error for quite a few other queries, for example, $CQ_0, CQ_3, CQ_{12}, CQ_{17}$ and $CQ_{19}$ on {\em ep}. For query $8N$ on {\em hu}, {\em ISO} is about 734$\times$ slower than {\em GM}, and it reports failure for queries $16N$ and $20N.$  This is because {\em hu} has a higher average degree and fewer distinct labels, which make it more challenging for subgraph isomorphism matching \cite{Sun020}.   Our results demonstrate that it is promising to extend {\em GM} to efficiently compute injective graph pattern matches by leveraging optimization techniques of {\em ISO}.

%While it is not an exact apple to apple comparison, we

\vspace*{1ex}\noindent\textbf{Evaluating D-queries.} %We have measured the performance of {\em JM}, {\em TM} and  {\em IJ} for evaluating ten random queries over Human, HPRD and Yeast. The ten queries for each data graph are ordered in  ascending number of query nodes, ranging from 4 to 20 for Human, and from 4 to 32 for HPRD and Yeast.
We have measured the performance of {\em JM}, {\em TM} and  {\em GM} for evaluating D-queries on different data graphs. Due to space limit, we show here the results for evaluating ten random D-queries over {\em hu, hp}, and {\em yt}. The ten queries for each data graph are ordered in  ascending number of query nodes. In Table~\ref{tab:dtime}, we record, for each algorithm, the number of unsolved queries in two categories: time out and out of memory.  We record also the number of solved queries as well as the average runtime of solved queries for each algorithm.

We observe that {\em GM} has the best performance overall among the three algorithms. It is able to solve all the given queries. In contrast, {\em JM} is only able to solve the first 2 or 4 queries on each data graph, and the number of nodes of solved queries are no more than 8.  While {\em TM} solves more and larger queries than {\em JM}, it is up to two orders of magnitude slower than {\em GM}. The results are consistent with those for H-queries and C-queries.

\begin{table}[!t]
\caption{Performance of {\em BJ, TJ} and {\em GM} for evaluating large D-queries on {\em hu, hp} and {\em yt}.}
\label{tab:dtime}
\begin{center}
\scalebox{.7}{\begin{tabular}{|l|c|c|c|c|c|}
\hline
Dataset  & Alg. & Time  &	Out of & Solved  &	Avg. time of  \\
 &  & out &	 memory &  queries &	solved queries (sec.) \\
\hline\hline
{\em hu}	 &{\em JM}  & 1	&7	&2	&1.51\\
	     &{\em TM}  & 3	&0	&7	&16.7\\
	     &{\em GM}  & 0   &0  &\textbf{10} &\textbf{0.53} \\
\hline
{\em hp}	 &{\em JM}  &	2 &4  &4 &1.86\\
	     &{\em TM}  & 1 &0  &9 &134.21\\
	     &{\em GM}   &	0 &0  &\textbf{10} &\textbf{0.58}\\
\hline
{\em yt}	 &{\em JM}  &	5 &3  &2 &0.14\\
	     &{\em TM}  &	3 &0  &7 &20.8\\
	     &{\em GM}   &	0 &0  &\textbf{10} &\textbf{0.34}\\
\hline
\end{tabular}}
\end{center}
%\vspace*{-2ex}
\end{table}

\begin{figure*}[!t]
       \center
       \subfigure[$HQ_2$]{ \scalebox{0.46}{ \label{fig:emhscaleq2} \epsfig{file=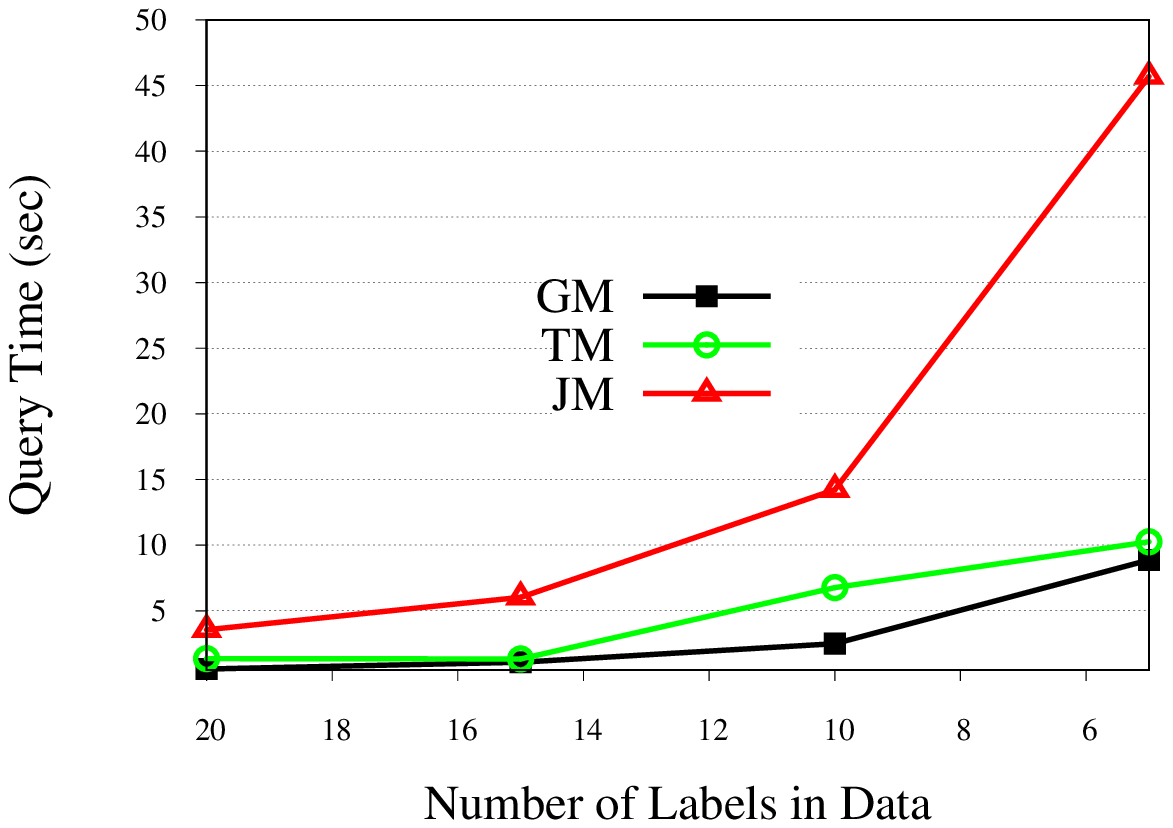} } }
       \subfigure[$HQ_4$]{ \scalebox{0.46}{ \label{fig:emhscaleq4} \epsfig{file=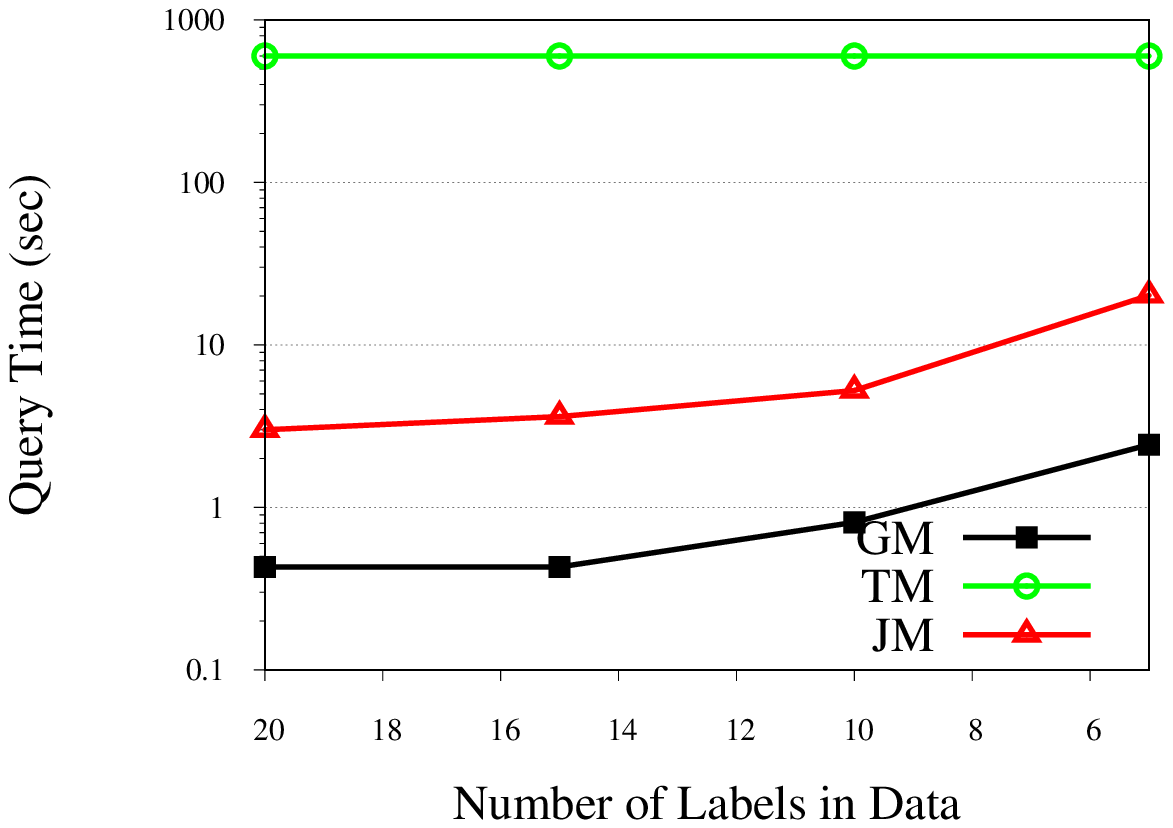} } }
       \subfigure[$HQ_7$]{ \scalebox{0.46}{ \label{fig:emhscaleq7} \epsfig{file=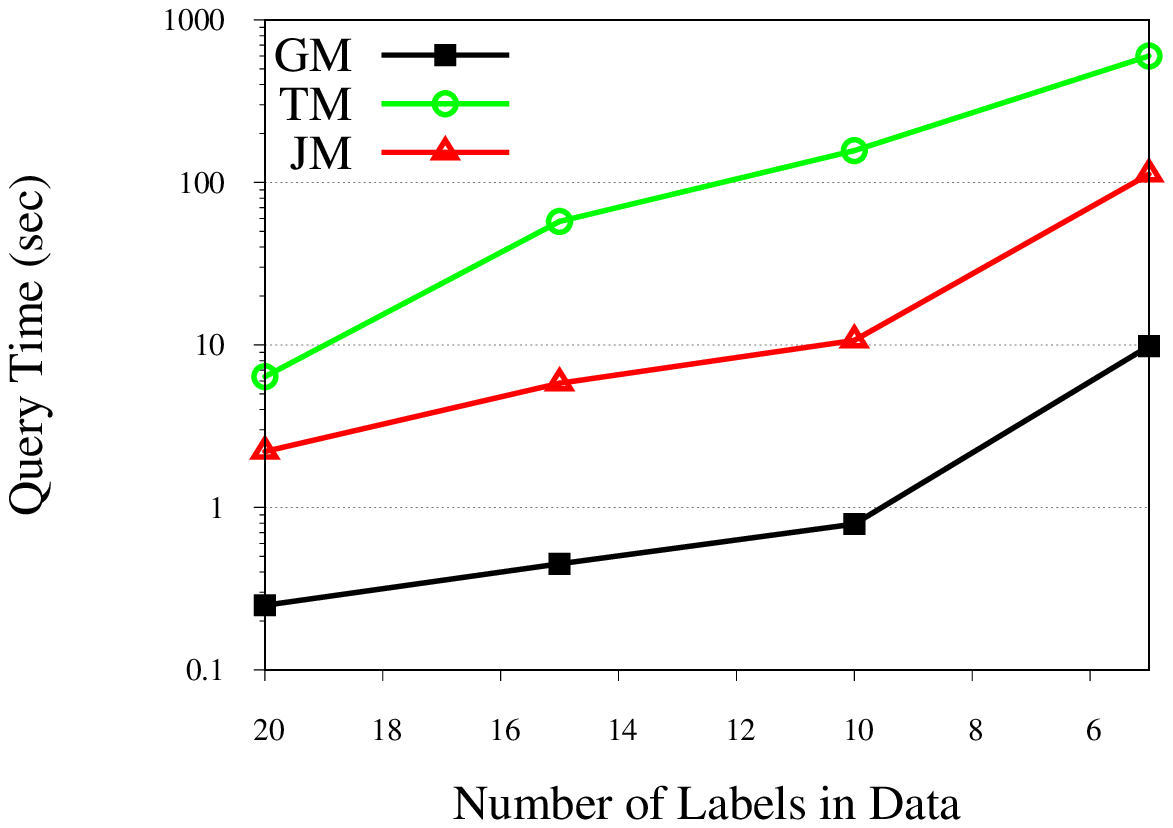}} }
       \subfigure[$HQ_{18}$]{ \scalebox{0.46}{ \label{fig:emhscaleq18} \epsfig{file=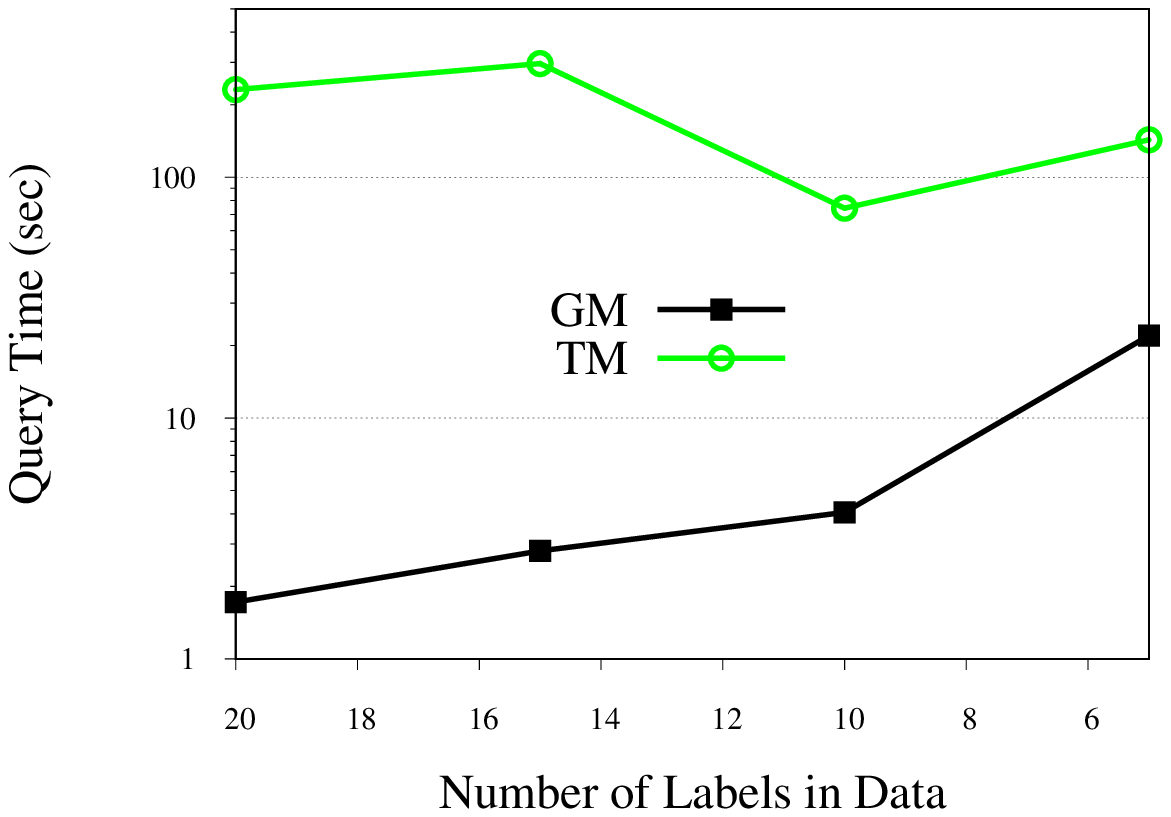} } }

        \caption{Elapsed time of H-queries when increasing number of labels on {\em em}.}
      \label{fig:emhscale}
      %\vspace*{-4ex}
\end{figure*}

\subsection{Scalability}

\vspace*{1ex}\noindent\textbf{Varying data labels.} In this experiment, we examine the impact of the total number of distinct graph labels on the query performance of the algorithms in comparison. We used the {\em em} graph, to produce 4 versions of it where the number of labels increases from 5 to 20 (the size of the graph is fixed). On these four versions of {\em em}, we evaluated one set of 20 hybrid query instances of the query templates of Fig. \ref{fig:graphQs}.

Fig.~\ref{fig:emhscale} reports on the query time of the three algorithms on the four queries: $HQ_2$, $HQ_4$, $HQ_7$, and $HQ_{18}$. We observer that the execution time of the algorithms tends to increase while decreasing the total number of graph labels. In particular, the increase rate becomes steeper when the number is close to 5. This is reasonable since the average cardinality of the input label inverted lists in a graph increases when the number of distinct labels in the graph decreases.

We can see that {\em GM} has the best performance in all the cases. While {\em TM} has compatible performance with {\em GM} on the tree pattern query $HQ_2$, it is outperformed by {\em GM} by up to orders of magnitude for the three graph pattern queries. In particular, it could not complete within 10 minutes for evaluating $HQ_4$ on all the four versions of {\em em}, due to the large number of intermediate results generated.  {\em JM} is up to 13$\times$ slower than {\em GM} for $HQ_2$, $HQ_4$, and $HQ_7$, but it failed to solve $HQ_{18}$ due to the out-of-memory error.

%mention the inaccuracy of the cost model used late

\begin{figure*}[!t]
       \center
       \subfigure[$HQ_8$]{ \scalebox{0.46}{ \label{fig:dphq8} \epsfig{file=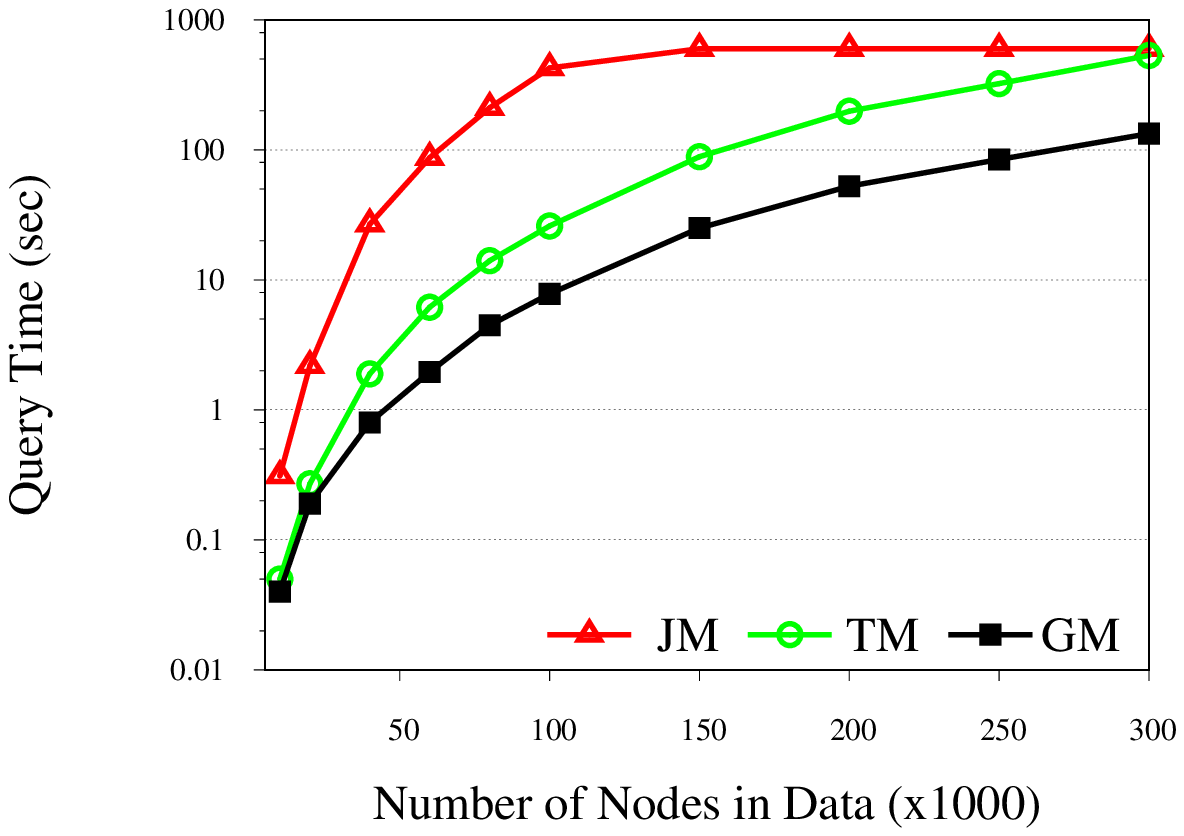} } }
       \subfigure[$HQ_{12}$]{ \scalebox{0.46}{ \label{fig:dphq12} \epsfig{file=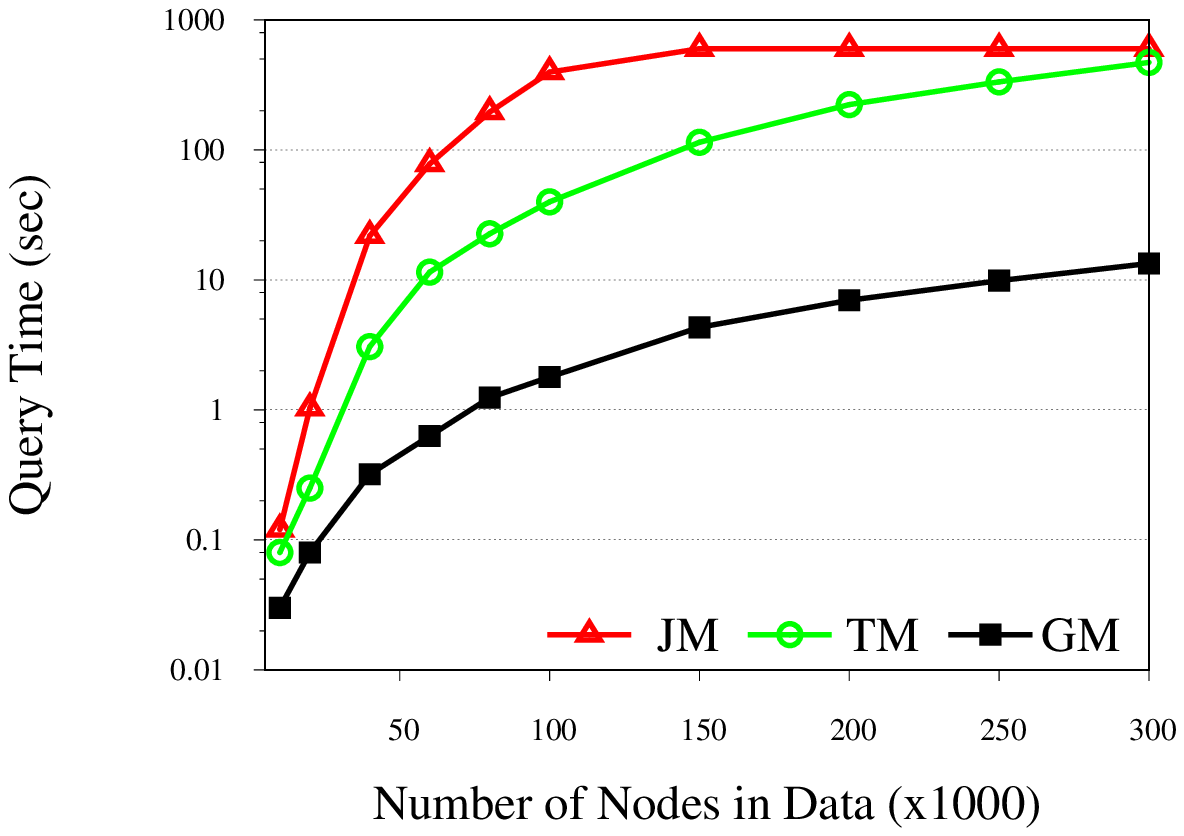} } }
       %\subfigure[$HQ_{15}$]{ \scalebox{0.46}{ \label{fig:dphq15} \epsfig{file=plots/dblpscaleq15.eps} } }
       %\vspace*{-3ex}
        \caption{Elapsed time of H-queries on increasingly larger subsets of {\em dp}.}
      \label{fig:dpscale}
      %\vspace*{-4ex}
\end{figure*}

\vspace*{.5ex}\noindent\textbf{Varying graph sizes.}  We evaluated the scalability of the three algorithms as the data set size grows. For this experiment, we recorded the elapsed query time on increasingly larger randomly chosen subsets of the DBLP data. Fig.~\ref{fig:dpscale} shows the results of the three algorithms evaluating instantiations of the query templates $HQ_8$ and $HQ_{12}$ shown in Fig.~\ref{fig:graphQs}. As expected, the execution time for all algorithms increases when the total number of graph nodes increases. {\em GM} scales smoothly compared to both {\em TM} and {\em JM} for evaluating the two queries.
%{\em GM} again provides significantly better performance than both {\em TM} and {\em JM} for evaluating the two queries.

\subsection{Effectiveness of New Framework}

In this subsection, we evaluate the effectiveness of our proposed techniques and strategies in reducing the overall querying time, including child constraint checking, RIG index construction and space consumption, pattern transitive reduction, and the join-based ordering strategy.

\vspace*{1ex}\noindent\textbf{Child constraints checking.} We compare three methods to check the satisfaction of child constraints specified in queries. Specifically, the operation checks whether a node pair $(v_{q_i}, v_{q_j})$ is an occurrence of a child query edge $(q_i, q_j).$

let $A_{v_q}$ denote the adjacency list of node $v_q$ in the graph. One method, denoted as {\em binSearch}, uses a binary search to check if $v_{q_j}$ is in $A_{v_{q_i}}$. The second method, denoted as {\em bitIter}, converts the child relationship checking into a set intersection operation, it checks if $v_{q_j}$ is in the intersection of $A_{v_q}$ with $cos(q_j)$, the candidate occurrence list of $q_j.$ Both the add $A_{v_q}$ and $cos(q_j)$ are stored as bit vectors, and the intersection is implemented using a bitwise AND operation. The third method, denoted as {\em bitBat}, is described in Section \ref{subsec:rigBuild}. It finds all the nodes in $cos(q_j)$ satisfying a child relationship with $v_{q_i}$ in one step.

\begin{figure}[!t]
       \center
       \subfigure[Child constraints checking]{ \scalebox{0.46}{ \label{fig:emcrelchk} \epsfig{file=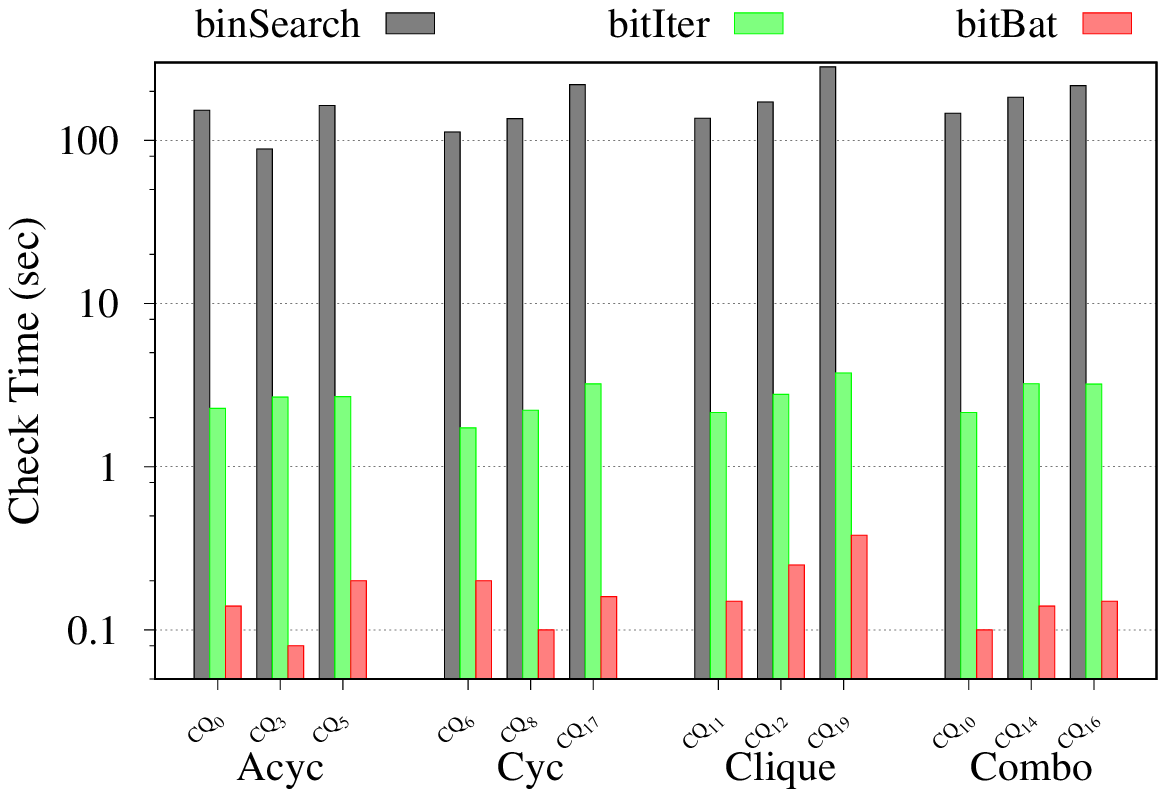} } }
       \subfigure[Simulation relation building]{ \scalebox{0.46}{ \label{fig:emhsimbuild} \epsfig{file=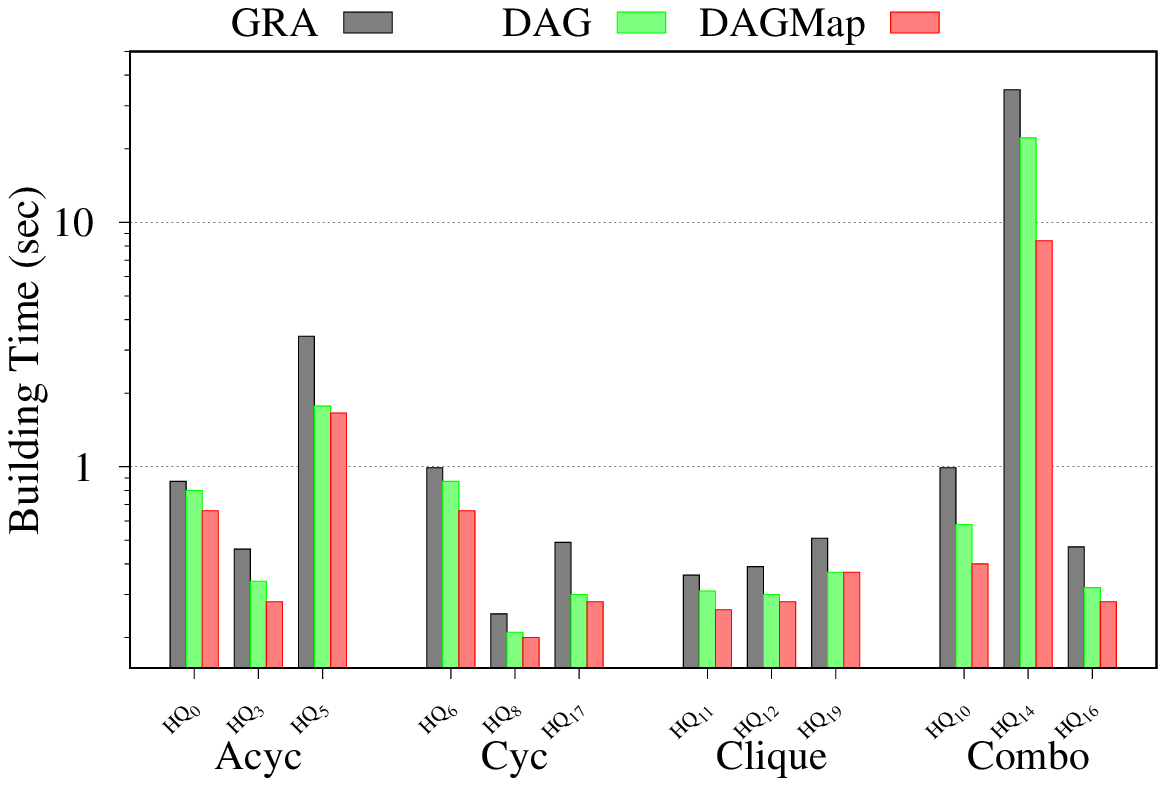}} }
       \caption{Child constraints checking and simulation relation building time on {\em em}.}
      \label{fig:emcrelsim}
      %\vspace*{-4ex}
\end{figure}

Fig.\ref{fig:emcrelchk} shows the running time of the three methods when evaluating C-queries using {\em GM} on graph {\em em}.  As before, due to space limitations, the figure only plots the results of three queries from each of the acyclic, cyclic, clique, and combo pattern classes.  As expected, {\em bitBat} consistently outperforms the other two methods by a large margin; {\em bitIter} comes the next and it is about 75$\times$ faster than {\em binSearch}.

\vspace*{1ex}\noindent\textbf{Simulation relation construction.} We compare three methods to construct ${\cal FB}$ double simulations.  One method, denoted here as {\em Gra}, is the baseline algorithm described in Section \ref{subsec:fbsimbas}. {\em Bas} first selects an arbitrary processing order on the query edges, then it builds ${\cal FB}$ by repeatedly visiting edges by that order and removing redundant nodes from candidate occurrence nodes of the query nodes of each query edge until a fixed point is reached. The other two methods,  denoted here as {\em Dag} and {\em DagMap}, are described in Section \ref{subsec:fbsimdag}. Both {\em Dag} and {\em DagMap} explore the pattern structure to construct ${\cal FB}$, which can stabilize faster by reducing the number of iteration passes of {\em Bas}. {\em DagMap} further applies the optimization techniques described in Section \ref{subsec:rigBuild} to speedup convergence for ${\cal FB}$ computation.

Fig.\ref{fig:emhsimbuild} shows the running time of the three methods when evaluating H-queries using {\em GM} on graph {\em em}.  The figure plots the results of three queries from each of the acyclic, cyclic, clique, and combo pattern classes.  As we can see, {\em DagMap} consistently outperforms the other two methods and {\em Dag} comes the next. The result demonstrates the effectiveness of our proposed techniques.

We also compared {\em Gra} with {\em Dag}+$\Delta$ described in Section \ref{subsec:fbsimdag} on computing simulations for queries with directed cycles. For this, we modified graph pattern queries used in the previous experiments by changing directions of a couple of edges. The results of the two methods for these queries on the data graphs in Table~\ref{tab:data_stat} show that {\em Dag}+$\Delta$ is faster than {\em Gra} by around 1.2$\times$ in every case. We omit the details due to page limits.

\begin{figure*}[!t]
       \center
       \subfigure[Size]{ \scalebox{0.46}{ \label{fig:epagsize} \epsfig{file=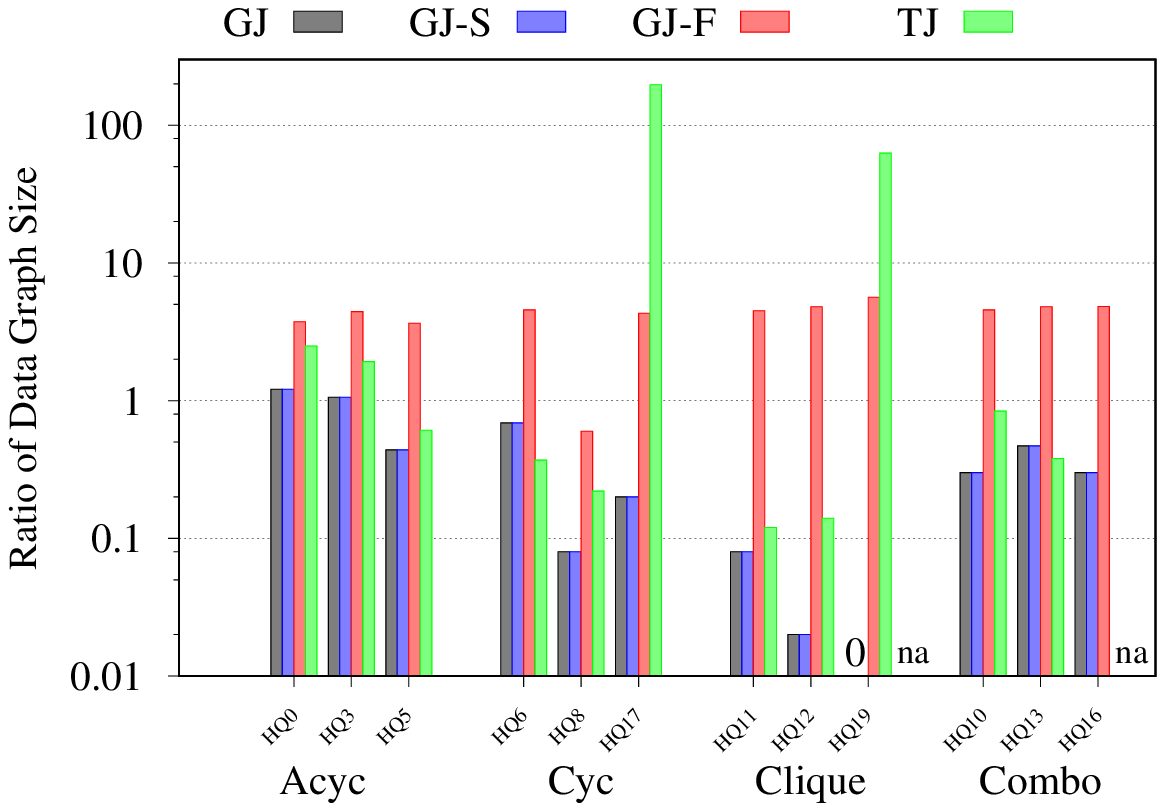} } }
       \subfigure[Construction time]{ \scalebox{0.46}{ \label{fig:epagtime} \epsfig{file=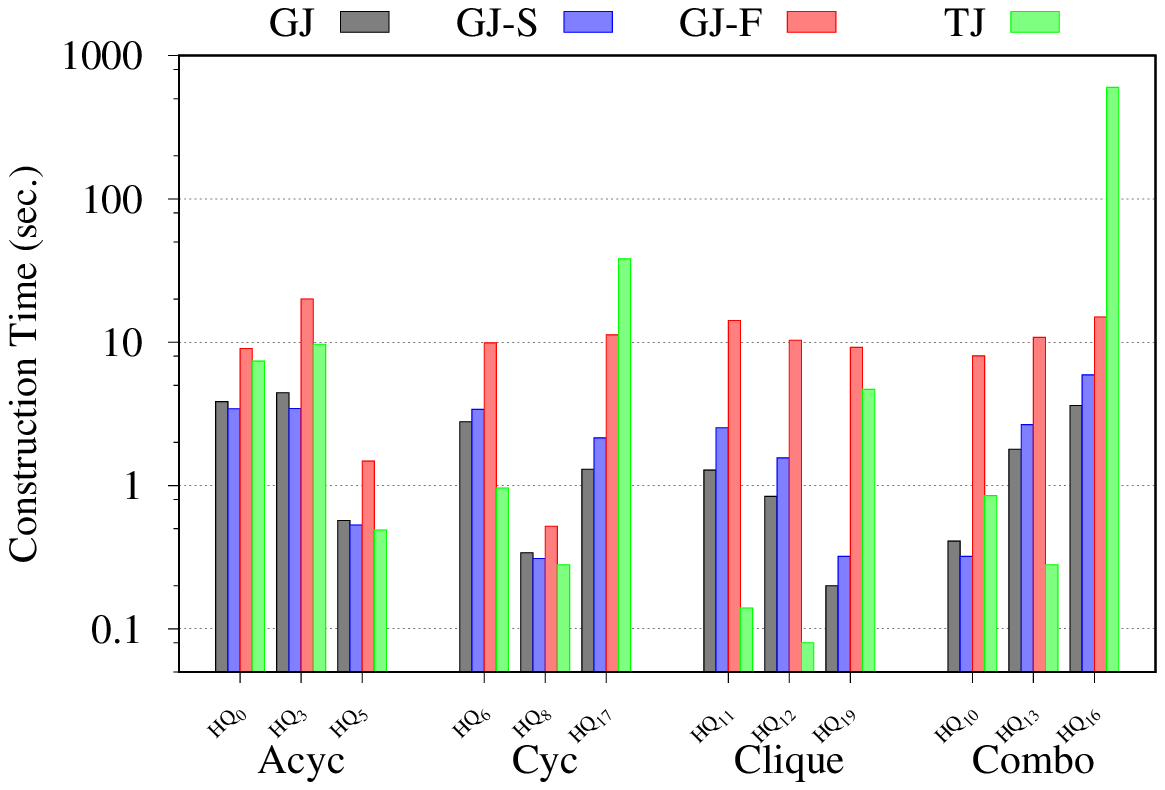}} }
       \subfigure[Query time]{ \scalebox{0.46}{ \label{fig:epetime} \epsfig{file=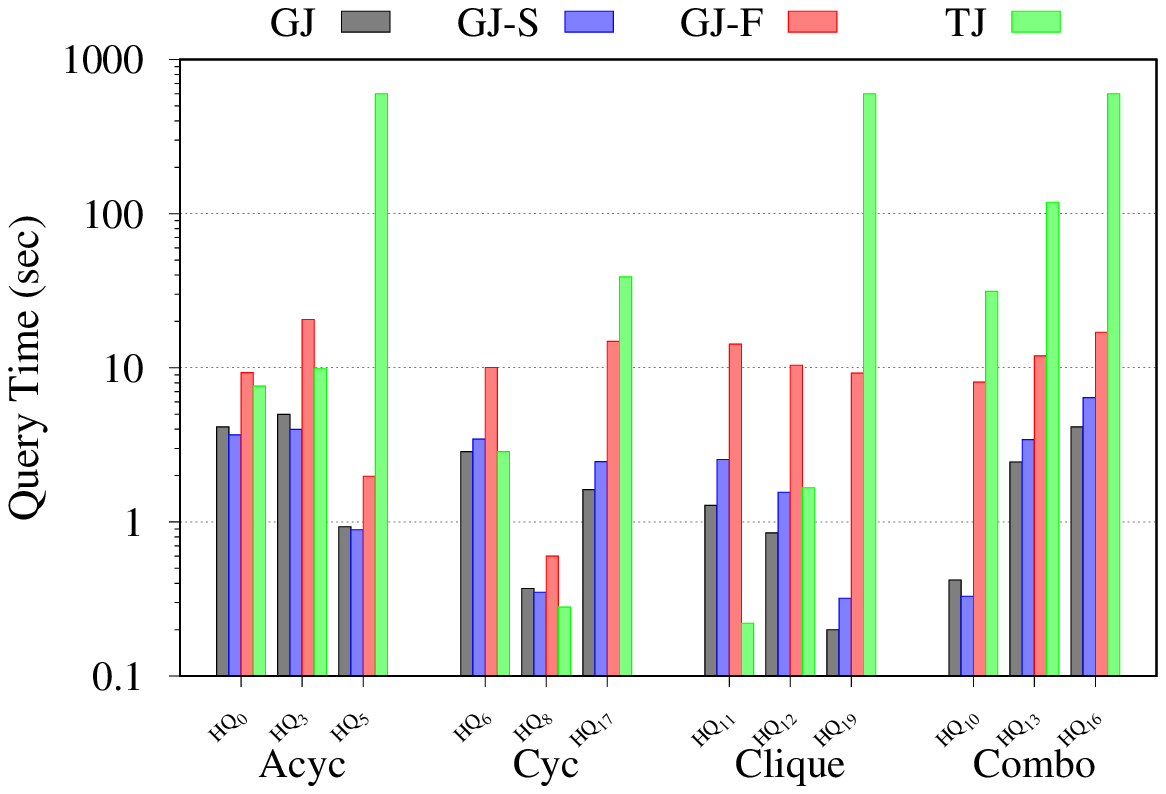} } }

        \caption{Size and construction time of summary query graphs and total query time by different algorithms on {\em ep}.}
      \label{fig:epagsizeandtime}
      \vspace*{-4ex}
\end{figure*}

\vspace*{1ex}\noindent\textbf{RIG size and construction time.} In this experiment, we examine the size and construction time of RIG graphs for different queries on data graphs. As usual, the size of a graph is measured by the total number of nodes and edges. For this, we design two variants of {\em GM}, referred to here as {\em GM-S} and {\em GM-F}, respectively. Unlike {\em GM}, {GM-S} does not apply the node pre-filtering procedure \cite{chen05dags,ZengH12} before computing ${\cal FB}$ simulation and constructing RIG, whereas {\em GM-F} does not compute ${\cal FB}$ simulation, but it applies the node pre-filtering to prune nodes from the inverted lists, then builds a RIG based on the pruned inverted lists. {\em TM} builds also an auxiliary data structure (it is called answer graph in \cite{WuTSL20}) for the tree of the given graph query.

We compare the resulted graph size and the construction time by the aforementioned four algorithms, and compare also the effectiveness of constructed graphs for assisting query evaluation by those algorithms. Fig.\ref{fig:epagsize}, Fig.\ref{fig:epagtime}, and Fig.\ref{fig:epetime} report the results on evaluating H-queries on graph {\em ep}, respectively. As before, only the results of three queries from each of the acyclic, cyclic, clique, and combo pattern classes are reported in the figures. The Y-axis of Fig.\ref{fig:epagsize} shows the constructed graph size as a percentage of the input data graph size.
%The reported construction time for each query does not include the node pre-filtering time and the simulation computation time.

Both {\em GM} and {\em GM-S} generate the smallest graph for all the cases, the average percentage of the generated graph size over the size of graph {\em ep} is around 0.4\%.  Notice that query $HQ_{19}$ has empty answer on {\em ep}, and this case is well detected by  {\em GM} and {\em GM-S} at the early stage of query evaluation, since the size of RIG constructed by the two is 0. Except on $HQ_{17}$ and $HQ_{19}$, {\em GM-F} produces the largest graph in all the other cases. The average percentage of RIG size over the size of graph {\em ep} is around 4.2\% for {\em GM-F} .  This demonstrates that the double simulation technique has much better pruning power than the node pre-filtering. In many cases, the size of graphs constructed by {\em TM} is much smaller than that of {\em GM-F}, around 0.79\% over the size of graph {\em ep}. However, the size of graphs constructed by {\em TM} on $HQ_{17}$ and $HQ_{19}$ is around 45$\times$ and 10$\times$ over that by {\em GM-F}, respectively. Also, the graph size on $HQ_{16}$ is not available for {\em TM}, since it reports time-out error when constructing the graph.

% comment on the building time of answer graph
% time and space complexity
% theoretic bound

From Fig.\ref{fig:epagtime}, we observe that {\em GM} and {\em GM-S} have similar graph construction time for each query, the average time being 1.8 sec and 2.2 sec, respectively. The average graph construction time of {\em GM-F} is about 4.5$\times$ larger than that of {\em GM}. Excluding $HQ_{16}$, the average construction time taken by {\em TM} is around 5.7 sec.

From Fig.\ref{fig:epetime}, we can see that {\em GM} has the overall best performance; {\em GM-S} runs a little bit slower than {\em GM}; whereas {\em GM-F} is around 4.3$\times$ slower than {\em GM}. The result demonstrates that smaller sized RIG can provide more speedup on query evaluation for the {\em GM} approach. {\em TM} gives the worst time performance as expected.

\begin{figure}[!t]
    \centering%
     \scalebox{.75}{ \epsfig{file=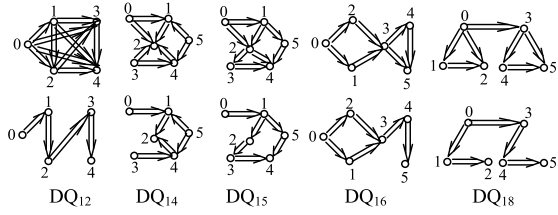}}%
       %\vspace*{-1ex}
     \caption{Examples of transitive reduction of D-queries.}
      \label{fig:transReductQs}
\vspace*{-1ex}
\end{figure}

\begin{figure}[!t]
	\center
    \subfigure[{\em em}]{ \scalebox{0.46}{ \label{fig:emtrtime} \epsfig{file=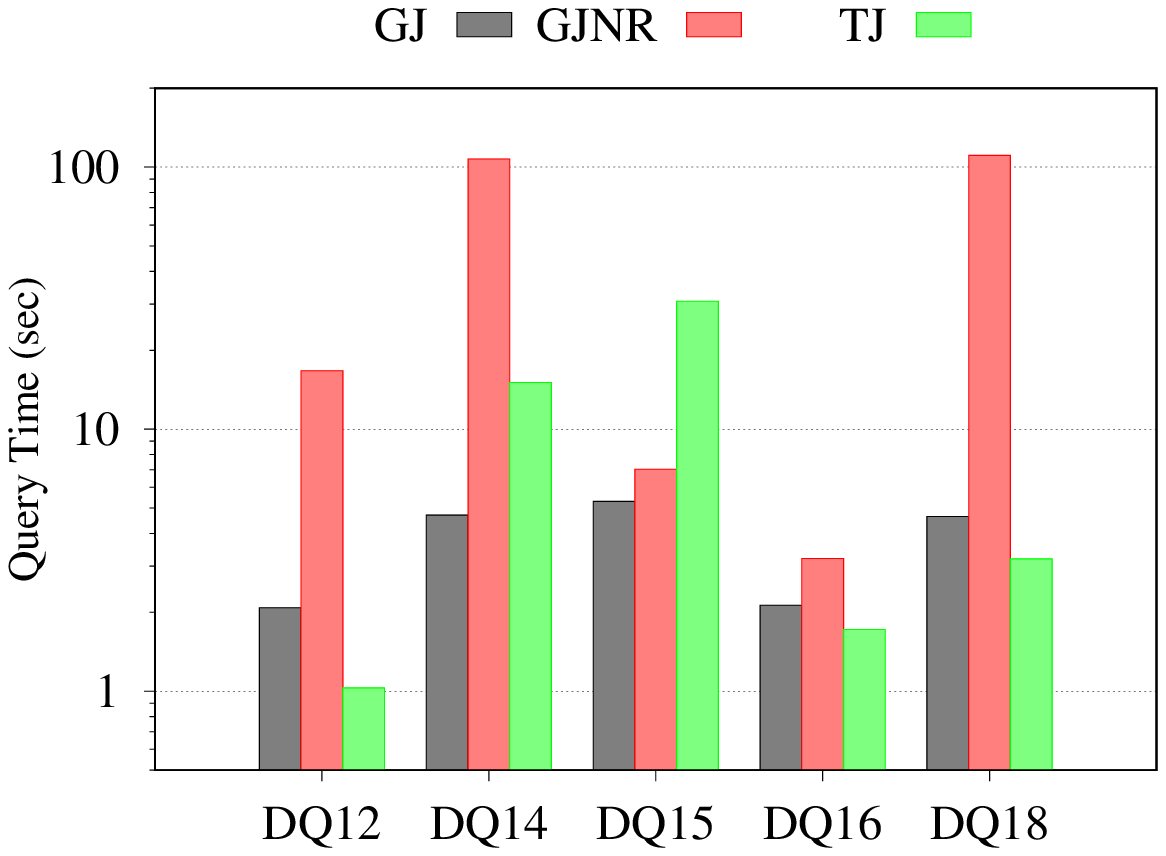} } }
    \subfigure[{\em ep}]{ \scalebox{0.46}{ \label{fig:eptrtime} \epsfig{file=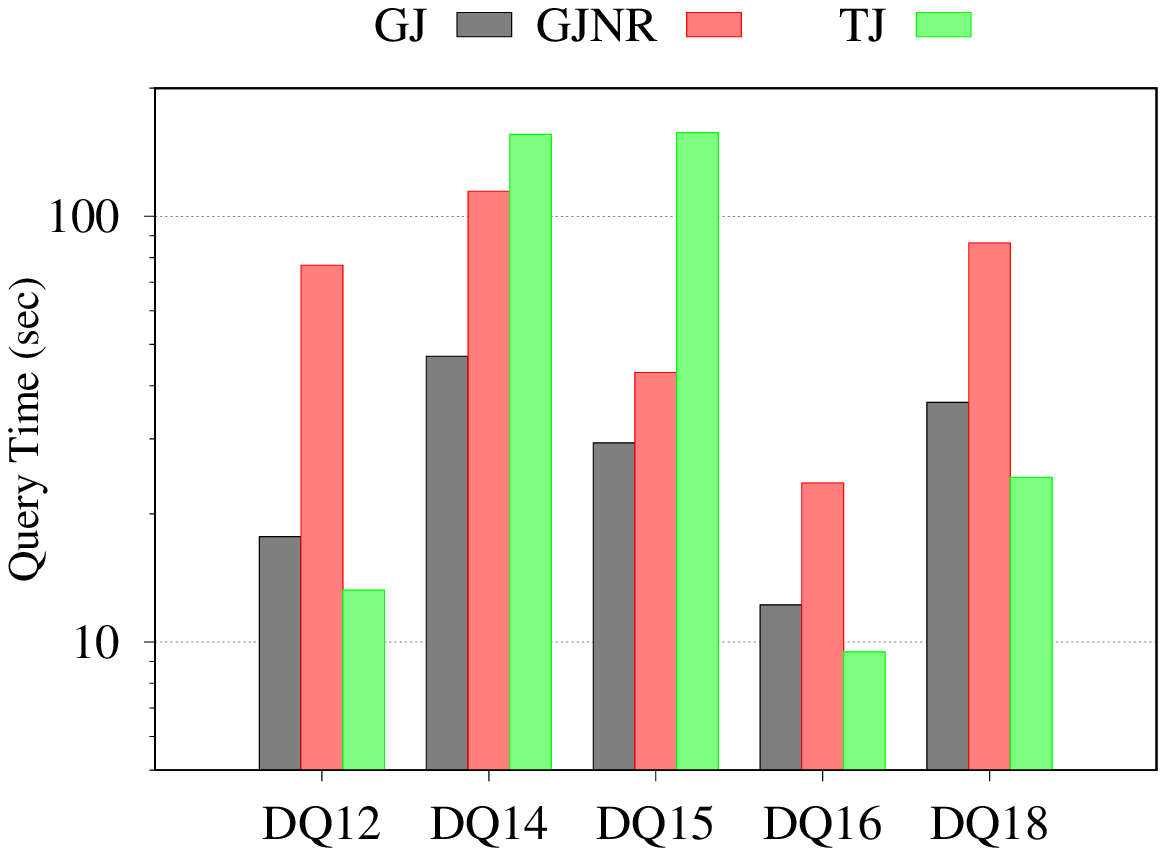} } }

   \caption{D-query evaluation time on {\em em} and {\em ep} with and without transitive reduction.}
	\label{fig:emeptrtime}
\end{figure}

\vspace*{1ex}\noindent\textbf{Pattern transitive reduction.} In this experiment, we examine the effectiveness of pattern transitive reduction (Section \ref{sec:reduction}) on reducing query evaluation time. Fig.\ref{fig:transReductQs} shows five D-queries and their respective transitive reduction form. We evaluate the D-queries and their reduced forms using the {\em GM} algorithm.

Fig. \ref{fig:emeptrtime} reports the query time of these queries on graphs {\em em} and {\em ep}, where the time for reduced (resp. no-reduced) queries is denoted as {\em GM} (resp. {\em GM-NR}). For comparison, we report also the query time of {\em TM} for the reduced queries. We observe that the pattern transitive reduction technique is very effective on reducing query evaluation time. On graph {\em em}, the average evaluation time of {\em GM-NR} is about 12$\times$ slower than {\em GM}, and it is even outperformed by {\em TM} by a factor of 3.7$\times$. The average speedup of {\em GM} over {\em GM-NR} on {\em ep} is about 1.4$\times$.

\vspace*{1ex}\noindent\textbf{Search order.} In this experiment, we compare the effectiveness of three search ordering strategies $JO$, $RI$, and $BJ$ for homomorphic pattern matching. Both $JO$ and $RI$ are described in Section \ref{subsec:alg}. $BJ$ finds an optimal left-deep join plan of the given query through dynamic programming. We integrate $JO$, $RI$, and $BJ$ with Algorithm {\em MJoin} which is used in {\em GM} to enumerate query occurrences (Section \ref{subsec:alg}).

Tab.\ref{tab:searchorder} shows the experiment results on the three ordering strategies. Except for query $HQ_{15}$ on graph {\em em}, {\em GM-JO} shows the best performance and {\em GM-BJ} comes the next. {\em GM-RI} does not perform well on most of these H-queries, even it is regarded as an effective technique in recent research on subgraph isomorphism matching \cite{Sun020}.

$BJ$ is able to find a good query plan, but it does not scale to large queries with tens of nodes \cite{NeumannR18}. Unlike $JO$ which uses cardinalities of node sets in RIG graphs to do cost estimation, $RI$ generates a search order based purely on the topological structure of the given query, independently of any target data graph. The experiment result demonstrates that an effective search ordering strategy for homomorphic pattern matching should take into account both query graph structure and data graph statistics.

\begin{table}[!t]
\caption{Effectiveness of search ordering methods.}
\label{tab:searchorder}
\vspace*{-3ex}
\begin{center}
\scalebox{.7}{\begin{tabular}{|l|c c c|c c c|}
\hline
Query &  \multicolumn{3}{|c|}{{\em em}} &	\multicolumn{3}{|c|}{{\em ep}} \\

   &   {\em GM-RI}	&{\em GM-JO} & {\em GM-BJ} & {\em GM-RI}&{\em GM-JO} & {\em GM-BJ} \\	

\hline\hline
$HQ_2$	&3.64	&1.88	&2.45	&7.00	&2.02	&2.09\\
$HQ_3$	&76.94	&53.75	&53.75	&90.92	&40.15	&41.71\\
$HQ_4$	&3.06	&1.05	&1.05	&4.67	&0.67	&0.88 \\
$HQ_{15}$	&1.33	&7.32	&1.79	&14.77	&2.22	&3.01 \\
$HQ_{18}$	&7.07	&0.99	&1.36	&441.94	&30.18	&38.15 \\
\hline
\end{tabular}}
\end{center}
\vspace*{-2ex}
\end{table}

\subsection{Comparison to Systems and Engines}

We compare the performance of {\em GM} with four recent query engines/systems including EmptyHeaded \cite{AbergerTOR16} (abbreviated as {\em EH}), {\em GraphFlow} \cite{MhedhbiKS21} (abbreviated as {\em GF}), {\em RapidMatch} \cite{SunSC0H20} (abbreviated as {\em RM}) and Neo4j. Neo4j is the most popular graph DBMS and {\em EH} is one of the most efficient graph database systems, whereas the evaluation results of \cite{MhedhbiKS21} show that {\em GF} largely outperforms Neo4j and {\em EH}.  All these three engines/systems were designed to process graph pattern queries whose edges are mapped with homomorphisms to edges in the data graph (therefore, they do not need a reachability index).

\vspace*{1ex}\noindent\textbf{Comparison with {\em GF}, {\em EH} and  Neo4j on C-queries.}  We first compare  {\em GM} with the three engines on evaluating C-queries. As a join-based method,  {\em GF} enumerates and optimizes join plans based on a cost model. In order to estimate join plan costs,  for each input data graph, {\em GF} constructs a catalog containing entries on cardinality information for subgraphs. When loading the data graph to main memory, {\em GF} builds some index on edges and nodes to facilitate query occurrence enumeration.

\begin{figure}[t]
\center
 \subfigure[Catalog building time in seconds of  {\em GF} on different data graphs.]{ \scalebox{0.9}{
\begin{tabular}{|c|c|c|c|c|c|c|c|}
\hline
{\em em}	& {\em ep}	&{\em hp} &{\em	yt}	& {\em hu} &{\em bs} &	{\em go} & {\em	am} \\
\hline
OM&	OM&	OM&	17686.46&	4603.86	&40.99	&1.33	&0.34\\
\hline
\end{tabular}
\label{tab:gfcat}
}
}

 \subfigure[Query evaluation time]{ \scalebox{.6}{ \label{fig:simvsgftime} \epsfig{file=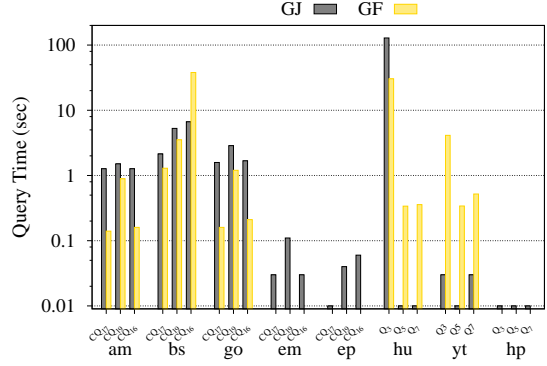}

 } }
 \caption{Performance comparison of {\em GM} and {\em GF} for C-query evaluation.}
      \label{fig:simvsgf}

  \vspace*{-2ex}
\end{figure}

Fig. \ref{tab:gfcat} shows the catalog building time in seconds of {\em GF} on different data graphs. We observe that, on graphs {\em em, ep} and {\em hp}, {\em GF} reported out-of-memory error when building catalogs and failed to evaluate any queries on these graphs. On graph {\em yt} and {\em hu}, the catalog construction took around 4.9 hours and 1.3 hours, respectively.

Fig. \ref{fig:simvsgftime} shows the query time of {\em GM} and {\em GF}  for evaluating C-queries on graphs {\em am, bs, go, hu} and {\em yt}. We did not compare with {\em EH} and  Neo4j here,  since the evaluation results of \cite{MhedhbiKS21} show that {\em GF} largely outperforms Neo4j and {\em EH} for evaluating similar graph queries on these graphs. In this experiment,  {\em GM} and {\em GF} enumerate all the matchings for each query.  We observe that, for queries on graphs {\em am, bs} and {\em go} which have a small number of labels ($\leq 5$), {\em GF} outperforms {\em GM} by 6.2$\times$, 1.2$\times$, and 11$\times$, respectively. However, on graphs {\em hu} and {\em yt} with a relatively large number of labels, {\em GF} is up to orders of magnitude slower than {\em GM}.

We now compare {\em GM} with {\em EH} and  Neo4j to evaluate C-queries on {\em em} and {\em ep}, where {\em GF} failed to evaluate any queries. The results are shown in Table \ref{tab:emvsehneo4j}. Our measurements for {\em EH} include the time on its precomputation step, without any disk IO that is done at this step. For reference, we also provide measurements for {\em EH} that exclude this precomputation step (abbreviated as {\em EH-Probe}). To compare with Neo4j, we used Neo4j v.4.2.1 and expressed queries in Cypher, its graph query language.

%In general, the performance advantage of worst-case optimal join plans rapidly increases as the complexity of the graph pattern queries grows. This is to be expected, as more complex pattern queries result in more intermediate results that can explode when using a binary join plan.
We observe that {\em GM} consistently exhibits the best runtime across all data sets and queries, outperforming {\em EH} by up to two orders of magnitude on the queries that can be completed by {\em EH}.  Although EmptyHeaded is a highly optimized system, its performance is negatively impacted by its optimization and compilation overhead as well as its expensive precomputation step.

As shown in Table \ref{tab:emvsehneo4j}, {\em GM} is significantly faster than Neo4j (up to 388$\times$) on the queries that can be finished by Neo4j.  Neo4j uses binary joins to evaluate queries and it is not optimized for complex graph pattern queries we study in this paper. The results clearly demonstrate the advantages of our proposed graph pattern evaluation approach.

\begin{table*}[!t]
\caption{Runtime (sec) of EmptyHeaded ({\em EH}) \cite{AbergerTOR16}, Neo4j and {\em GM} for C-queries on graphs {\em em} and {\em ep}. (Notations: OM = out of memory, FA = failure, TO = timeout)}
\label{tab:emvsehneo4j}
\vspace*{-2ex}
\begin{center}
\scalebox{.66}{\begin{tabular}{|l|l|c c c|c c c|c c c|c c c|}
\hline
Dataset &  Algorithm  &\multicolumn{3}{|c|}{Acyc} &	\multicolumn{3}{|c|}{Cyc} 	&  \multicolumn{3}{|c|}{Clique} & 	\multicolumn{3}{|c|}{Combo}\\

   &   & $CQ_0$	&$CQ_3$	&$CQ_5$ &$CQ_6$	&$CQ_8$	&$CQ_{17}$	&$CQ_{11}$	&$CQ_{12}$	&$CQ_{19}$	&$CQ_{10}$	&$CQ_{13}$	&$CQ_{16}$\\	

\hline\hline
{\em em}	& EH-probe &0.25 &0.28 &0.28 &0.16 &0.24 &OM &0.16 &0.16 &OM &0.26 &0.17 &FA\\
        & EH	&4.09 &10.66 &10.67 &2.89 &3.89	&OM	&4.61 &93.30 &OM &4.18 &20.19 &FA\\
        & Neo4j	&0.33 &13.80 &34.92	&1.08 &0.47	&0.49 &3.30	&TO	&4.09 &0.29	&0.31 &2.20\\
        & GM	&0.10 &	0.12 &0.09	&0.11 &0.10	&0.13 &0.16	&0.12 &0.39	&0.12 &0.14	&0.14\\
\hline
{\em ep}	& EH-probe &0.12 &0.12	&0.13 &0.06	&0.12 &TO &0.06	&0.06 &TO &0.11	&0.07 &OM\\
        & EH	&4.00  &10.44 &10.49 &2.84 &3.81 &TO  &4.58	&90.95	&TO	&4.12 &20.26 &OM\\
        & Neo4j	&0.09  &1.07  &0.31	&0.07 &0.09	&0.41 &0.07	&0.50 &0.96	&0.07 &0.10	&0.18\\
        & GM	&0.07  &0.10  &0.03	&0.08 &0.08	&0.05 &0.05	&0.09 &0.14	&0.07 &0.06	&0.06\\
\hline
\end{tabular}}
\end{center}
\vspace*{-2ex}
\end{table*}

\vspace*{1ex}\noindent\textbf{Comparison with {\em RM} on C-queries.} {\em RM} is a recent graph query engine which outperforms the state of the art graph matching approaches CFL \cite{BiCLQZ16}, DAF \cite{HanKGPH19} and {\em GF} \cite{MhedhbiS19,MhedhbiKS21}. {\em RM} is a tree-based graph query engine that adopts worst-case optimal (WCO) style joins in its result enumeration. It proposes a sophisticated search order method based on the nucleus decomposition of query graphs. To improve its enumeration efficiency, {\em RM} adopts several optimizations, including advanced set intersection methods \cite{Veldhuizen14,Han0Y18},  the intersection caching \cite{MhedhbiS19,MhedhbiKS21}, and the failing set pruning \cite{HanKGPH19}.

We compare {\em RM} with two variants of {\em GM}, denoted as {\em GM-JO} and {\em GM-RI}, which use $JO$ and $RI$ as their respective search order method. We use the recommended configuration for {\em RM}\footnotemark[\value{footnote}] to compute homomorphic matches. We follow also the experimental setting described in \cite{SunSC0H20} by setting the time limit as 5 minutes and the number of matches as $10^5$. We run the experiment on the data graphs and query workloads used in \cite{SunSC0H20}. As {\em RM} considers undirected graphs, to compare with {\em RM}, we store each edge of data graphs in both directions and use them as input to {\em GM} for the experiment. Each data graph has one dense query set (the degree of each query node is at least 3) and one sparse query set (the degree of each query node is less than 3). Each set contains 200 connected*** query graphs with the same number of nodes.

\begin{figure}[!t]
	\center
 %\captionsetup[subfigure]{aboveskip=-1pt,belowskip=-1pt}
 \hspace{-1.5em}%
    \subfigure[Dense query sets]{ \scalebox{0.35}{ \label{fig:huctimedenudg} \epsfig{file=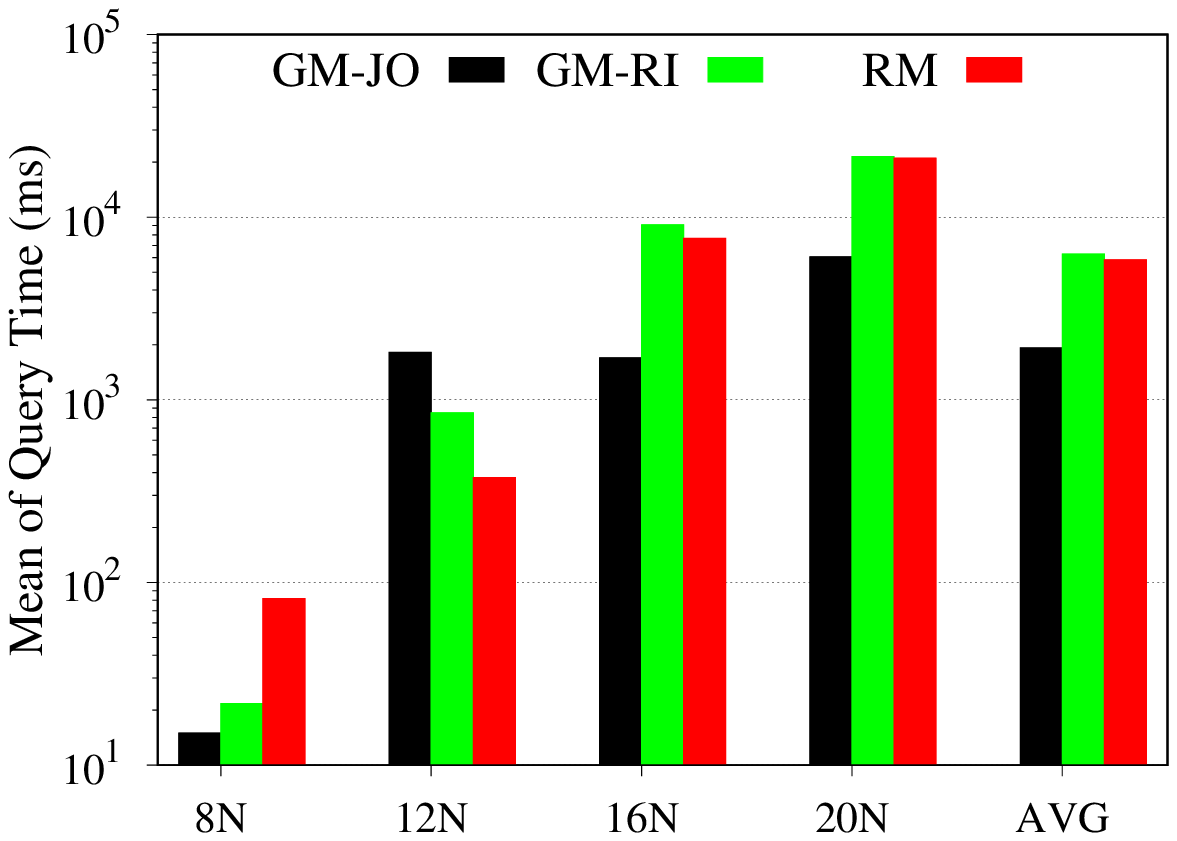} } } \hspace{-1.5em}%
    \subfigure[Sparse query sets]{ \scalebox{0.35}{\label{fig:huctimespaudg} \epsfig{file=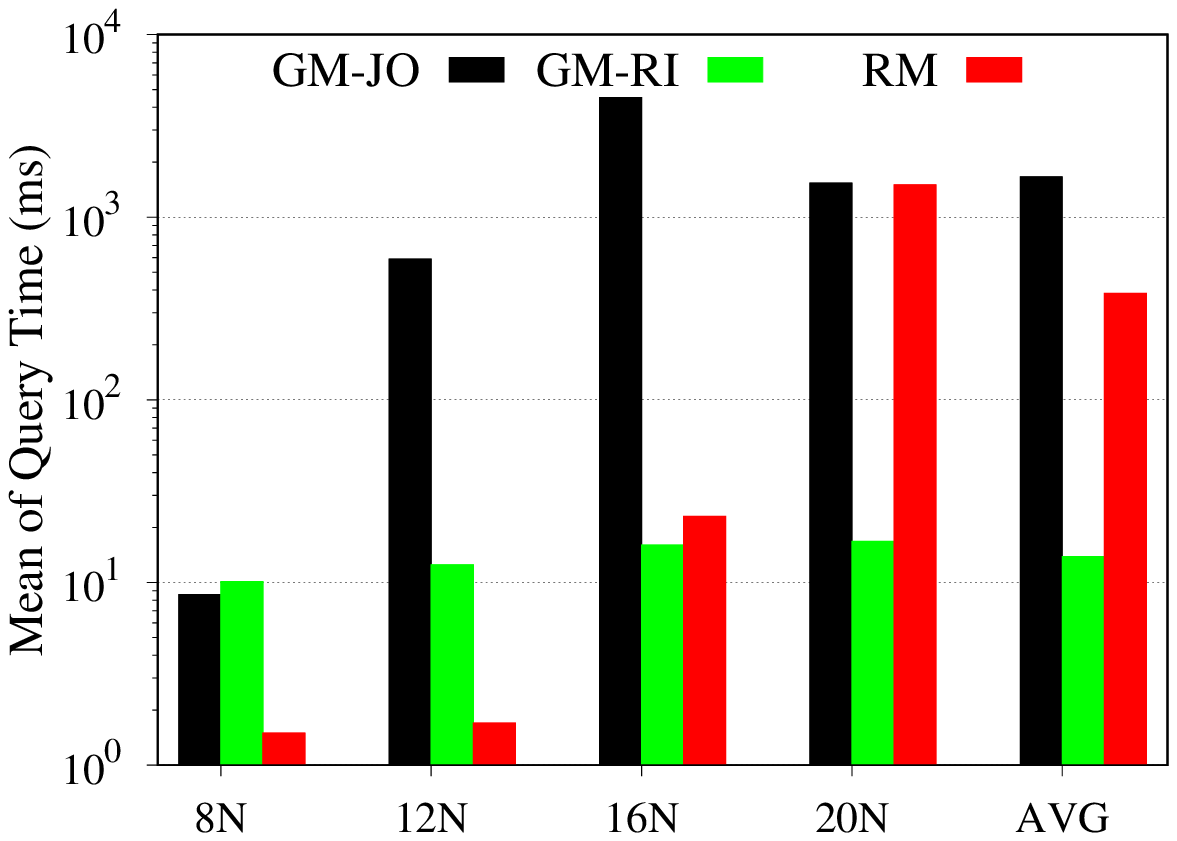} } }
 \hspace{-1.5em}%
     \vspace*{-2ex}
     \caption{Performance comparison of {\em GM} with {\em RM} for large queries on the undirected Human data graph.}
\vspace*{-2ex}
	\label{fig:huctimeudg}
\end{figure}

Figs.~\ref{fig:huctimedenudg} and \ref{fig:huctimespaudg} present the mean of query time on different query sizes of dense and sparse query sets, respectively, on the Human data graph. We choose the Human data graph since it is a real dataset and it is very dense and most of its nodes have the same label making the graph matching particular challenging \cite{Sun020,SunSC0H20}. The number of nodes for queries on the graph varies from 8 to 20. Query sets with $i$ nodes are denoted as $i$N.  For the dense query sets, {\em GM-JO} has, overall, the best performance. It runs more than 2 times faster than {\em RM} on average. {\em GM-RI} runs slightly slower than {\em RM} on average. In contrast, for the sparse query sets, {\em GM-JO} gives the worst performance, while {\em GM-RI} achieves up to two order of magnitude average speedup than the other two algorithms.

We notice that the execution time of the three algorithms is dominated by the enumeration time, which in turn is determined by selected search orders for queries. One algorithm runs much slower than competing algorithms because it generates ineffective search orders for a number of queries. Both $RI$ and the search order method of {\em RM} prioritize the dense sub-structure of a query $Q$ since they assume that the dense sub-structures of $Q$ generally appear less frequently in the data graph. When this assumption in the heuristic search rule does not hold on the workloads, they can generate ineffective searching orders. This often happens for dense queries as our results demonstrate. $JO$ performs well on dense queries, but worse on sparse ones, because the cardinality differences of candidate occurrence sets among query nodes tend to be very small for sparse queries, this makes it difficult to choose an effective search order. Advanced subgraph cardinality estimation methods \cite{ParkKBKHH20} can help $JO$ to improve its search order quality.

In summary, our experimental results on Human graph demonstrate that {\em GM} with two simple search order methods beats the highly optimized {\em RM} that comes with a sophisticated search order method in most tested workloads.

\vspace*{1ex}\noindent\textbf{Comparison with {\em GF} and Neo4j on D-queries.} We also ran experiments to compare the performance of these approaches for evaluating D-queries (reachability graph pattern queries). Before evaluating D-queries, {\em GM} builds a reachability index on the input data graph $G$ using the BFL (Bloom Filter Labeling) algorithm \cite{SuZWY17}. To express reachability relationships for descendant query edges in Neo4j, we use one Neo4j APOC (Awesome Procedures On Cypher) procedure, which expands to subgraph nodes reachable from the start node following relationships to max-level adhering to the label filters.

%Neo4j can express descendant query edges using the infinite variable-length pattern matching construct of Cypher. However, Cypher's mode of expansion with the infinite variable-length pattern matching construct will attempt to find all possible paths matching a descendant edge.  This is not ideal for a reachability query on  highly connected graphs, as the number of possible unique paths in the graph to every other node in the graph can become huge.

% Reachability queries characterize connectivity by the existence of a path of arbitrary length

%To express reachability relationships for descendant query edges in Neo4j, we use the APOC (Awesome Procedures On Cypher) procedures.  Specifically, to express a descendant query edge, we use the APOC procedure that expands to subgraph nodes reachable from the start node following relationships to max-level adhering to the label filters.

%Specifically, we express descendant query edges in Cypher with {\em the expand to nodes in  subgraph procedure}, which expands to subgraph nodes reachable from the start node following relationships to max-level adhering to the label filters.

Since {\em GF} was unable to match edges to paths in data graph, we designed an indirect way for it to evaluate D-queries: first generate the transitive closure $G'$ on the input graph $G$ and then use $G'$ as the input data graph. In the experiments, we used the Floyd–Warshall algorithm to compute the transitive closure of the data graph. We omit the comparison with {\em EH} due to its worse performance on C-queries than {\em GM}.

Fig. \ref{fig:gmgfneo4jdqs} presents the results on fragments of {\em em} graphs.  Specifically, Table \ref{fig:gmgfneo4jdqs}(a) shows the time for constructing the BFL reachability index, the graph transitive closure and the catalog on Email graphs with different numbers of labels and nodes. As we can see, the time for constructing BFL indices remains very small for graphs of different sizes (it is 0.38 sec. for the original {\em em} graph with 265K nodes). In contrast, the transitive closure construction time grows very fast as the number of graph nodes increases. For a graph with 3k nodes, the transitive closure construction takes more than one hour. We observe also that the catalog construction is affected enormously by the growing cardinality of the label and node sets of the graph.

We used only 1k-sized Email graphs (with different numbers of labels) to compare the query time of {\em GM}, {\em GF} and Neo4j, because: (1) the time for building transitive closures and catalogs on large-sized graphs for {\em GF} is prohibitive large, and (2) Neo4j was unable to finish within 1 hour for most queries on larger graphs. We evaluated five representative D-queries $DQ_0$, $DQ_4$, $DQ_{14}$, $DQ_{15}$, and $DQ_{16}$. All the three approaches are configured to find all the query matchings.%, while Neo4j uses the count function of Cypher to return the number of solution tuples.

%Neo4j was unable to solve any query within 10 minutes.

Table \ref{fig:gmgfneo4jdqs}(b) shows only the query time of the three for $DQ_4$, $DQ_{15}$, and $Q_{16}$. The results for the other queries are similar.  We observe that {\em GF} performs better than {\em GM} on the {\em em} graph with 5 labels. However, {\em GM} greatly outperforms {\em GF} when the number of labels increases from 10 to 20. Note that in reporting the query times of  {\em GF}, the transitive closure and the catalog construction times and ignored since otherwise {\em GF} underperforms {\em GM} by several orders of magnitude if feasible at all. Neo4j shows the worst query performance in all the cases. However, unlike {\em GF}, Neo4j can evaluate reachability queries directly.

\begin{figure}[!t]
\centering
  \subfigure[Building time (sec.) of BFL, transitive closure (TC) and catalog (CAT).]{
 \scalebox{0.62}{
    \begin{tabular}{|c|c|c|c|c|c|c|}
\hline
\#lbs  & \#nodes & BFL(sec.)& TC(sec.) &CAT(sec.) \\
\hline\hline
5	 &1k  &0.01& 22.95	&5.52	\\
10   &1k  &0.01& 22.67	&10.84	\\
15   &1k  &0.01& 23.07	&55.97	\\
20   &1k  &0.01& 23.58	&323.92	\\
20   &2k  &0.01& 207.93	&outOfMemory\\
20   &3k  &0.02& 765.65	&outOfMemory\\
20   &5k  &0.03& 4042.62	&outOfMemory\\
\hline
\end{tabular}
       }}
    \hskip 0.5ex
  \subfigure[Query time (sec.) on Email graphs of 1K nodes.]{
   \scalebox{0.62}{
    \begin{tabular}{|c|c|c|c|c|c|}
\hline
Query  & Alg. & \#lbs=5 &\#lbs=10 &\#lbs=15 &\#lbs=20 \\
\hline\hline
$DQ_4$	 &Neo4j &981.64	&93.462	&5.599	&3.852	\\
         &{\em GF} &0.27	&0.12	&0.09	&0.09	\\
         &{\em GM} &1.12	&0.1	&0.01	&0.01\\
\hline \hline
$DQ_{15}$	 &Neo4j &$>1h$	&2395.089	&573.407 &48.188\\
         &{\em GF} &2.69	&0.26	&0.38	&0.39	\\
         &{\em GM} &13.84	&0.31	&0.03	&0.03\\
\hline \hline
$DQ_{16}$	 &Neo4j &$>1h$	&2944.943 &542.899 &30.705\\
         &{\em GF} &0.70	&0.25	&0.20	&0.36	\\
         &{\em GM} &4.34	&0.11	&0.07	&0.01\\
\hline
\end{tabular}

    }}

\vspace*{-2ex}
 \caption{Comparison of {\em GM}, {\em GF} and Neo4j for reachability graph pattern queries on Email graphs.}
    \label{fig:gmgfneo4jdqs}
    %\vspace*{-1ex}
\end{figure}

\vspace*{1ex}\noindent\textbf{Comparison with Neo4j on H-queries.} Finally, we compare {\em GM} and Neo4j on evaluating hybrid graph pattern queries. Neither {\em RM}, {\em GF} nor {\em EH} is able to evaluate hybrid queries.  As for D-queries, we use the Neo4j APOC procedure to express H-queries in Cypher.   The results on a fragment of {\em em} graph with 30K nodes are shown in Table \ref{tab:gjneo4jhq}. As we can see, {\em GM} is again significantly faster than Neo4j on all the queries.

\begin{table*}[!t]
\caption{Runtime (sec) of Neo4j and {\em GM} for H-queries on a fragment of {\em em} graph with 30K nodes.}
\label{tab:gjneo4jhq}
\vspace*{-2ex}
\begin{center}
\scalebox{.6}{\begin{tabular}{|l|l|c c c|c c c|c c c|c c c|}
\hline
Dataset &  Alg.  &\multicolumn{3}{|c|}{Acyc} &	\multicolumn{3}{|c|}{Cyc} 	&  \multicolumn{3}{|c|}{Clique} & 	\multicolumn{3}{|c|}{Combo}\\

   &   & $HQ_0$	&$HQ_3$	&$HQ_5$ &$HQ_6$	&$HQ_8$	&$HQ_{17}$	&$HQ_{11}$	&$HQ_{12}$	&$HQ_{19}$	&$HQ_{10}$	&$HQ_{13}$	&$HQ_{16}$\\	

\hline\hline
{\em em}& Neo4j	&51.952	& 457.034 & $>1h$	&60.119	&35.86	&118.709 &54.104 & $>1h$	& $>1h$ &319.064 & $>1h$& 476.426\\
        & GM	&0.29	& 0.22	&0.32	&0.09	&0.05	&0.02    &0.02	 & 0.02	& 0.04	&0.04 &1.31 &0.16\\
\hline
\end{tabular}}
\end{center}
\vspace*{-2ex}
\end{table*}

Overall, {\em GM} is much more efficient than {\em GF}, {\em EH} and Neo4j on evaluating graph pattern queries. To determine node reachability in graphs, it does not need to compute the graph transitive closure. Instead, it uses a reachability index (BFL) which can be computed efficiently.  Also, unlike {\em GF} which relies on statistics (i.e., catalogs) that are prohibitively expensive to compute, {\em GM} uses a RIG graph that can be built efficiently on-the-fly during query processing and does not have to be materialized on disk.

% \label{sec:experiments}
\section{Related Work}
\label{sec:related}

We review related work on graph pattern query evaluation algorithms. Our discussion focuses on in-memory algorithms that find all occurrences from a graph pattern to a single large data graph. We categorize the related work by the type of morphism used to map the patterns to the data structure. The morphism determines how a pattern is mapped to the data graph: a homomorphism is a mapping from pattern nodes to data graph nodes, while an isomorphism is a one-to-one function from pattern nodes to data graph nodes.
An isomorphism must be a homomorphism, but not vice versa. Note that when each query node has a distinct label, both are identical due to the label constraint in the definition. We also briefly review recent work on graph simulations.

%\vspace*{.5ex}\noindent\textbf{Isomorphic mapping algorithms.}
\vspace*{.5ex}\noindent\textbf{Isomorphic graph matching algorithms.} A recent survey \cite{Sun020} studies the performance of representative in-memory isomorphic mapping algorithms. It puts them in a common framework to compare them on their individual techniques including methods of nodes filtering, matching order, result enumeration and optimization techniques.

Specifically, the majority of algorithms for isomorphic mapping adopt a backtracking method to find the query answer \cite{Ullmann76}, which recursively extends partial matches by mapping query nodes to data nodes. Many of the earlier algorithms directly explore the input data graph to enumerate all results. Several recent ones \cite{HanLL13,BiCLQZ16,HanKGPH19,BhattaraiLH19} adopt the preprocessing-enumeration framework, which performs preprocessing to find, for each query node, the set of possible data nodes (called candidates), and builds an auxiliary data structure (ADS) to maintain edges between candidates. These algorithms usually first generate a BFS tree of the query, and use the tree as the processing unit for building ADS. Then, they use ADS to generate a good matching order. Finally, they enumerate query results with the assistance of ADS along with the matching order.

Isomorphic mapping algorithms employ a variety of filtering methods, exploiting the one-to-one node mapping constraint of isomorphisms,  to reduce the candidates for each query node. Some algorithms design optimization strategies to further reduce the search space during the enumeration. These pruning or optimization techniques however can not be applied to homomorphic pattern match queries since their pruning rules are based on the necessary condition of subgraph isomorphism.

\vspace*{.5ex}\noindent\textbf{Join-based homomorphic graph matching algorithms.} The majority of homomorphic graph matching algorithms consider edge-to-edge mapping. A large number of those algorithms use {\em JM}, the join-based approach,  which converts graph pattern matching to a series of Selinger-style, pair-wise joins.  This is the method used by \cite{NeumannW10,ChengYY11,ZouCOZ12,ZengYWSW13} and in database management systems such as PostgreSQL, MonetDB and Neo4j. Recent theoretical results suggest that Selinger-style, pair-wise join algorithms are asymptotically suboptimal for graph-pattern queries \cite{NguyenABKNRR14}. The suboptimality lies in the fact that Selinger-style algorithms only consider pairs of joins at a time. Consequently, the intermediate results can be more than the maximum output size of a query.

Recently, the development of the worst-case optimal join (WCOJ) changes the landscape. By scanning all relations simultaneously, the running time of WCOJ matches the worst-case output size of a given join query \cite{NgoRR13}. The recent systems utilizing WCOJ \cite{AbergerTOR16,FreitagBSKN20,MhedhbiKS21} significantly outperform the classical relational systems as well as native graph databases such as Neo4j. These WCOJ algorithms compute multiway joins using multiway intersections. The pair-wise join algorithms as well recent WCOJ algorithms consider almost exclusively edge-to-edge mappings between queries and data.

While the design of our query answer enumeration algorithm {\em MJoin} was inspired by WCOJ algorithms, one main difference is that, {\em MJoin} performs multiway intersections on runtime index graphs (RIG), whereas WCOJ algorithms directly enumerates results on data graphs.
%The new algorithms evaluate the multiway join operator in a worst-case optimal manner [7,8,51,52,66], which is provably asymptotically better than the one-pair-at-a-time join paradigm.

\vspace*{.5ex}\noindent\textbf{Homomorphic edge-to-path mapping algorithms.} Homomorphisms for mapping graph patterns similar to those considered in this paper were introduced in \cite{FanLMWW10} (called $p$-hom), which however did not address the problem of efficiently computing graph pattern matches; instead, it used the notion of $p$-hom to resolve a graph similarity problem between two graphs.

Cheng et al. \cite{ChengYY11} proposed an algorithm called R-Join, which is a {\em JM} algorithm. An important challenge for {\em JM} algorithms is finding a good join order. To optimize the join order, R-Join uses dynamic programming to exhaustively enumerate left-deep tree query plans. Due to the large number of potential query plans, R-Join is efficient only for small queries (less than 10 nodes). As is typical with {\em JM} algorithms, R-Join suffers from the problem of numerous intermediate results. As a consequence, its performance degrades rapidly when the graph becomes larger~\cite{LiangZJZH14}. R-Join was used in \cite{ZouCOZ12} as the underlying processing unit for finding matches of bounded graph patterns, where each edge denotes the reachability of a pair of nodes and carries a bound on the length of the paths in a data graph.

A graph pattern matching algorithm called DagStackD was developed in \cite{chen05dags}, which belongs to the {\em TM} category. Given a graph pattern query $Q$, DagStackD first finds a spanning tree $Q_T$ of $Q$, then evaluates $Q_T$, and filters out tuples that violate the reachability relationships specified by the edges of $Q$ missing in $Q_T$. To evaluate $Q_T$, a tree pattern evaluation algorithm is presented. This algorithm decomposes the tree query into a set of root-to-leaf paths, evaluates each query path, and merge-joins their results to generate the tree-pattern query answer. A tree pattern query evaluation method called TPQ-3Hop is presented in \cite{ZengJZ12,LiangZJZH14}. It is designed on top of a hop-based reachability indexing scheme. Reference \cite{WuTSL22} adopts homomorphisms for findings the matches of graph patterns on data graphs, but like the aforementioned algorithms R-Join and DagStackD,  it considers only patterns with reachability edges and does not leverage simulation to prune the graph pattern search space.

\vspace*{.5ex}\noindent\textbf{Graph simulation-based algorithms.} Graph simulation-based graph pattern matching algorithms \cite{FanLMTWW10,FanLMTW11,MaCFHW14} typically work as follows: for each pattern node, initially compute match sets of query nodes. Then, repeatedly refine the match set of each query node by joining with the match sets of all its children and/or parents until the fixpoint of the match set for each query node is reached. Unlike morphisms-based graph pattern matching which is NP-complete,  simulation-based graph pattern matching can be performed in cubic-time. However, simulation and its extensions \cite{FanLMTWW10,FanLMTW11} do not preserve the structural properties in graph pattern matching and therefore, may return excessive undesirable matches.

Simulation-based pruning was proposed very recently as a powerful node pruning technique in graph matching  \cite{MennickeKNKB19}. It uses subgraph simulation or some variants to filter unnecessary tuples before answering queries. Nevertheless, existing simulation-based pruning techniques consider only edge-to-edge matching, and thus cannot support hybrid pattern queries like those we consider in this paper.

\vspace*{.5ex}\noindent\textbf{Regular path query evaluation on graphs.} A regular path query (RPQ) specifies a regular expression $R$ over the edge or node labels in a data graph $G$.  Variations and extensions of RPQs are supported in query languages such as SPARQL\footnote{https://www.w3.org/TR/sparql11-overview/.} and the Cypher language of the Neo4j graph database\footnote{https://neo4j.com/.}. The evaluation of $R$ on $G$ can be defined in several ways, including finding all paths in $G$ \cite{KoschmiederL12} and finding all paths between two fixed nodes in $G$ \cite{WadhwaPRBB19}, such that the sequence of labels along each path forms a word in the language recognized by $R$. The evaluation problem of RPQ on an arbitrary graph is NP-hard \cite{MendelzonW95}.

The work of \cite{FanLMTW11} studied a restricted form of RPQ. Beside considering a restricted subclass of regular expressions,  the restricted RPQ defines the query answer differently from the general RPQ.  Given two nodes in $G$, the general RPQ finds all the paths in $G$ between the two nodes satisfying $R$ (the path finding semantics), whereas the restricted RPQ finds the existence of a path between the two nodes satisfying $R$ (the path existence semantics, also known as node reachability in $G$).  As shown in \cite{FanLMTW11}, the restricted RPQ allows for evaluation in cubic time.  The same work proposed a class of graph patterns where edges are restricted RPQs, and defined graph pattern matching based on a revised notion of graph simulation. Two cubic algorithms were devised to evaluate the proposed graph patterns.

The hybrid graph pattern queries considered in this paper are a subclass of the graph patterns defined in \cite{FanLMTW11}, where each edge is either without any modifier or with a ’*+’ modifier (the latter is mapped to a path of unbounded length in the graph). Unlike \cite{FanLMTW11} however, our graph pattern matching is defined in terms of homomorphisms, which is a NP-complete problem.

%Regular graph pattern queries are queries whose edges are augmented with regular expressions as constraints \cite{FanLMTW11}.
\vspace*{.5ex}\noindent\textbf{Subgraph searching on a graph database.} The problem of subgraph searching on a graph database is defined differently from the problem of the subgraph (or query) matching to a single large graph reviewed above. Given a graph database $D$ containing a collection of small data graphs and a query graph $Q$, the subgraph searching problem aims to retrieve all data graphs in $D$ that contain $Q$. The containment is usually defined as the isomorphism test from $Q$ to a data graph. Most of subgraph searching algorithms proposed in the literature follow an {\em indexing-filtering-verification} paradigm: (1) build an index on substructures or features of data graphs, (2) filter out from $D$ those data graphs whose features do not contain $Q$ as a subgraph; and (3) a isomorphism test is conducted against each remaining data graphs in the verification phase. More recent subgraph searching algorithms \cite{Sun019,KimCP0HH21} utilize efficient subgraph matching techniques to improve the subgraph searching performance.

\vspace*{.5ex}\noindent\textbf{Comparison to our approach.} The problem we address in this paper is  different than those addressed by existing graph pattern matching approaches. We consider general graph patterns and not simply paths or trees. Our patterns contain child and descendant edges, allowing for both edge-to-edge and edge-to-path matches to the data graph. Patterns are mapped to the data graph using homomorphisms which relax the strict one-to-one mapping entailed by isomorphisms and, unlike graph simulation,  preserve the topology of the data graph.

Unlike both {\em JM} and {\em TM}, our graph pattern matching approach is holistic in the sense that it does not decompose the given query into subpatterns, but it rather tries to match the query against the input graph as a whole.  Our approach also follows the preprocessing-enumeration framework, but the specific techniques we propose are largely different from those used by the recent isomorphism algorithms \cite{HanLL13,AbergerTOR16,HanKGPH19,BhattaraiLH19}, since their techniques are based on the necessary condition of subgraph isomorphism.

 	% \label{sec:related}
%\vspace*{-2ex}
\section{Conclusion}
\label{sec:conclusion}

We have addressed the problem of efficiently evaluating hybrid graph patterns using homomorphisms over a large data graph. By allowing \emph{edge-to-path} mappings, homomorphisms can extract matches ``hidden'' deep within large graphs which might be missed by \emph{edge-to-edge} mappings of subgraph isomorphisms.
We have introduced the concept of runtime index graph (RIG) to compactly encode the pattern matching search space. %The RIG can be efficiently built {\em on-the-fly} and does not have to be persisted to disk.
To further reduce the search space,  we have designed a novel graph simulation-based node-filtering technique to prune nodes that do not contribute to the final query answer. We have also designed a novel join-based query occurrence enumeration algorithm which leverages multi-way joins realized as intersections of adjacency lists and node sets of RIG.  We have conducted extensive experimental evaluation to verify the efficiency and scalability of our approach and showed that it largely outperforms state-of-the-art approaches.

Our future work involves investigating pattern matching in a dynamic data graph setting where the matches of the graph pattern are computed incrementally. We are also exploring query optimization techniques to further improve the performance of the hybrid pattern matching approach.

 % \label{sec:conclusion}

\bibliographystyle{abbrv}
%\bibliography{C:/Users/dth/Documents/PAPERS/biblio1}
%\bibliography{mybib}
\small{\bibliography{D:/Research/writing/bib/mybib}}
%\small{\bibliography{mybib}}
%\bibliography{mybib}
\newpage\section*{APPENDIX}  % use *-form to suppress numbering
\label{sec:appendix}

\vspace{1.5ex}\noindent\textbf{Theorem 4.1} {\em Algorithm {\em FB\-SimDag} correctly computes double simulation ${\cal FB}$ of dag pattern $Q$ by data graph $G$.}

\vspace{1.5ex}\noindent\textbf{Proof.} Initially, for each node $q\in V_Q$, its $FB(q)$ is set to be $ms(q).$  Algorithm  {\em FB\-SimDag} incrementally refines $FB$ by disqualifying pairs of nodes violating its definition in a bottom-up way. Nodes in $FB(q)$ for every leaf node $q\in V_Q$ trivially satisfy the forward simulation conditions, whereas nodes in $FB(q)$ for internal nodes $q\in V_Q$ are discarded only when they have no required child/descendant nodes in $V_G$.

%When {\em forwardSim} in the current pass terminates, ${\cal FB}$ is the forward simulation of $Q$ by $G$.  The reason is,
Once the processing on the current pattern node $q$ is done in the current pass, {\em forwardSim} guarantees that every node in $FB(q)$ forward simulates $q$.  Since pattern $Q$ is a dag and nodes of $Q$ are visited in a bottom-up way, node removal from $FB(q)$ will not make nodes in $FB(c)$  of each child node $c$ of $q$ violate the forward simulation condition. Therefore, the resulted  simulation $FB$ is by far the largest binary relation satisfying the forward simulation conditions.

After the bottom-up traversal, {\em FB\-Sim} proceeds the top-down traversal traverse of the nodes of $Q$ to find pairs of nodes of $FB$ violating the conditions of backward simulation. For every source node $q\in V_Q$,  nodes in $FB(q)$ trivially satisfy the backward simulation conditions; for other $q\in V_Q$, nodes in $FB(q)$ are discarded only when they have no required parent/ancestor nodes in $V_G$.

%When {\em backwardSim} in the current pass terminates, ${\cal FB}$ is the backward simulation of $Q$ by $G$.  The reason is that,
Once the processing on pattern node $q$ completes, {\em backwardSim} guarantees that every node in $FB(q)$ backward simulates $q$ in the current pass.  Since pattern $Q$ is a dag and nodes of $Q$ are visited in a top-down way,  node removal from $FB(q)$ will not make nodes in $FB(p)$ of each parent node $p$ of $q$ violate the backward simulation condition. Thus, the resulted simulation $FB$ is by far the largest binary relation satisfying the backward simulation conditions. However, When $q$ has multiple parents in $Q$, the removal of node $v_q \in FB(q)$ can  make some node in $FB(p)$ of a parent $p$ of $q$ fail to satisfy forward simulation conditions.

In general, the reduction of $FB(q)$ in {\em forwardSim} (resp. {\em backwardSim}) can possibly render some nodes in $FB(q)$ violating backward (resp. forward) simulation conditions. Hence, {\em FB\-Sim} needs to repeat the above two processes until $FB$ becomes stable, that is, when no more false positive nodes can be removed from any $FB(q)$. The resulted $FB$ is the largest binary relation satisfying the double simulation conditions.\hfill$\Box$

%\end{proof}

\vspace{1.5ex}\noindent\textbf{Theorem 5.1} {\em The time complexity of Algorithm {\em MJoin} is in $O(nm\prod_{e_j\in E(Q)}|R_j|^{x_j})$ and its space complexity is in $O(n\times MaxCos)$, where $MaxCos$= $max\{|cos(q)|\}$, for all query nodes $q\in V(Q)$, and $cos(q)$ is the candidate occurrence set of node $q$ in $G_Q.$}

\vspace{1.5ex}\noindent\textbf{Proof.} Let $U\subseteq V(Q)$, we denote $E_U$ as the set $\{e_j\in E(Q)|U\cap e_j \neq \emptyset\}$, and denote $\textbf{t}_U$ as the projection of a query occurrence tuple $\textbf{t}$ of $Q$ on $U.$ To prove Theorem \ref{them:complexity}, we utilize the {\em query decomposition lemma} proved by Ngo et al. in \cite{NgoRR13} which can be stated as follows.

\begin{mylemma}
\label{lemma:qdl}
Let $Q$ be the graph pattern query describing the natural join query $R_1\Join \ldots \Join R_m.$ Let $U\uplus W$ be an arbitrary partition of $V(Q)$ with $1 \leq |U| < n$ and $L:=\Join_{e_j\in E_U}\pi_{U}(R_j).$ Then

\begin{equation}
\label{eq:1}
 %\mathlarger{\sum_{t_U\in L}\prod_{e_j\in E_W}} |R_j \leftsemijoin t_U|^{x_j}\leq \mathlarger\prod_{e_j\in E(Q)}|R_j|^{x_j}.
  \displaystyle \sum_{\textbf{t}_U\in L}\left(\displaystyle \prod_{e_j\in E(Q)} |R_j \leftsemijoin \textbf{t}_U|^{x_j}\right)~\leq~\displaystyle \prod_{e_j\in E(Q)}|R_j|^{x_j}.
\end{equation}

\end{mylemma}

%Query $Q$ can be defined as a natural join $\Join_{e_j\in E(Q)}R_j.$
We prove the time complexity of Algorithm {\em MJoin} by induction over the length of $V(Q).$ The base case is when $Q$ has a single node $q$, that is, $n=|V(Q)|$ =1. In this case, $Q$ can be viewed as having one unary relation $R_j$ over the single node $q$, such that $R_j$ equals the candidate occurrence node set $cos(q)$ in $G_Q$. Algorithm {\em MJoin} returns each element in $cos(q)$ as an occurrence tuple of $Q,$ which clearly can be done in time $O(|cos(q)|)= O(|R_j|^{x_j})$, where $x_j \geq 1$ by the definition of fractional edge cover. Hence Theorem \ref{them:complexity} holds trivially.

In the inductive case $n = |V(Q)| \geq 2.$  Let $q_1, \ldots, q_n$ be a search order picked by  Algorithm {\em MJoin}. At the search step $i$,  $1 \leq i \leq n$, we define a partition $U\uplus W$ of $V(Q)$ as $U = \{q_1, \ldots, q_i\}$ and $W = \{q_{i+1}, \ldots, q_n\}.$  Define $L:=\Join_{e_j\in E_U}\Pi_{U}(R_j)$ as in Lemma \ref{lemma:qdl}. Define also $U':= U -\{q_i\},$ and $L':=\Join_{e_j\in E_{U'}}\Pi_{U'}(R_j).$

For each tuple $\textbf{t}_U\in L$ we define a new join query:

\begin{equation}
\label{eq:2}
\begin{split}
 Q[\textbf{t}_U] & := \Join_{e_j\in E_W}\Pi_{W}(R_j \leftsemijoin \textbf{t}_U)
\end{split}
\end{equation}

Then, the original query $Q$ can be written as:

\begin{equation}
\label{eq:3}
\begin{split}
 %Q[\textbf{t}_U] & := \Join_{e_j\in E_W}\Pi_{W}(R_j \leftsemijoin \textbf{t}_U)
   Q & = \displaystyle{\bigcup_{\textbf{t}_U\in L}}(\{\textbf{t}_U\}\times Q[\textbf{t}_U])
\end{split}
\end{equation}

Algorithm {\em MIJoin} computes set $L'$ at the search step $i-1.$  By induction, it takes time $O(m|U'|\prod_{e_j\in E_{U'}}|R_j|^{x_j})$  to compute $L'.$  Using $L'$, {\em MIJoin} computes $L$ at the search step $i.$ For this, {\em MIJoin} first produces set $cos_i$ using the loop in lines 5-7, the time of which is bounded by

\begin{equation}
\label{eq:4}
\begin{split}
        \displaystyle{\bigcap_{e_j\in E_{\{q_i\}}}}\Pi_{\{q_i\}}(R_j)
\end{split}
\end{equation}

By construction this set intersection is computed in time proportional to

%$$|E_U|\mathlarger{\min_{e_j\in E_U}} |\pi_{U}(R_j)|\leq m(\min_{e_j\in E_U}|R_j|)^{\sum_{e_j\in E(Q)}x_j}$$

\begin{equation}
\label{eq:5}
\begin{split}
|E_{\{q_i\}}|\min_{e_j\in E_{\{q_i\}}} |\Pi_{\{q_i\}}(R_j)| & \leq m\left(\min_{e_j\in E_{\{q_i\}}}|R_j|\right)^{\sum_{e_j\in E(Q)}x_j}\\
                                          & = m\displaystyle\prod_{e_j\in E(Q)}\left(\min_{e_j\in E_{\{q_i\}}}|R_j|\right)^{x_j} \\
                                          & \leq m\displaystyle\prod_{e_j\in E(Q)}|R_j|^{x_j}
\end{split}
\end{equation}
since $|R_j|\geq 1$ and $x_j >0$.

%By induction, it takes time $O(m|U|\prod_{e_j\in E_U}|R_j|^{x_j})$  to compute $L$, the answer of query $Q_U.$ 
%By induction, it takes time $O(m|U'|\prod_{e_j\in E_{U'}}|R_j|^{x_j})$  to compute $L'$ at the search step $i-1.$ 
Algorithm {\em MIJoin} then produces $L$ using $L' \times cos_i$ (Lines 8-9). Concretely, for a tuple $\textbf{t}_U'\in L'$ and a $v_i\in cos_i,$  {\em MIJoin} produces a tuple $\textbf{t}_u \in L$  by combining $\textbf{t}_u'$ with $v_i$ (Line 9).

The loop in lines 8-10 computes $Q[\textbf{t}_U]$ for each tuple $\textbf{t}_U\in L.$ Applying the inductive hypothesis, the runtime of the recursive call in line 10 is $O(m(n-|U|)\prod_{e_j\in E_W}|R_j \leftsemijoin \textbf{t}_U|^{x_j})$. By Lemma \ref{lemma:qdl}, the total runtime of the loop in lines 8-10 is proportional to:

\begin{equation}
\label{eq:6}
\begin{split}
   & \displaystyle m(n-|U|)\sum_{\textbf{t}_U\in L}\prod_{e_j\in E_W}|R_j \leftsemijoin \textbf{t}_U|^{x_j}\\
   & \leq m(n-|U|) \displaystyle \prod_{e_j\in E(Q)}|R_j|^{x_j}
\end{split}
\end{equation}

Putting together, from equations (\ref{eq:5}) and (\ref{eq:6}), the total runtime of Algorithm {\em MIJoin} is:

\begin{equation}
\label{eq:7}
\begin{split}
    &\displaystyle{m|U'|\prod_{e_j\in E_{U'}}|R_j|^{x_j} + m(n-|U|)\prod_{e_j\in E(Q)}|R_j|^{x_j} + m\prod_{e_j\in E(Q)}|R_j|^{x_j}} \\
    & = \displaystyle{m|U'|\prod_{e_j\in E_{U'}}|R_j|^{x_j} + m(n-|U|+1)\prod_{e_j\in E(Q)}|R_j|^{x_j}} \\
    %& \leq  mn\displaystyle\prod_{e_j\in E(Q)}|R_j|^{x_j} + m\displaystyle\prod_{e_j\in E(Q)}|R_j|^{x_j} \\
     & = \displaystyle{m|U'|\prod_{e_j\in E(Q)}|R_j|^{x_j} + m(n-|U|+1)\prod_{e_j\in E(Q)}|R_j|^{x_j}} \\
    & \leq  mn\displaystyle\prod_{e_j\in E(Q)}|R_j|^{x_j} 
\end{split}
\end{equation}
since $|R_j|\geq 1$, $x_j >0$ and $|U'| = |U| -1.$

Finally, Algorithm $MJoin$ uses $O(MaxCos)$ space for set $cos_i$ for each node $q_i \in V(Q).$ This completes the proof. \hfill$\Box$

\end{document}